\documentclass[11pt]{article}
\usepackage[margin=1in]{geometry}
\usepackage[T1]{fontenc}
\usepackage[utf8]{inputenc}
\usepackage{graphicx}
\usepackage{booktabs}
\usepackage{longtable}
\usepackage[hidelinks]{hyperref}
\usepackage{float}
\usepackage{caption}
\usepackage{subcaption}
\usepackage{amsmath}
\usepackage{setspace}
\usepackage{pdflscape}
\title{\textbf{Governing Artificial Intelligence:\\ Public Preferences and Regulatory Options}}
\author{Magnus Lundgren\thanks{Department of Political Science, University of Gothenburg. Corresponding author: \texttt{magnus.lundgren@gu.se}} \and Jonas Tallberg\thanks{Department of Political Science, Stockholm University}}
\date{}

\begin{document}
\thispagestyle{empty}
\begin{titlepage}
\thispagestyle{empty}
\setlength{\parskip}{0pt}
\centering
\vspace*{\stretch{2}}

{\LARGE\bfseries Governing Artificial Intelligence:\\[0.35em]
Public Preferences and Regulatory Options\par}

\vspace{\stretch{1.2}}

{\large Magnus Lundgren\textsuperscript{1}\qquad Jonas Tallberg\textsuperscript{2}\par}
\vspace{1.1em}
{\small
\textsuperscript{1}Department of Political Science, University of Gothenburg\\[0.25em]
\textsuperscript{2}Department of Political Science, Stockholm University\par}

\vspace{\stretch{1.6}}

\begin{minipage}{0.86\textwidth}
{\centering\bfseries\large Abstract\par}
\vspace{0.7em}
\setlength{\parskip}{0pt}
\normalsize\noindent Artificial intelligence (AI) is rapidly transforming economies, societies, and polities, raising fundamental questions about how it should be regulated. Policymakers face choices over whether to prioritize innovation or safety, rely on public oversight or private self-regulation, and govern nationally or internationally. Yet little is known about how citizens evaluate these competing priorities. Here we report a conjoint survey experiment conducted in seven countries with diverse political and economic profiles. We find that citizens strongly support regulating AI and generally prioritize safety over innovation, public governance over private self-regulation, and international over national approaches. The preference for safety is strongest among those who perceive AI as risky, unpredictable, and personally consequential. These findings reveal a systematic misalignment between dominant regulatory approaches and citizen preferences.
\end{minipage}

\vspace*{\stretch{2.4}}
\end{titlepage}

\setcounter{page}{2}

Artificial intelligence (AI) is rapidly reshaping economies, societies, and polities, confronting policymakers with a set of fundamental policy choices (1,2). As its impact expands, the question is no longer whether AI should be governed, but how. Should policymakers prioritize technological innovation or public safety? Should governance rely on public oversight or private self-regulation? Should regulation be organized at the national level or through international collaboration? States have begun to articulate distinct approaches (3). The European Union (EU) has advanced a model centered on precaution (4), public authority, and international cooperation; the United States (US) has emphasized innovation, market-led governance, and national flexibility (5); and China has pursued state-led development that combines technological advancement with centralized political control (6). These divergent models reflect deeper tensions in the governance of AI, yet little is known about how citizens evaluate these competing priorities.

Understanding public preferences is critical to the legitimacy and effectiveness of AI governance. Regulation aligned with citizen preferences is more likely to command compliance, sustain trust, and achieve policy goals (7--12). This is particularly important in the context of AI, where regulatory choices shape not only technological development but also fundamental individual outcomes such as privacy, security, and opportunity (13). As AI moves from a specialist policy concern into the arena of mass politics, public preferences increasingly shape both mobilization for or against regulation and the regulatory options that elected officials can credibly pursue (14,15).

Existing studies suggest that publics support management of AI risks, but they offer only limited insight into how citizens evaluate specific features of governance (16--21). Four limitations are particularly important. First, most studies focus on general attitudes toward AI risks and benefits rather than preferences over concrete regulatory choices. Second, they are largely confined to single-country contexts, leaving open how preferences vary across political regimes, stages of economic development, and positions in the global AI race. Third, there is little systematic evidence on how preferences vary across different domains of AI application, despite these posing distinct risks and dilemmas. Finally, we know surprisingly little about how individual perceptions of AI risks and consequences shape preferences over regulatory design.

Here we investigate how citizens across diverse political and economic contexts evaluate competing approaches to AI governance. We fielded a conjoint survey experiment in seven countries in December 2025---Brazil, China, Germany, India, South Africa, the United Kingdom (UK), and the US---spanning major regions, regime types, and positions in the global AI landscape. Respondents evaluate hypothetical regulatory proposals that vary along three dimensions: policy objective (innovation versus safety), governance mode (public versus private), and level of authority (national versus international). The experiment was implemented across three salient domains---AI in the workplace, policing, and warfare---with variation in perceived risks and benefits.

We find that citizens in all seven countries strongly support regulating AI. Moreover, they prioritize safety over innovation---a pattern that holds across countries and domains of AI application. This preference is strongest among individuals who perceive AI as risky, unpredictable, and personally consequential. Citizens also tend to prefer public governance to private self-regulation, and international collaboration to national authority. Variation across countries and domains in the strength of these patterns suggests that public preferences over AI governance are responsive to national and sectoral contexts.

Our results provide systematic evidence on how citizens weigh fundamental priorities in AI governance. By uncovering the structure and determinants of citizen preferences across countries and application domains, this study contributes to a deeper understanding of the political foundations of emerging AI regulatory regimes. The findings suggest that current approaches to AI governance are often out of sync with public preferences.

\section*{Assessing Regulatory Preferences}

We conceptualize regulatory preferences as structured evaluative judgments about the design of governance arrangements, including the objectives pursued, the actors involved, and the level at which authority is exercised. In line with established definitions, preferences are understood as comparative evaluations that are sufficiently stable to guide decisions, yet sensitive to context and information (22,23).

To capture these preferences, we focus on three dimensions of AI regulatory design. The first concerns the objective of governance, reflecting the balance between promoting technological innovation and ensuring public safety (20,24,25). The second concerns the mode of governance, distinguishing between public authority and private self-regulation (26--28). The third concerns the level of governance, contrasting national and international approaches (29--31). These dimensions are grounded in research on institutional design, international cooperation, and technology governance, and correspond to core debates about the goals, agents, and scope of regulation.

These dimensions are also politically contested. Governments, firms, and civil society actors disagree along all three dimensions (3,32--36). Reflecting such divisions, citizens may hold divergent preferences depending on their perceptions of risks and benefits, their attitudes toward markets and state authority, and their views on international cooperation.

We examine how citizens navigate these priorities across three domains of AI application: the workplace, policing, and warfare. These domains feature prominently in public debates and involve different combinations of risks and benefits (37--44). In the workplace, AI promises productivity gains, innovation, and relief from repetitive tasks, while also raising concerns about job displacement and algorithmic management. In policing, AI may improve crime detection, emergency response, and people's sense of safety, yet also prompts worries about surveillance, bias, and civil liberties. In warfare, AI could enhance precision, reduce risks to military personnel, and strengthen defensive capabilities, but also fuels concerns about the erosion of human control. Comparing across domains lets us assess whether individuals favor a general regulatory approach that applies similar principles across contexts, or a more differentiated approach tailored to specific applications.

To do so, we fielded a cross-national conjoint survey experiment embedded in a public opinion survey (Supplementary Information). The survey was carried out by Dynata in seven countries: Brazil, China, Germany, India, South Africa, the UK, and the US. Country selection was designed to capture variation along four key dimensions: world region, regime type, economic development, and position in the global AI landscape (see Methods). In each of the seven countries, a sample of at least 2,000 adult respondents was recruited for a total N of 14,239. Samples were quota-stratified to achieve national representativeness by age, gender, and education.

Respondents evaluated pairs of hypothetical regulatory proposals varying along the three governance dimensions (Fig. S1). We use hypothetical proposals since this enables us to vary the design features with greater precision and to test them without having respondents think about specific proposals (45,46). We restrict them to three dimensions to keep the comparison cognitively manageable. For each pair, respondents indicated which proposal they preferred and rated each proposal individually, enabling us to capture both comparative choices and levels of support. The experiment was administered in thematic blocks (workplace, policing, warfare), each prefaced by a balanced presentation of benefits and risks, with block ordering randomized to mitigate context and sequence effects.

The conjoint design lets us estimate the causal effect of regulatory features on citizens' choices. We use established methods for conjoint experiments, focusing on marginal means, which capture the average level of support for each attribute level across all possible designs (47,48). We further examine how preferences vary across individuals with different perceptions of AI and its consequences. All statistical tests are two-tailed and evaluate significance at $\alpha$ = 0.05.

\section*{Results}

We begin by assessing public support for the regulation of AI. Figure 1(a) shows that respondents in all seven countries express strong support for regulating this technology. Support is highest in China, the UK, and the US, somewhat lower in Germany and Brazil, and lowest---though still substantial---in South Africa and India.

Figure 1(b) reports the effects of individual-level characteristics on support for AI regulation. Among demographic factors, support increases with age, education, and income, while gender and occupation show no significant association. The absence of occupational differences is notable given current debates about job displacement (38), suggesting that such concerns are broadly shared rather than concentrated in specific groups. Among AI-related attitudes, individuals who perceive AI as a high risk and who are more knowledgeable about AI are more supportive of regulation, whereas trust in technology firms is not significantly associated with support. Finally, neither political ideology nor trust in government shape preferences for AI regulation.

\begin{figure}[htbp]
\centering
\includegraphics[width=\textwidth]{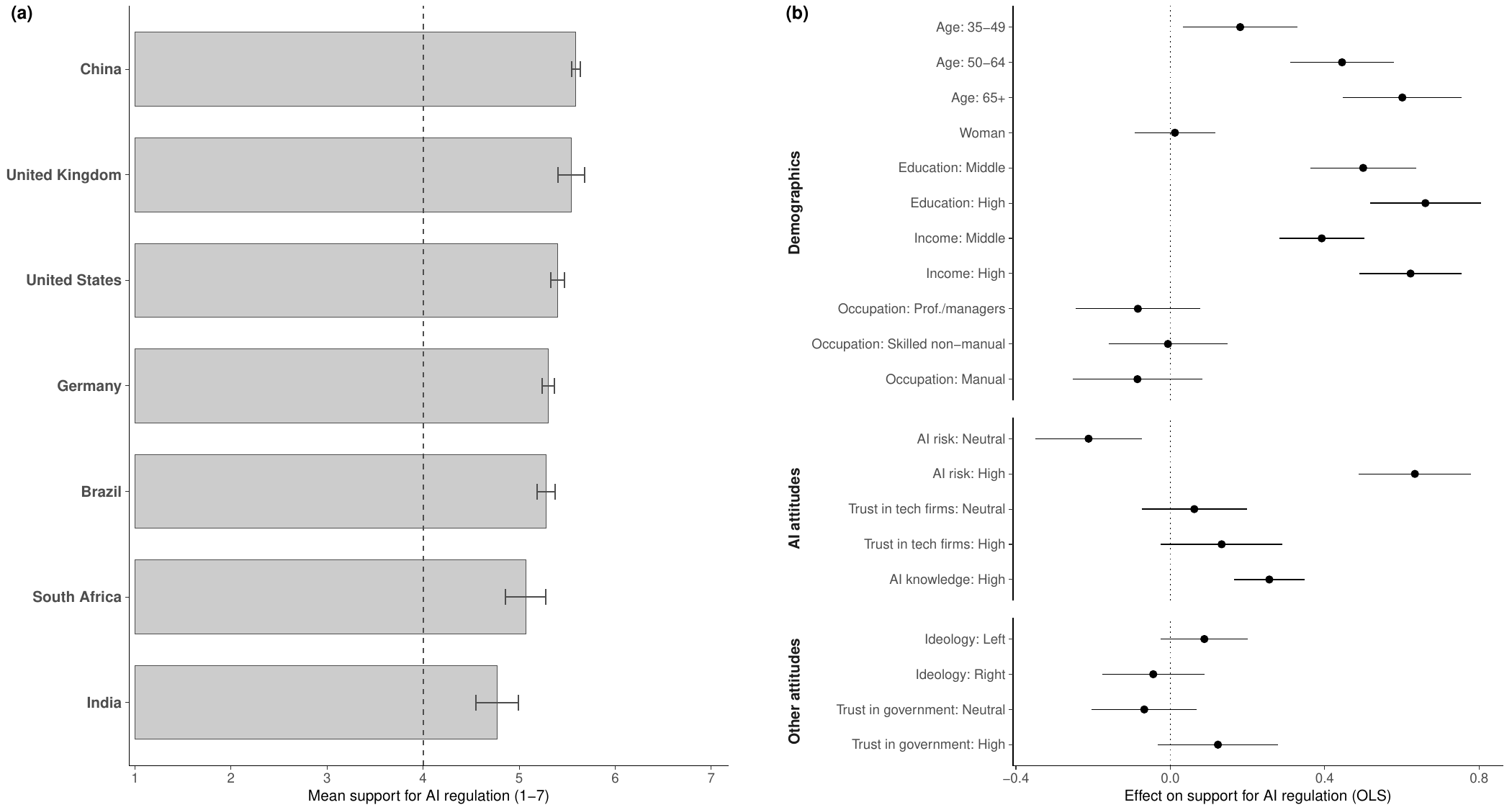}
\caption{\textbf{Support for AI regulation.} Panel (a) shows weighted mean support for regulating AI across seven countries (7-point scale), based on the question: ``Generally speaking, to what extent do you agree or disagree that AI requires regulation?'' (1 = strongly disagree, 7 = strongly agree). Error bars indicate 95\% confidence intervals; the dashed line marks the midpoint of the scale (4 = ``Neither agree nor disagree''). Panel (b) shows estimated coefficients predicting individual-level support for AI regulation from weighted OLS with country fixed effects and robust (HC1) standard errors (full output in Table S12). Reference categories in Panel (b) are: Age 18--34, Man, Education (Low), Income (Low), Not in labor force, AI risk (Low), Trust in tech firms (Low), AI knowledge (Low), Ideology (Center), and Trust in government (Low). Points show coefficient estimates with 95\% confidence intervals. N = 14,239 (Brazil = 2,066; China = 2,036; Germany = 2,010; India = 2,060; South Africa = 2,026; United Kingdom = 2,029; United States = 2,012).}
\end{figure}

Having established that citizens are broadly supportive of AI regulation, we examine their preferences over alternative governance designs. Figure 2 reports the results from our conjoint experiment, pooling respondents across all seven countries. The choice tasks (panel (a)) reveal clear preferences: respondents favor safety over innovation, governmental rather than private authority, and international over national governance. While preferences are pronounced across all three dimensions, the emphasis on safety stands out, with a gap in choice probability of 12 percentage points, or roughly four times the differences on mode and level.

The rating tasks (panel (b)) reinforce this pattern. At the same time, even the less preferred options receive ratings above the midpoint of the scale (4 on a 1--7 scale), indicating that support for AI regulation is broad-based rather than conditional on specific institutional designs.

\begin{figure}[htbp]
\centering
\includegraphics[width=\textwidth]{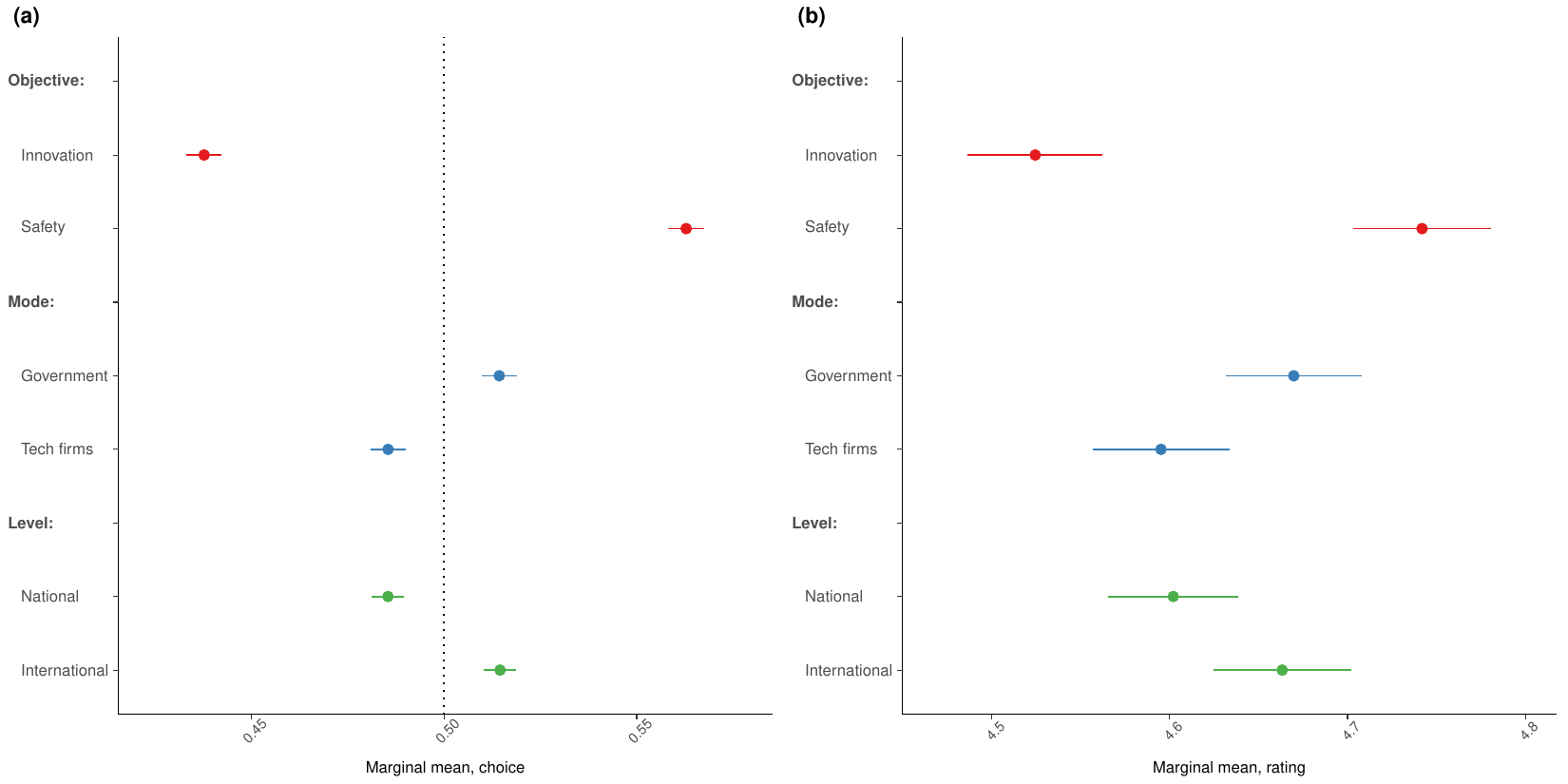}
\caption{\textbf{Support for AI regulatory attributes.} Panels show pooled marginal means for the conjoint outcomes (a) profile choice and (b) profile rating. Each point is the weighted mean outcome among profiles that include the indicated attribute level (Objective, Mode, Level), estimated using respondent-clustered standard errors and survey weights; horizontal error bars indicate 95\% confidence intervals.}
\end{figure}

Next, we examine whether preferences vary across three domains of AI application: the workplace, policing, and warfare. Figure 3 shows that preferences are remarkably consistent across domains. Irrespective of application area, respondents on average favor safety over innovation, government authority over private authority, and international over national regulation.

This consistency is most pronounced on the innovation--safety dimension, where preferences are strong and uniform across all three domains. By contrast, variation is somewhat greater on the other two dimensions. For regulatory authority, respondents show no clear preference between government and private authority in the workplace domain, whereas they favor government authority in policing and warfare. A similar pattern emerges for the level of governance: preferences for international over national regulation are modest in the workplace and policing domains, but more pronounced in military AI.

\begin{figure}[htbp]
\centering
\includegraphics[width=\textwidth]{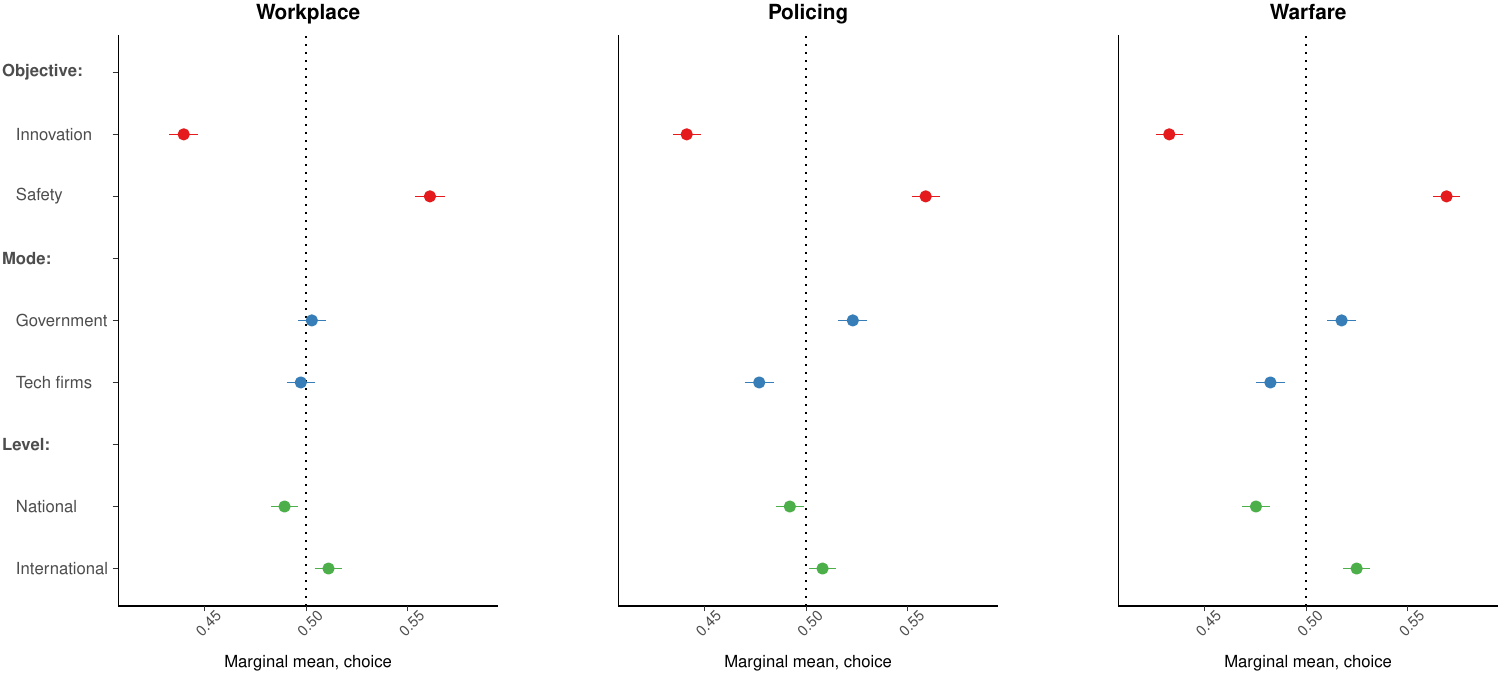}
\caption{\textbf{Support for AI regulatory attributes, by domain of application.} Panels show marginal means for the conjoint outcome profile choice, estimated separately for each domain. Each point is the weighted mean probability that a profile is chosen when it includes the indicated attribute level (Objective, Mode, Level). Estimates use survey weights and respondent-clustered standard errors; horizontal error bars indicate 95\% confidence intervals.}
\end{figure}

As a next step, we assess the consistency of the aggregate findings in national contexts. Figure 4 presents the results for choice tasks for each of the seven countries. On the objective dimension, citizens in all countries prefer safety over innovation. This preference is strongest in the UK, Germany, and the US, somewhat weaker in Brazil and South Africa, and more modest in India. In China, the estimated effect points in the same direction but does not reach statistical significance (p=0.072).

On the mode dimension, respondents in four countries---China, Germany, the UK, and the US---favor government rather than firm authority, thereby driving the corresponding result in the pooled analysis. By contrast, respondents in Brazil show a slight preference for self-regulation by technology firms, while those in India and South Africa express no significant preference.

On the level dimension, the support for international collaboration is likewise concentrated in a subset of countries. Respondents in Brazil, Germany, India, South Africa, and the UK favor international over national approaches to AI regulation. In contrast, respondents in the two major AI powers---China and the US---do not, with Chinese respondents preferring national regulation and US respondents showing no clear preference.

\begin{figure}[htbp]
\centering
\includegraphics[width=\textwidth]{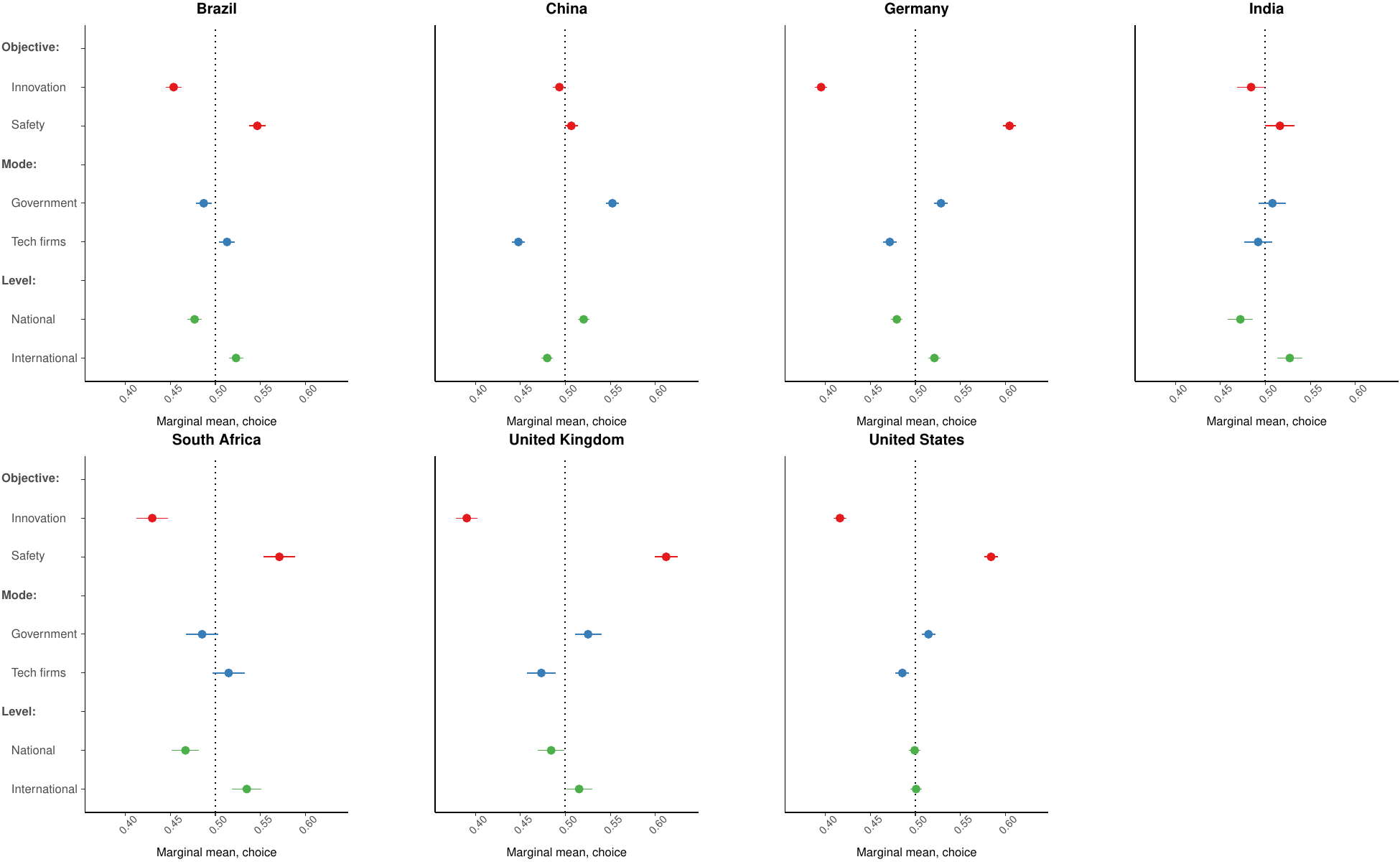}
\caption{\textbf{Support for AI regulatory attributes, by country.} Panels show marginal means for the conjoint outcome profile choice, estimated separately for each of the seven countries. Each point is the weighted mean probability that a profile is chosen when it includes the indicated attribute level (Objective, Mode, Level). Estimates use survey weights and respondent-clustered standard errors; horizontal error bars indicate 95\% confidence intervals.}
\end{figure}

Having established these regulatory preferences, we next examine how they vary across respondents with different individual-level characteristics. Figure 5 reports the results of six sub-group analyses designed to identify heterogeneous treatment effects, each comparing respondents with contrasting values on a specific individual-level factor.

The first analysis shows that perceptions of AI as a risk shape preferences along the innovation--safety dimension. Respondents who perceive AI as more risky express a stronger preference for safety over innovation (panel a). Similar patterns emerge for perceived unpredictability in AI's societal consequences (b) and expected personal affectedness (c). Taken together, these findings indicate that citizens who are more concerned about AI's potential effects exhibit stronger support for safety-oriented rather than innovation-oriented regulation.

We further find that respondents with more internationalist attitudes---favoring collaboration to solve global problems even when this reduces national sovereignty---display a stronger preference for international over national governance (d). In contrast, left-right ideological orientation does not shape preferences over regulatory design (e). Finally, an exploratory comparison shows that preferences differ by social position (f), as elite respondents (combining high education, high income, and managerial or professional occupations) favor safety over innovation less strongly, and government authority over tech firm authority more strongly, than non-elites (see also Fig. S26).

\begin{figure}[htbp]
\centering
\includegraphics[width=\textwidth]{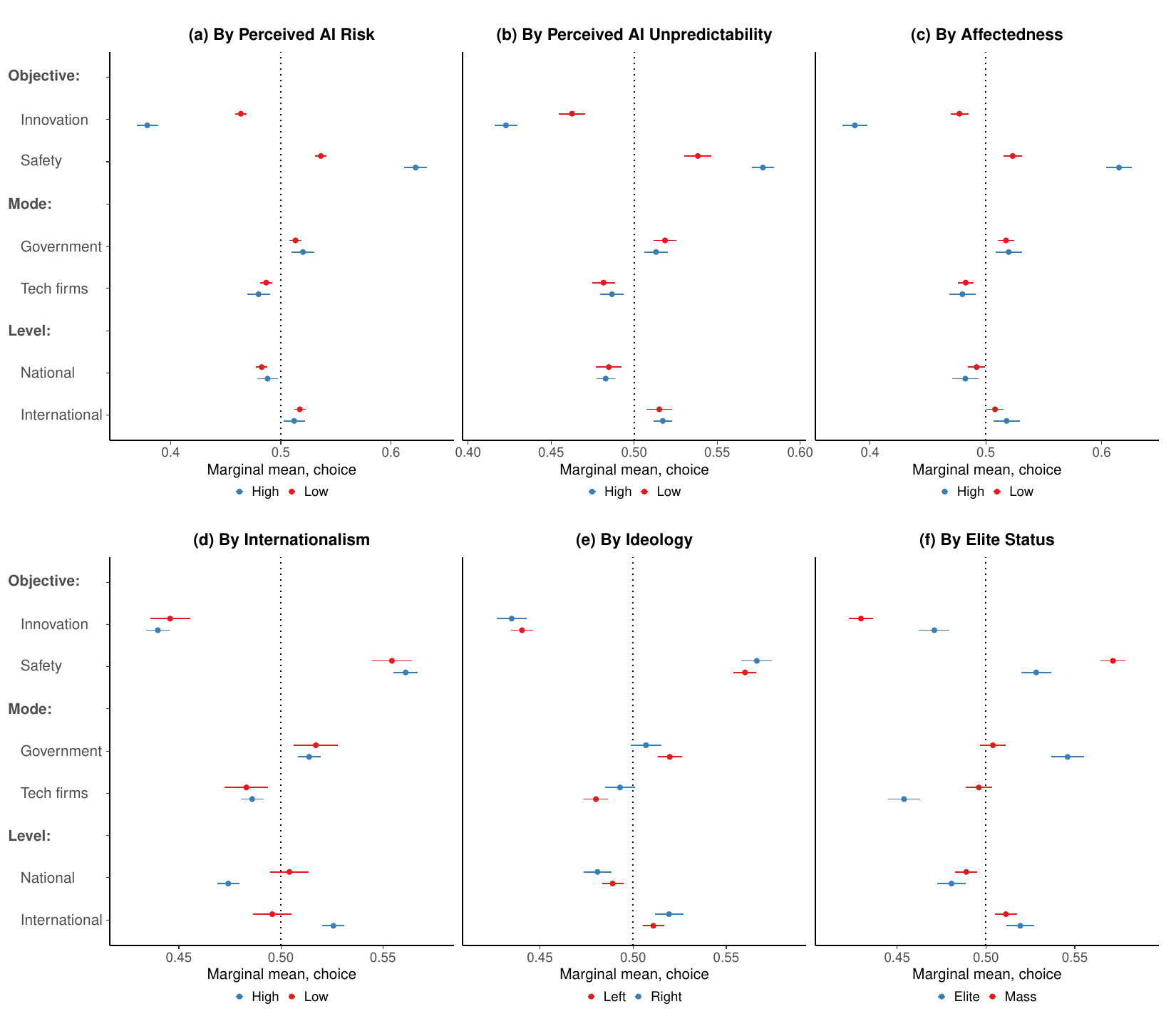}
\caption{\textbf{Heterogeneous treatment effects.} Panels show marginal means for the conjoint outcome profile choice, estimated separately for respondents with contrasting values on (a) perceived AI risk, (b) perceived AI unpredictability, (c) personal AI affectedness, (d) internationalism, (e) left--right ideology, and (f) elite-mass status. Each point is the weighted mean probability that a profile is chosen when it includes the indicated attribute level. Estimates use survey weights and respondent-clustered standard errors; horizontal error bars indicate 95\% confidence intervals}
\end{figure}

\section*{Discussion}

This study provides new evidence on the political foundations of AI governance by revealing how citizens evaluate fundamental regulatory options (14,15). Rather than expressing diffuse or fragmented attitudes, citizens across diverse political and economic contexts display broadly similar preferences regarding AI regulation. They strongly support regulation of AI and tend to favor approaches prioritizing safety over innovation, public authority over private self-regulation, and international over national governance. These patterns travel across countries and domains of application, even as their strength varies in meaningful ways, suggesting that public opinion on AI governance is organized around identifiable evaluative principles.

Our findings further point to an important mechanism underlying these preferences. Evaluations of AI governance are not primarily structured by ideological orientation, but by how individuals perceive the risks, unpredictability, and personal relevance of AI. Citizens who view AI as more risky, uncertain, and consequential express stronger support for safety-oriented regulation. This pattern is consistent with recent psychological research on the factors that shape resistance to AI systems, including perceived opacity, autonomy losses, and risk, and extends earlier cross-national evidence that attitudes toward AI regulation are organized around perceived consequences (16,49). This, in turn, suggests that public responses to AI are similar to those to other high-risk technologies, such as nuclear energy, where preferences over regulation have proven highly sensitive to perceived risks and salient events (50,51). Preferences over AI governance thus appear grounded in experiential and forward-looking assessments of technological change. As a result, public preferences in this domain are likely to be dynamic and endogenous to AI development itself, evolving as individuals encounter its effects in practice.

Beyond risk perceptions, our findings also reveal systematic differences between elite and non-elite respondents. While the broad preference structure on AI regulation is shared across socioeconomic groups, the intensity of preferences varies: elite respondents are less likely to prioritize safety over innovation, and more likely to prefer government over private authority, while showing no significant difference to non-elite respondents on the national--international dimension. This pattern is consistent with findings from other surveys on attitudes toward AI regulation (52) and other studies of elite--mass gaps in political behavior and attitudes (53,54).

Although the broad pattern of preferences holds across countries, there is some evidence that preferences over AI governance are shaped not only by individual attitudes and social position but also by national and sectoral context. Citizens in China and India display the weakest preference for safety over innovation, plausibly reflecting a broader preference for tech-fueled growth in countries with significant AI sectors and earlier stages of economic development, where economic considerations play a more central role in public AI assessments (16,21). Citizens in Brazil and South Africa---both more peripheral in the AI economy---display greater openness to private regulatory authority, consistent with research showing that where states are less well equipped to govern complex technologies, publics favor non-state alternatives (55). Citizens in China and the US, the two dominant AI powers, are less willing to embrace international governance, echoing findings from trade and environmental politics where preferences reflect national competitive positions (56). Across application domains, similar patterns emerge: citizens show greater support for private authority in the workplace, where firms are the primary operators, and stronger support for international governance in warfare, where cross-border externalities are most salient.  These patterns suggest that citizens' general preferences toward AI regulation are moderated by the specific conditions of different domains.

The global preference structure we identify stands in contrast to the regulatory approaches currently pursued by major powers (3,6,57). While the EU's model---emphasizing safety, public authority, and international collaboration---comes closest to the observed preference profile, the dominant approaches in the US and China diverge in systematic ways. The US emphasis on innovation, market-led governance, and national flexibility is least consistent with public preferences, including among US respondents. China's model, which combines rapid innovation with state-led risk management, aligns with citizen preferences on public authority but diverges on the level and, to some extent, the objective of regulation. More broadly, these patterns indicate that AI governance is not only geopolitically fragmented, but also systematically misaligned with citizen demand (see Fig. S24).

This misalignment has implications for both the legitimacy and effectiveness of AI governance. Regulatory arrangements aligned with citizen preferences are more likely to command public trust, sustain compliance, and endure politically (7--12). By contrast, governance models that diverge from widely held preferences risk generating contestation, undermining confidence in regulatory institutions, and weakening policy implementation. Approaches to AI regulation emphasizing public safety, government authority, and international coordination are thus more likely to prove politically sustainable, while strategies prioritizing rapid innovation, private self-regulation, and national approaches may face greater public resistance. From this perspective, the politics of AI is shaped by a triple misalignment: horizontal, as major powers pursue contrasting regulatory models; vertical, as several prominent approaches diverge from public preferences; and temporal, as short-term incentives for innovation stand in tension with long-term public risk concerns. Governments wishing to secure durable public support for AI governance may thus need to rebalance toward safety-oriented regulation, strengthen public oversight mechanisms, and invest in international coordination frameworks.

This study has three main limitations. First, the conjoint design simplifies the regulatory choices to three dimensions with two levels each. While these dimensions capture the most politically contested features of AI regulation, the design necessarily abstracts away from additional complexities to maintain respondent comprehension (45). Second, although the seven-country sample spans variation in region, regime type, economic development, and position in the global AI landscape, it cannot capture all relevant contexts. Third, the survey measures preferences at a single point in time. Given our finding that preferences are shaped by perceived risk and personal exposure---factors likely to evolve as AI technologies mature---tracking their temporal dynamics is an important avenue for future research.

\section*{References and Notes}
\begingroup
\setlength{\parskip}{0pt}
\begin{enumerate}\setlength{\itemsep}{2pt}
\item D. Acemoglu, S. Johnson, Power and progress: our thousand-year struggle over technology and prosperity (PublicAffairs, 2023).
\item Y. Bengio, G. Hinton, A. Yao, D. Song, P. Abbeel, T. Darrell, Y. N. Harari, Y. Q. Zhang, L. Xue, S. Shalev-Shwartz, G. Hadfield, J. Clune, T. Maharaj, F. Hutter, A. G. Baydin, S. McIlraith, Q. Gao, A. Acharya, D. Krueger, A. Dragan, P. Torr, S. Russell, D. Kahneman, J. Brauner, S. Mindermann, Managing extreme AI risks amid rapid progress. Science 384, 842--845 (2024).
\item A. Bradford, Digital empires: the global battle to regulate technology (Oxford university press, 2023).
\item European Parliament and Council of the European Union, Regulation (EU) 2024/1689 laying down harmonised rules on artificial intelligence (Artificial Intelligence Act). Off. J. Eur. Union L 2024/1689 (2024).
\item The White House, ``Ensuring a National Policy Framework for Artificial Intelligence'' (Executive Order, 2025).
\item H. Roberts, J. Cowls, E. Hine, J. Morley, V. Wang, M. Taddeo, L. Floridi, Governing artificial intelligence in China and the European Union: Comparing aims and promoting ethical outcomes. The Information Society 39, 79--97 (2023).
\item A. Chilton, K. Linos, Preferences and Compliance with International Law. Theoretical Inquiries in Law 22, 247--298 (2021).
\item L. Dellmuth, J. A. Scholte, J. Tallberg, S. Verhaegen, Citizens, Elites, and the Legitimacy of Global Governance (Oxford University Press, 2022).
\item K. Ingold, I. Stadelmann-Steffen, L. Kammermann, The acceptance of instruments in instrument mix situations: Citizens' perspective on Swiss energy transition. Research Policy 48, 103694 (2019).
\item T. Bernauer, S. Mohrenberg, V. Koubi, Do citizens evaluate international cooperation based on information about procedural and outcome quality?. The Review of International Organizations 15, 505--529 (2020).
\item T. M. Franck, The Power of Legitimacy Among Nations (Oxford University Press, 1990).
\item T. R. Tyler, Why People Obey the Law: Procedural Justice, Legitimacy, and Compliance (Yale University Press, 1990).
\item M. Lünich, C. Starke, ``The Public Legitimacy of Artificial Intelligence Governance'' in Handbook on the Global Governance of Artificial Intelligence, M. Furendal, M. Lundgren, Eds. (Edward Elgar, 2026).
\item S. Feldman, Structure and Consistency in Public Opinion: the Role of Core Beliefs and Values. American Journal of Political Science 32, 416 (1988).
\item P. E. Converse, ``The Nature of Belief Systems in Mass Publics'' in Ideology and Discontent, D. E. Apter, Ed. (Free Press, 1964), pp. 206--261.
\item S. Ehret, Public preferences for governing AI technology: Comparative evidence. Journal of European Public Policy 29, 1779--1798 (2022).
\item B. Magistro, S. Borwein, R. M. Alvarez, B. Bonikowski, P. J. Loewen, Attitudes toward artificial intelligence (AI) and globalization: Common microfoundations and political implications. American Journal of Political Science 70, 348--365 (2026).
\item B. Zhang, ``Public Opinion toward Artificial Intelligence'' in The Oxford Handbook of AI Governance, J. B. Bullock, Y. C. Chen, J. Himmelreich, V. M. Hudson, A. Korinek, M. M. Young, B. Zhang, Eds. (Oxford University Press, 2024), pp. 553--571.
\item P. D. König, S. Wurster, M. B. Siewert, Sustainability challenges of artificial intelligence and Citizens' regulatory preferences. Government Information Quarterly 40, 101863 (2023).
\item H. Machado, S. Silva, L. Neiva, Publics' views on ethical challenges of artificial intelligence: a scoping review. AI and Ethics 5, 139--167 (2025).
\item N. Dreksler, H. Law, C. Ahn, D. S. Schiff, K. J. Schiff, Z. Peskowitz, ``What Does the Public Think About AI? An Overview of the Public's Attitudes Toward AI and a Resource for Future Research'' (Centre for the Governance of AI, 2025).
\item K. J. Arrow, Social Choice and Individual Values (Yale University Press, 2012).
\item J. A. Frieden, ``Actors and Preferences in International Relations'' in Strategic Choice and International Relations, D. A. Lake, R. Powell, Eds. (Princeton University Press, 1999), pp. 39--76.
\item D. Vogel, The politics of precaution: regulating health, safety, and environmental risks in Europe and the United States (Princeton University Press, 2012).
\item M. Nitzberg, J. Zysman, Algorithms, data, and platforms: the diverse challenges of governing AI. Journal of European Public Policy 29, 1753--1778 (2022).
\item K. W. Abbott, P. Genschel, D. Snidal, B. Zangl, Eds., The Governor's Dilemma: Indirect Governance Beyond Principals and Agents (Oxford University Press, 2020).
\item G. Majone, Regulating Europe (Routledge, 1996).
\item T. Büthe, W. Mattli, The New Global Rulers: The Privatization of Regulation in the World Economy (Princeton University Press, 2011).
\item L. Hooghe, G. Marks, T. Lenz, J. Bezuijen, B. Ceka, S. Derderyan, Measuring International Authority: A Postfunctionalist Theory of Governance, Vol. III (Oxford University Press, 2017).
\item B. Koremenos, The Continent of International Law: Explaining Agreement Design (Cambridge University Press, 2016).
\item L. L. Martin, Interests, Power, and Multilateralism. International Organization 46, 765--792 (1992).
\item J. Tallberg, M. Lundgren, J. Geith, AI regulation in the European Union: examining non-state actor preferences. Business and Politics 26, 218--239 (2024).
\item C. Djeffal, M. B. Siewert, S. Wurster, Role of the state and responsibility in governing artificial intelligence: a comparative analysis of AI strategies. Journal of European Public Policy 29, 1799--1821 (2022).
\item G. Auld, A. Casovan, A. Clarke, B. Faveri, Governing AI through ethical standards: learning from the experiences of other private governance initiatives. Journal of European Public Policy 29, 1822--1844 (2022).
\item R. Radu, Steering the governance of artificial intelligence: national strategies in perspective. Policy and Society 40, 178--193 (2021).
\item A. Jobin, M. Ienca, E. Vayena, The global landscape of AI ethics guidelines. Nature Machine Intelligence 1, 389--399 (2019).
\item E. Brynjolfsson, T. Mitchell, What can machine learning do? Workforce implications. Science 358, 1530--1534 (2017).
\item D. Acemoglu, D. Autor, J. Hazell, P. Restrepo, Artificial Intelligence and Jobs: Evidence from Online Vacancies. Journal of Labor Economics 40, S293--S340 (2022).
\item M. C. Horowitz, Public opinion and the politics of the killer robots debate. Research \& Politics 3, 2053168015627183 (2016).
\item S. Miller, Lethal autonomous weapon systems (LAWS): meaningful human Control, collective moral responsibility and institutional design. Ethics and Information Technology 27, 63 (2025).
\item S. Brayne, Predict and surveil: data, discretion, and the future of policing (Oxford University Press, 2020).
\item M. Smith, S. Miller, The ethical application of biometric facial recognition technology. AI \& SOCIETY 37, 167--175 (2022).
\item S. Zuboff, The age of surveillance capitalism: the fight for a human future at the new frontier of power (Profile books, 2019).
\item P. Scharre, Army of none: autonomous weapons and the future of war (W.W. Norton \& Company, 2018).
\item R. Brutger, J. D. Kertzer, J. Renshon, D. Tingley, C. M. Weiss, Abstraction and Detail in Experimental Design. American Journal of Political Science 67, 979--995 (2023).
\item L. Dellmuth, J. Tallberg, Legitimacy Politics: Elite Communication and Public Opinion in Global Governance (Cambridge University Press, 2023).
\item J. Hainmueller, D. J. Hopkins, T. Yamamoto, Causal inference in conjoint analysis: Understanding multidimensional choices via stated preference experiments. Political Analysis 22, 1--30 (2014).
\item T. J. Leeper, S. B. Hobolt, J. Tilley, Measuring Subgroup Preferences in Conjoint Experiments. Political Analysis 28, 207--221 (2020).
\item J. De Freitas, S. Agarwal, B. Schmitt, N. Haslam, Psychological factors underlying attitudes toward AI tools. Nature Human Behaviour 7, 1845--1854 (2023).
\item E. A. Rosa, R. E. Dunlap, Nuclear Power: Three Decades of Public Opinion. Public Opinion Quarterly 58, 295--324 (1994).
\item Y. Kim, M. Kim, W. Kim, Effect of the Fukushima nuclear disaster on global public acceptance of nuclear energy. Energy Policy 61, 822--828 (2013).
\item Pew Research Center, ``How the U.S. Public and AI Experts View Artificial Intelligence'' (2025).
\item J. D. Kertzer, Re-Assessing Elite-Public Gaps in Political Behavior. American Journal of Political Science 66, 539--553 (2022).
\item L. Dellmuth, J. A. Scholte, J. Tallberg, S. Verhaegen, The Elite-Citizen Gap in International Organization Legitimacy. American Political Science Review 116, 283--300 (2022).
\item H. V. Milner, D. L. Nielson, M. G. Findley, Citizen preferences and public goods: comparing preferences for foreign aid and government programs in Uganda. The Review of International Organizations 11, 219--245 (2016).
\item E. D. Mansfield, D. C. Mutz, Support for Free Trade: Self-interest, Sociotropic Politics, and Out-group Anxiety. International Organization 63, 425--457 (2009).
\item E. Hine, L. Floridi, Artificial intelligence with American values and Chinese characteristics: a comparative analysis of American and Chinese governmental AI policies. AI \& SOCIETY 39, 257--278 (2024).
\item S. O. Hansson, T. Grüne-Yanoff, ``Preferences'' in Stanford Encyclopedia of Philosophy (Spring 2022 Edition), E. N. Zalta, Ed. (Metaphysics Research Lab, Stanford University, 2022).
\item J. Cheng, J. Zeng, Shaping AI's future? China in global AI governance. Journal of Contemporary China 32, 794--810 (2022).
\item A. Dafoe, B. Zhang, D. Caughey, Information equivalence in survey experiments. Political Analysis 26, 399--416 (2018).
\item A. J. Berinsky, M. F. Margolis, M. W. Sances, Separating the shirkers from the workers? Making sure respondents pay attention on self-administered surveys. American Journal of Political Science 58, 739--753 (2014).
\item J. Hainmueller, D. Hangartner, T. Yamamoto, Validating vignette and conjoint survey experiments against real-world behavior. Proc. Natl. Acad. Sci. U.S.A. 112, 2395--2400 (2015).
\item OECD, ``Recommendation of the Council on Artificial Intelligence'' (OECD/LEGAL/0449, OECD, 2019).
\item G7, ``Hiroshima AI Process Comprehensive Policy Framework'' (G7 Digital and Tech Ministers' Meeting, 2023).
\end{enumerate}
\endgroup

\clearpage
\setcounter{figure}{0}
\setcounter{table}{0}
\renewcommand{\thefigure}{S\arabic{figure}}
\renewcommand{\thetable}{S\arabic{table}}

\thispagestyle{empty}
\begin{center}
{\large\textbf{Supplementary Materials for}}\\[1.2em]
{\large\textbf{Governing Artificial Intelligence: Public Preferences and Regulatory Options}}\\[1.2em]
Magnus Lundgren, Jonas Tallberg\\[0.6em]
Corresponding author: magnus.lundgren@gu.se
\end{center}

\vspace{1.5em}
\noindent\textbf{This PDF file includes:}\\[0.4em]
Materials and Methods\\
Supplementary Text\\
Figs. S1 to S33\\
Tables S1 to S16

\clearpage

\section{Materials and Methods}

\subsection{Survey instrument}

The survey was fielded by Dynata, a global provider of online survey panels, in December 2025. The full questionnaire was designed in English and professionally translated into each country's official language(s). The English-language master questionnaire is reproduced in Section~\ref{sec:survey_full}. Question numbering throughout this supplement follows the pre-registered instrument.

\subsection{Theoretical foundations}

The outcome of interest in this study is citizens' regulatory preferences, reflecting their evaluative judgments about how AI should be governed. Conceptually, regulatory preferences involve structured attitudes about the appropriate design of governance arrangements, including the actors involved in designing regulation, which instruments should be used, and which policy goals should be prioritized. In line with standard definitions, preferences are understood as comparative evaluations that are stable enough to guide choice, yet context-sensitive and open to influence through information (\textit{22,23,58}).

In this study, preferences are expressed through comparative choices and support ratings for different regulatory proposals with attributes that vary along three dimensions: one pertaining to the objective of regulation (technological innovation or people's safety), one pertaining to the mode of regulation (public-led or private-led), and one pertaining to the level of regulation (national or international).

The selection of regulatory dimensions is grounded in research on international cooperation, institutional design, and technology governance. The objective dimension reflects perceived options in the regulation of technology generally (\textit{24}) and AI specifically (\textit{20,25}). The mode dimension draws on a longstanding debate about the legitimacy and effectiveness of public versus private rulemaking (\textit{27,28}). The level dimension invokes a classic debate about national versus international governance (\textit{29}).

These regulatory dimensions are also politically contested. Governments, firms, and civil society actors frequently disagree about whether AI regulation should prioritize innovation over precaution (\textit{32}). Internationally, these disagreements map onto different governance models promoted by leading powers, with the EU, US, and China advancing divergent regulatory visions rooted in their industrial, institutional and ideological traditions (\textit{3,59}).

Finally, these are dimensions over which reasonable people may disagree, in the sense that they involve trade-offs between goals that are independently desirable but difficult to attain simultaneously. Citizens are likely to hold divergent preferences based on, for instance, their understanding of AI, attitudes toward international cooperation, and ideological orientation.

The experiment investigates how individuals navigate key regulatory options and which governance configurations they find most appealing across three salient AI issue areas: (1)~AI in the workplace, (2)~AI in policing, and (3)~AI in warfare. Each of these domains has featured prominently in public debates and policy discussions about the risks and benefits of AI. Each has also been studied extensively in isolation---for instance, research on algorithmic management in labor, algorithmic bias in surveillance, and lethal autonomous weapons---but there is yet little research on how individuals evaluate regulatory options across areas. By comparing responses across these three issue areas, we can assess whether individuals tend to support a horizontal regulatory approach to AI (a unified set of rules applied across domains) or favor a sectoral approach, tailored to the specific risks and applications of each domain.

\subsection{Hypotheses}

The experiment tests three main hypotheses about the effect of regulatory attributes on preferences:

\begin{itemize}
\item \textbf{H1:} Proposals to strengthen people's safety will receive greater support than those to strengthen technological innovation.
\item \textbf{H2:} Proposals developed by governments will receive greater support than those developed by private firms.
\item \textbf{H3:} Proposals for regulation at the international level will receive greater support than those for regulation at the national level.
\end{itemize}

\noindent To evaluate how respondents' preferences are conditioned by prior attitudes, the experiment tests five heterogeneous treatment effect hypotheses:

\begin{itemize}
\item \textbf{H4:} The preference for prioritizing people's safety will be stronger among respondents who view AI as a societal risk compared to those who view it as an opportunity.
\item \textbf{H5:} The preference for prioritizing people's safety will be stronger among respondents who view the consequences of AI as unpredictable compared to those who view them as predictable.
\item \textbf{H6:} The preference for prioritizing people's safety will be stronger among respondents who expect to be more negatively personally affected by AI innovations compared with those who expect to be more positively personally affected.
\item \textbf{H7:} The preference for international-level regulation will be stronger among respondents who support international cooperation than among those favoring national autonomy.
\item \textbf{H8:} The preference for governmental regulation will be stronger among left-leaning respondents than among right-leaning respondents.
\end{itemize}

\subsection{Survey design}

\subsubsection{Pre-treatment items}

Respondents first answered four questions measuring their attitudes toward AI:

\begin{itemize}
\item \textbf{Q1 (AI risk):} ``Generally speaking, do you think that AI presents an opportunity or a risk to human society?'' (1 = Always an opportunity, 7 = Always a risk)
\item \textbf{Q2 (AI predictability):} ``Generally speaking, how predictable or unpredictable do you think that the consequences of AI are for human society?'' (1 = Very unpredictable, 7 = Very predictable). \emph{Note:} This variable is reverse-coded to create an unpredictability scale (8 $-$ Q2), where higher values indicate greater perceived unpredictability.
\item \textbf{Q3 (Support for regulation):} ``Generally speaking, to what extent do you agree or disagree that AI requires regulation?'' (1 = Strongly disagree, 7 = Strongly agree)
\item \textbf{Q4 (Internationalism):} ``Generally speaking, to what extent do you agree or disagree that [country] should work together with other countries to solve global problems, even if this reduces [country]'s national sovereignty?'' (1 = Strongly disagree, 7 = Strongly agree)
\end{itemize}

\subsubsection{Attention check}

After the pre-treatment items, respondents were presented with a question about their news sources (Q5). The question was designed as a demanding instructed response item (attention check): the question text explicitly asked respondents to select only ``Radio'' and ``TikTok'' from a list of eleven news sources, and to ignore the surface-level question about media consumption. Respondents who selected exactly these two options were classified as having passed the attention check. All main analyses include both attentive and inattentive respondents; robustness to restricting the sample to attentive respondents only is demonstrated in Section~\ref{sec:F}.

\subsubsection{Conjoint experiment}

The conjoint experiment was administered in three thematic blocks (workplace, policing, warfare), with block order randomized across respondents. Each block contained three paired conjoint tasks. In each task, respondents evaluated two hypothetical regulatory proposals varying along three dimensions:

\begin{itemize}
\item \textbf{Objective:} Strengthening people's safety against AI harms \emph{vs.} Strengthening technological innovation in AI
\item \textbf{Mode:} Government authorities \emph{vs.} Technology firms
\item \textbf{Level:} National level \emph{vs.} International level
\end{itemize}

\begin{figure}[H]
\centering
\includegraphics[width=0.9\textwidth]{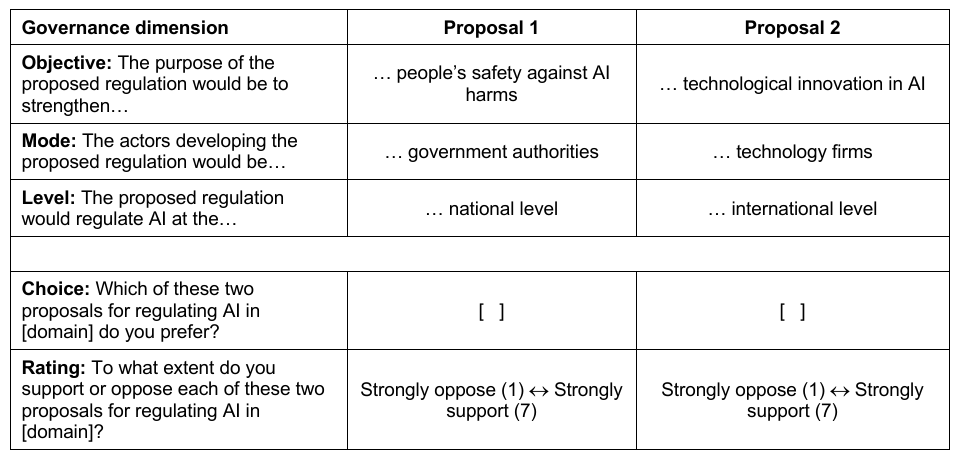}
\caption{\textbf{Example choice task as shown to respondents.} A single conjoint task pairing two regulatory proposals that vary on the three attributes (objective, mode, level). Respondents chose the proposal they preferred and rated each on a 7-point scale.}
\label{fig:example_screen}
\end{figure}

For each pair, respondents indicated which proposal they preferred (forced choice) and rated each proposal on a 7-point scale (1 = Strongly oppose, 7 = Strongly support). Attribute order was randomized across respondents but held constant within respondent. An example choice task as shown to respondents is presented in Fig.~\ref{fig:example_screen}.

\textbf{Information equivalence.} To support information equivalence across attribute levels (\textit{60}), the experiment was designed with three features. First, each thematic block opens with a short preamble that presents both the prospective benefits of AI (such as productivity and economic growth in the workplace, efficiency and security in policing, and battlefield advantage in warfare) and the corresponding risks of comparable length and specificity, so that respondents enter the conjoint tasks with concrete instantiations of both dimensions made salient (see Section~\ref{sec:survey_full}). This also addresses the possibility that ``innovation'' is read more abstractly than ``safety'' (\textit{45}). Second, all attribute labels were described in parallel form with comparable length and in neutral wording. Third, the four pre-treatment AI items (risk/opportunity, predictability, support for regulation, internationalism) used bipolar 7-point scales with balanced anchors, so that respondents did not enter the conjoint primed in one direction.

The randomization of attribute levels was subject to a constraint imposed by the survey platform: no task presented two fully identical proposals (i.e., proposals matching on all three attributes). Tasks could differ on one, two, or all three attributes. As a result, each individual attribute matched across the two proposals approximately 43\% of the time (rather than the 50\% expected under full independence). This constrained design ensures that every choice task presents a meaningful comparison. The marginal distribution of each attribute level remains balanced at 50\% (see Section~\ref{sec:B}).
\subsubsection{Post-treatment items}

After the conjoint experiment, the survey measured:
\begin{itemize}
\item AI knowledge: Three factual questions (Q35--Q37). Respondents were coded as ``high knowledge'' if they correctly answered all three: Q35 = Face recognition (option 3), Q36 = Anthropic (option 1), Q37 = Deepfakes (option 2).
\item Ideology: Three items (Q38--Q40) measuring attitudes on income equality, public vs.\ private ownership, and government responsibility. Averaged and rounded to create a 1--7 index (1--3 = Left, 4 = Center, 5--7 = Right).
\item Trust in government (Q41), Trust in UN (Q42), Trust in technology companies (Q43): Each on a 7-point scale (1 = Very little confidence, 7 = Very much confidence).
\item Interpersonal trust (Q45), economic satisfaction (Q46), and geographic identity (Q44).
\end{itemize}

\subsubsection{Demographics}
Gender (Q7), age (HQ6, recoded into 4 bands), education (H\_Q8, recoded into Low/Middle/High), income (Q9, country-specific brackets recoded into Low/Middle/High), and occupation (Q10, recoded into 4 categories).

\subsection{Sample and data collection}

The survey was conducted in seven countries selected to capture variation along four key dimensions: world region, regime type, economic development, and position in the global AI landscape. The sample includes two countries from Europe (Germany, United Kingdom), two from the Americas (Brazil, United States), two from Asia (China, India), and one from Africa (South Africa). These countries span the range from liberal democracies to closed autocracies, from high-income to lower-middle-income economies, and from leading AI powers (China, United States, United Kingdom) to countries with less developed AI sectors. Table~\ref{tab:country_chars} summarizes the key characteristics of the sample countries. Table~\ref{tab:B1} provides summary statistics for each country sample.

\begin{table}[H]
\centering
\caption{\textbf{Country sample characteristics.}}
\label{tab:country_chars}
\begin{tabular}{llllll}
\toprule
Country & N & Region & AI rank & Regime type & Economic development \\
\midrule
Brazil & 2,066 & Americas & Not top-10 & Electoral democracy & Upper-middle income \\
China & 2,036 & Asia & Top-10 & Closed autocracy & Upper-middle income \\
Germany & 2,010 & Europe & Top-10 & Liberal democracy & High income \\
India & 2,060 & Asia & Top-10 & Electoral autocracy & Lower-middle income \\
South Africa & 2,026 & Africa & Not top-10 & Liberal democracy & Upper-middle income \\
United Kingdom & 2,029 & Europe & Top-10 & Liberal democracy & High income \\
United States & 2,012 & Americas & Top-10 & Liberal democracy & High income \\
\bottomrule
\end{tabular}

\smallskip
{\scriptsize\noindent Notes: AI rank based on Tortoise Media Global AI Index; Regime type from V-Dem Regimes of the World classification; Economic development based on World Bank income categories.}
\end{table}

In each of the seven countries, a sample of at least 2,000 adult respondents was recruited through Dynata's online panels, for a total N = 14,239. Dynata recruits panelists through a combination of online channels and partnership networks. Samples were quota-stratified by age, gender, and education to approximate national representativeness.

\textbf{Survey weights.} To adjust for remaining imbalances between the achieved sample and the target population, Dynata constructed post-stratification weights using rim weighting (iterative proportional fitting) on age, gender, and education within each country. All main analyses employ these weights unless otherwise noted. We demonstrate robustness of the findings to the removal of weights in Section~\ref{sec:E}.

\subsection{Estimation approach}

Each of the 14,239 respondents evaluated nine pairs of proposals (three blocks of three tasks each), yielding a total of 256,302 evaluated regulatory profiles. We report two types of estimates from the conjoint experiment. The main text focuses on marginal means; AMCEs are reported in Section~\ref{sec:C} of this supplement.

\textbf{Marginal means (MMs)} capture the average probability that a profile is chosen (or the average rating) when it includes a given attribute level, averaging over all combinations of the remaining attributes (\textit{48}). Formally, the marginal mean for attribute level $l$ is $\text{MM}_l = E[Y_i | X_k = l]$, where $Y_i$ is the outcome (choice or rating) and $X_k = l$ indicates that attribute $k$ takes level $l$. Differences in MMs between attribute levels (e.g., Safety vs.\ Innovation) indicate the direction and magnitude of preferences. All MMs are estimated via weighted regression of the outcome on an intercept only, within each attribute-level subset, with standard errors clustered at the respondent level. We use MMs as the primary quantity of interest following the recommendation of Leeper, Hobolt, and Tilley (\textit{48}), who demonstrate that MMs are more appropriate than AMCEs for assessing the overall level of support for specific attribute levels and for comparing preferences across subgroups.

\textbf{Average marginal component effects (AMCEs)} estimate the causal effect of changing an attribute from its baseline level to an alternative level on the probability of a profile being chosen (\textit{47}). AMCEs are estimated via weighted linear regression of the choice or rating outcome on indicator variables for each attribute level, with one level per attribute omitted as the reference category, and standard errors clustered at the respondent level. The baseline categories are Innovation (Objective), Government (Mode), and National (Level).

\textbf{Heterogeneous treatment effects.} To test hypotheses H4--H8 about preference heterogeneity, we split the sample into subgroups based on pre-treatment moderator variables and estimate separate MMs and AMCEs within each subgroup. For continuous moderators measured on 7-point scales, we classify respondents scoring 1--3 as ``Low'' and those scoring 5--7 as ``High,'' excluding respondents at the midpoint (4) to ensure clear separation. For the ideology index (averaged across three items and rounded), we classify scores of 1--3 as ``Left'' and 5--7 as ``Right,'' with ``Center'' (4) excluded. For AI knowledge, which is constructed as a binary variable from three factual questions (coded as ``High'' if all three are answered correctly, ``Low'' otherwise), the subgroup split follows this constructed binary. For the attention check, the split follows the pass/fail classification described in Section~1.4.2.

\textbf{Predicted support profiles.} In the predicted-support analysis (Fig.~\ref{fig:predsupport}), we estimate the predicted probability of choosing each of the eight possible regulatory profiles (all combinations of the three binary attributes). These predictions are derived from a weighted linear probability model: $\text{Pr}(\text{chosen}_i) = \beta_0 + \beta_1 \text{Safety}_i + \beta_2 \text{TechFirms}_i + \beta_3 \text{International}_i + \epsilon_i$, with standard errors clustered by respondent. The additive specification is supported by the absence of significant attribute interactions (see Section~\ref{sec:I}).

\subsection{Pre-registration}

The study was pre-registered on OSF (\url{https://osf.io/5rz9p/}). The pre-registered hypotheses (H1--H8) are presented in Section~1.3. The main treatment-effect tests (H1--H3) and the heterogeneous-effects tests (H4--H8; Figure 5 of the main text) follow this pre-registered plan. Three analyses are exploratory and were not pre-specified: the determinants of general support for AI regulation (Figure 1b), the elite--mass comparison (Figure 5f), and the predicted support for composite governance models (Fig.~\ref{fig:predsupport}).

\subsection{Ethical approval}

The study received ethical approval from the Swedish Ethical Review Authority (Etikpr\"{o}vningsmyndigheten), Dnr 2025-06636-01.

\clearpage

\section{Supplementary Text}

\subsection{Sample and Weighting Diagnostics}
\label{sec:B}

\begin{table}[H]
\centering
\caption{\textbf{Sample diagnostics by country.} Weighted mean support for AI regulation (7-point scale) and standard deviation. N = 14,239.}
\label{tab:B1}
\small
\scriptsize
\begin{tabular}{lrrr}
\toprule
\textbf{Country} & \textbf{N} & \textbf{Mean support} & \textbf{SD} \\
\midrule
Brazil & 2066 & 5.28 & 1.66 \\
China & 2036 & 5.59 & 1.00 \\
Germany & 2010 & 5.30 & 1.44 \\
India & 2060 & 4.77 & 1.97 \\
South Africa & 2026 & 5.07 & 1.89 \\
United Kingdom & 2029 & 5.54 & 1.47 \\
United States & 2012 & 5.40 & 1.59 \\
Pooled & 14239 & 5.28 & 1.63 \\
\bottomrule
\end{tabular}

\end{table}

Table~\ref{tab:B1} reports the sample size and weighted mean support for AI regulation by country. Table~\ref{tab:B2} reports the demographic composition of each country sample. Table~\ref{tab:B3} reports weighting diagnostics.

\begin{table}[H]
\centering
\caption{\textbf{Sample composition by country (\%).} All values are unweighted shares of respondents.}
\label{tab:B2}
\scriptsize
\begin{tabular}{l rrrr rr rrr rrr}
\toprule
 & \multicolumn{4}{c}{\textbf{Age}} & \multicolumn{2}{c}{\textbf{Gender}} & \multicolumn{3}{c}{\textbf{Education}} & \multicolumn{3}{c}{\textbf{Income}} \\
\cmidrule(lr){2-5} \cmidrule(lr){6-7} \cmidrule(lr){8-10} \cmidrule(lr){11-13}
\textbf{Country} & 18--34 & 35--49 & 50--64 & 65+ & Man & Woman & Low & Mid & High & Low & Mid & High \\
\midrule
Brazil & 35.6 & 28.9 & 21.4 & 14.0 & 47.1 & 52.7 & 9.7 & 57.8 & 32.5 & 46.0 & 26.2 & 27.7 \\
China & 32.2 & 29.0 & 25.2 & 13.6 & 49.5 & 49.3 & 14.6 & 39.4 & 46.0 & 11.4 & 53.6 & 35.0 \\
Germany & 24.0 & 22.3 & 27.5 & 26.2 & 48.9 & 50.9 & 11.9 & 63.2 & 24.9 & 42.1 & 45.1 & 12.7 \\
India & 46.7 & 28.4 & 16.3 & 8.7 & 51.7 & 47.7 & 5.2 & 36.4 & 58.4 & 31.6 & 20.1 & 48.3 \\
South Africa & 44.2 & 29.7 & 17.0 & 9.0 & 47.3 & 52.6 & 3.8 & 74.5 & 21.7 & 41.0 & 34.4 & 24.7 \\
United Kingdom & 27.9 & 24.4 & 24.5 & 23.1 & 48.3 & 51.4 & 3.0 & 54.1 & 42.9 & 54.3 & 32.2 & 13.5 \\
United States & 29.6 & 24.1 & 24.9 & 21.5 & 48.3 & 51.4 & 7.3 & 52.0 & 40.7 & 49.5 & 37.2 & 13.3 \\
\bottomrule
\end{tabular}

\end{table} As is common with online panel surveys, the achieved samples in some countries under-represent lower-education respondents, who are less likely to participate in online surveys. This is particularly the case in the United Kingdom, India, and South Africa, where the weighting procedure assigns larger weights to the relatively small number of low-education respondents to correct for this imbalance. The robustness of the main findings to the removal of survey weights is demonstrated in Section~\ref{sec:E}.

\begin{table}[H]
\centering
\caption{\textbf{Survey weight diagnostics by country.}}
\label{tab:B3}
\small
\scriptsize
\begin{tabular}{lrrrr}
\toprule
\textbf{Country} & \textbf{Mean} & \textbf{SD} & \textbf{Min} & \textbf{Max} \\
\midrule
Brazil & 1.00 & 0.87 & 0.48 & 4.16 \\
China & 1.00 & 0.24 & 0.75 & 2.05 \\
Germany & 1.00 & 0.16 & 0.90 & 1.46 \\
India & 1.00 & 2.39 & 0.28 & 17.36 \\
South Africa & 1.00 & 2.37 & 0.29 & 16.35 \\
United Kingdom & 1.00 & 1.89 & 0.34 & 16.37 \\
United States & 1.00 & 0.23 & 0.86 & 1.89 \\
\bottomrule
\end{tabular}

\end{table}

Table~\ref{tab:B4} reports the randomization balance across conjoint attribute levels. All attributes are balanced at 50/50, confirming successful randomization.

\begin{table}[H]
\centering
\caption{\textbf{Randomization balance of conjoint attribute levels (pooled across all countries and tasks).} Expected proportion under uniform randomization: 50\%.}
\label{tab:B4}
\small
\scriptsize
\begin{tabular}{lrrr}
\toprule
\textbf{Attribute} & \textbf{Level} & \textbf{N} & \textbf{\%} \\
\midrule
Objective & Innovation & 128148 & 50.00 \\
Objective & Safety & 128154 & 50.00 \\
Mode & Government & 128152 & 50.00 \\
Mode & Tech firms & 128150 & 50.00 \\
Level & National & 128153 & 50.00 \\
Level & International & 128149 & 50.00 \\
\bottomrule
\end{tabular}
\end{table}

\begin{table}[H]
\centering
\caption{\textbf{Attention check pass and fail rates by country.} Respondents passed if they correctly followed the instructed response item (selecting only ``Radio'' and ``TikTok'').}
\label{tab:B5}
\small
\scriptsize
\begin{tabular}{lrrr}
\toprule
\textbf{Country} & \textbf{Att Check} & \textbf{N} & \textbf{\%} \\
\midrule
Brazil & Failed & 857 & 41.50 \\
Brazil & Passed & 1209 & 58.50 \\
China & Failed & 631 & 31.00 \\
China & Passed & 1405 & 69.00 \\
Germany & Failed & 807 & 40.10 \\
Germany & Passed & 1203 & 59.90 \\
India & Failed & 1083 & 52.60 \\
India & Passed & 977 & 47.40 \\
South Africa & Failed & 577 & 28.50 \\
South Africa & Passed & 1449 & 71.50 \\
United Kingdom & Failed & 778 & 38.30 \\
United Kingdom & Passed & 1251 & 61.70 \\
United States & Failed & 875 & 43.50 \\
United States & Passed & 1137 & 56.50 \\
\bottomrule
\end{tabular}
\end{table}


\clearpage

\subsection{Alternative Estimators: AMCE Results}
\label{sec:C}

\begin{figure}[H]
\centering
\includegraphics[width=\textwidth]{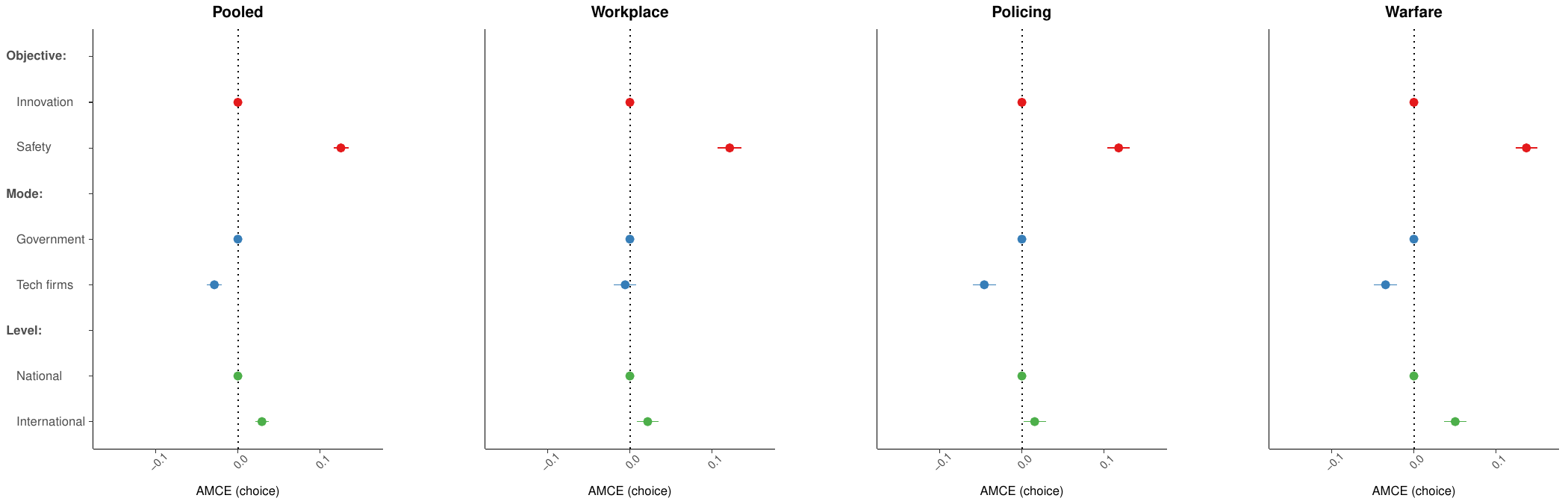}
\caption{\textbf{Average marginal component effects (AMCEs) for profile choice, pooled and by domain of application.} Each point shows the estimated change in the probability of a profile being chosen when the attribute level is changed from the baseline. Baseline categories: Innovation, Government, National. Estimates use survey weights and respondent-clustered standard errors; horizontal error bars indicate 95\% confidence intervals.}
\label{fig:C1}
\end{figure}

The main paper reports marginal means. Here we present the corresponding average marginal component effects (AMCEs), which express the effect of each attribute as a change in the probability of being chosen relative to a baseline category: Innovation (Objective), Government (Mode), and National (Level). With binary attributes, AMCEs are directly proportional to the MM differences reported in the main text, but they provide a useful complementary perspective by expressing effect sizes in percentage-point terms. Safety-oriented regulation increases the probability of being chosen by approximately 12 percentage points relative to innovation-oriented regulation in the pooled sample (fig.~\ref{fig:C1}).

\begin{figure}[H]
\centering
\includegraphics[width=\textwidth]{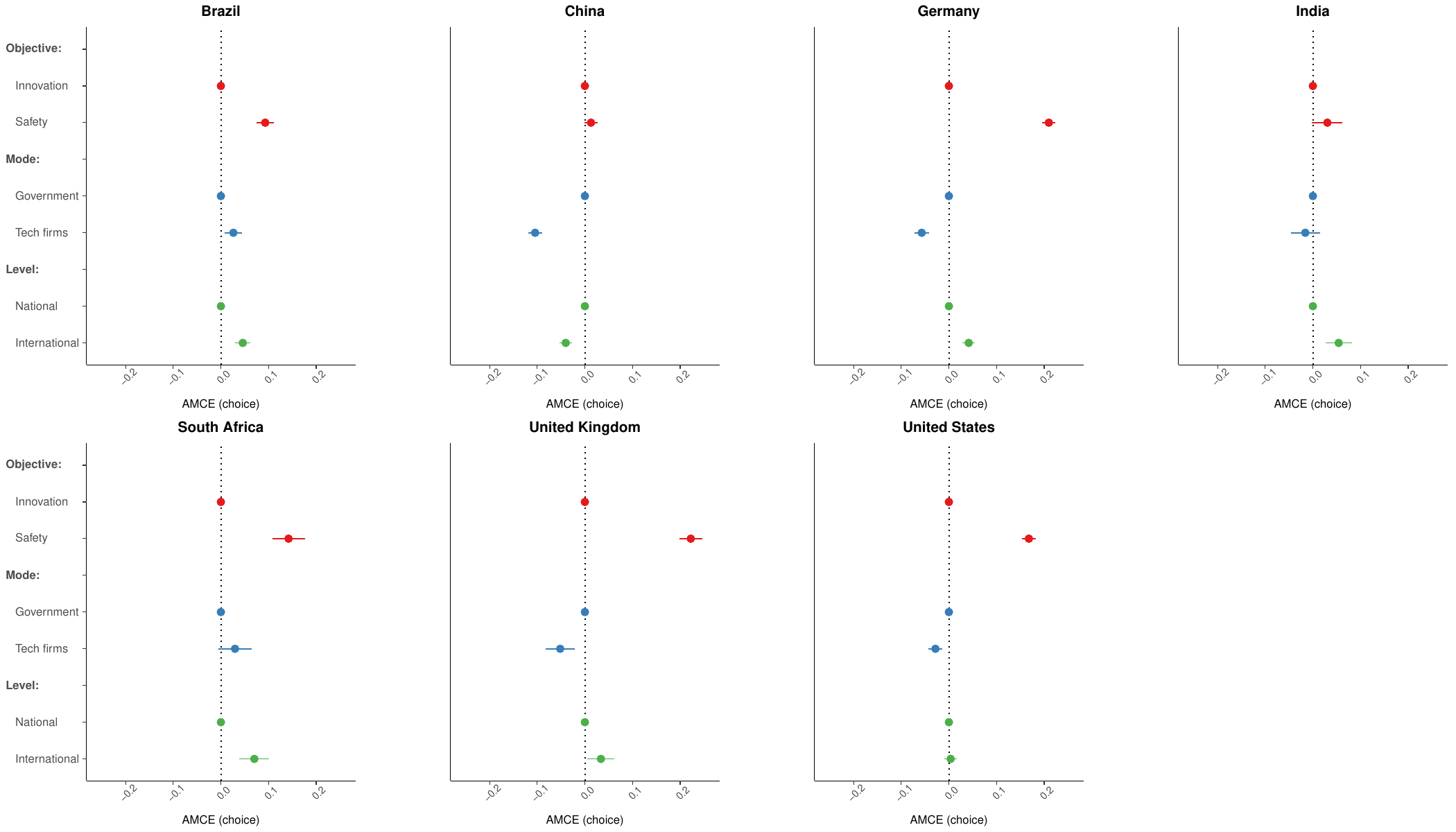}
\caption{\textbf{Average marginal component effects (AMCEs) for profile choice, by country.} Baseline categories and estimation as in fig.~\ref{fig:C1}.}
\label{fig:C2}
\end{figure}

At the country level (fig.~\ref{fig:C2}), the AMCE estimates mirror the marginal means reported in Figure 4 of the main text. The safety effect is largest in the United Kingdom and Germany and smallest in China and India, while the governance mode effect is largest in China.

\clearpage

\subsection{Rating Outcome Results}
\label{sec:D}

The main paper focuses on the binary choice outcome (which proposal the respondent preferred). Here we replicate the main analyses using the 7-point rating outcome (how much the respondent supported each proposal individually). The rating outcome provides a complementary measure that captures the intensity of support rather than just comparative preference.

\begin{figure}[H]
\centering
\includegraphics[width=\textwidth]{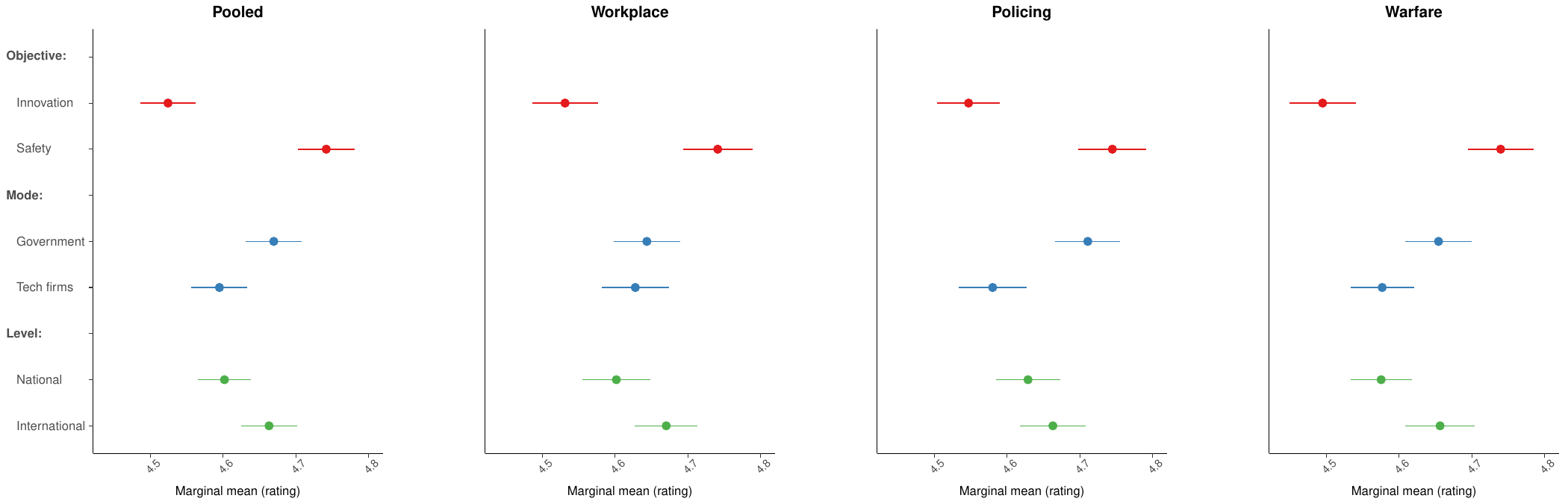}
\caption{\textbf{Marginal means for profile rating (7-point scale), pooled and by domain of application.} Each point shows the average rating assigned to profiles that include the indicated attribute level. Estimates use survey weights and respondent-clustered standard errors; horizontal error bars indicate 95\% confidence intervals.}
\label{fig:D1}
\end{figure}

\begin{figure}[H]
\centering
\includegraphics[width=\textwidth]{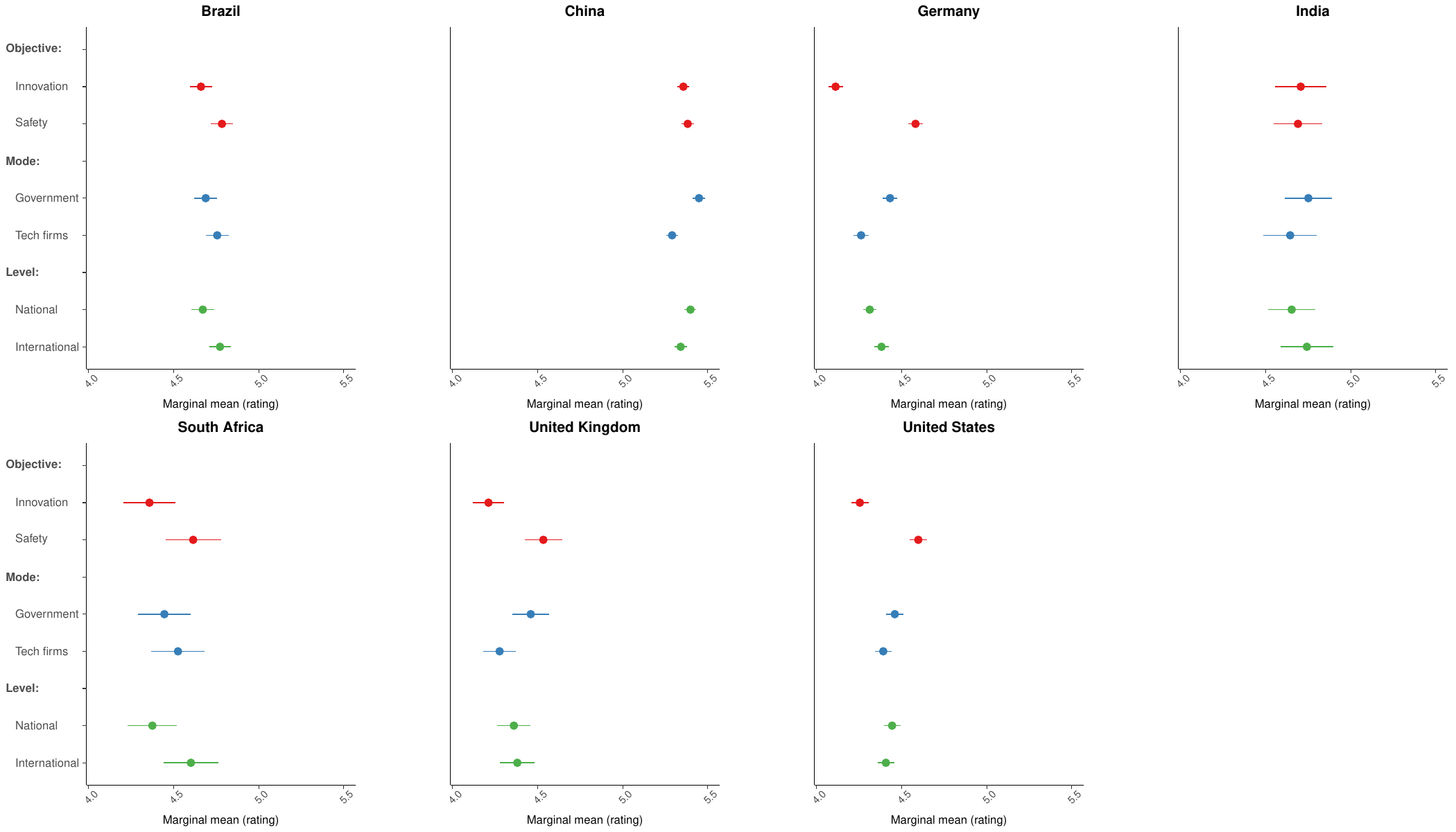}
\caption{\textbf{Marginal means for profile rating, by country.} Estimation as in fig.~\ref{fig:D1}.}
\label{fig:D2}
\end{figure}

The rating-based results closely parallel the choice-based findings. As shown in fig.~\ref{fig:D1}, safety-oriented proposals receive substantially higher ratings than innovation-oriented proposals across all domains. The pattern extends to the country level (fig.~\ref{fig:D2}), where the ranking of countries by the size of the safety--innovation gap closely mirrors the choice-based results reported in the main text. One notable difference is that average rating levels vary across countries, with Chinese respondents rating all proposals higher than respondents in other countries. This likely reflects cross-national differences in response styles rather than substantive differences in preferences, and illustrates why the forced choice outcome---which eliminates such scale-use heterogeneity---serves as the more informative primary measure.

\clearpage

\subsection{Robustness: Unweighted Estimates}
\label{sec:E}

\begin{figure}[H]
\centering
\includegraphics[width=\textwidth]{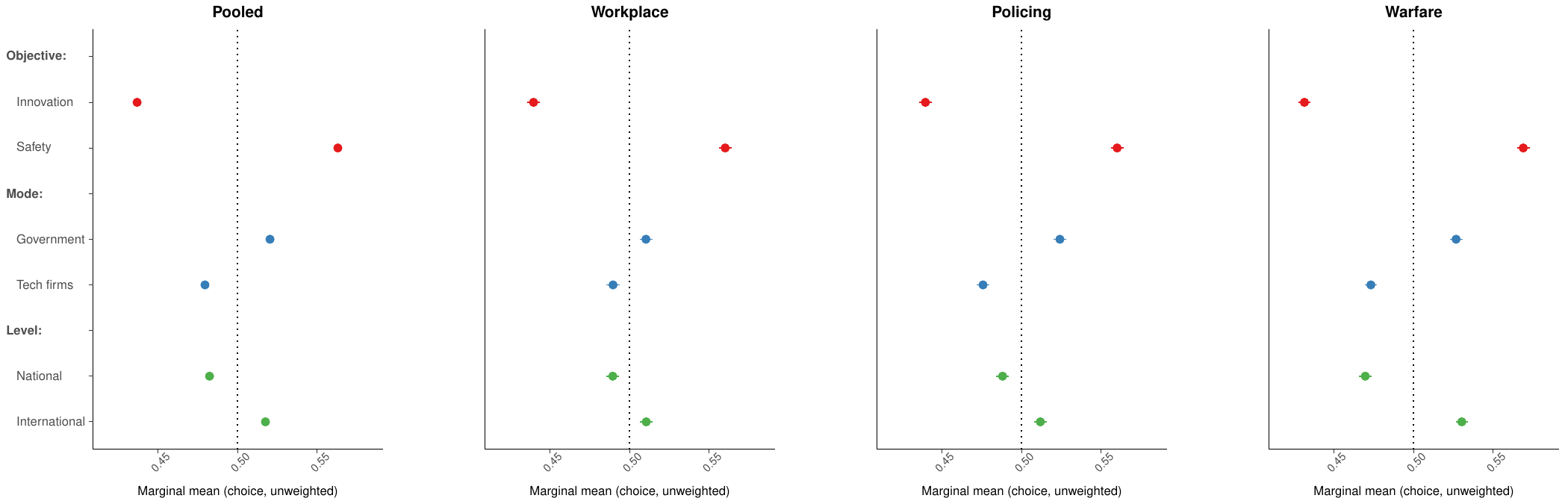}
\caption{\textbf{Marginal means for profile choice without survey weights, pooled and by domain of application.} Estimation as in the main text but with all weights set to 1.}
\label{fig:E1}
\end{figure}

\begin{figure}[H]
\centering
\includegraphics[width=\textwidth]{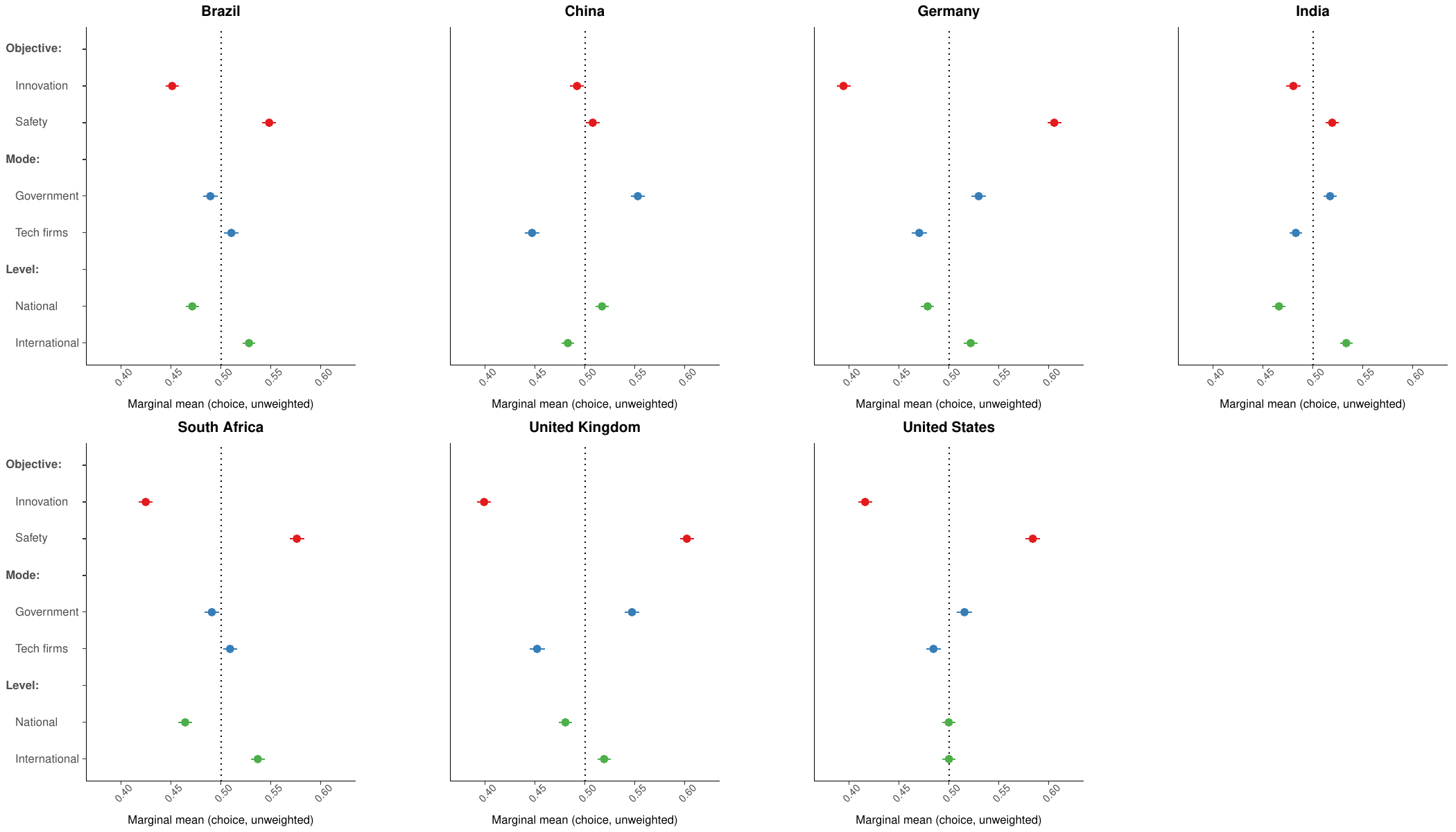}
\caption{\textbf{Marginal means for profile choice without survey weights, by country.}}
\label{fig:E2}
\end{figure}

All main analyses use survey weights to adjust for sampling imbalances. To assess whether the results are sensitive to the application of these weights, we re-estimate all marginal means without weighting. As shown in figs.~\ref{fig:E1} and~\ref{fig:E2}, the unweighted estimates point in the same direction as the weighted results and are similar or slightly larger in magnitude. Confidence intervals are tighter without weights, particularly in India and South Africa, where the weighting procedure assigns large weights to a small under-represented low-education subgroup. As a result, several preferences that fall just short of conventional significance in the weighted analysis---including the safety preference in China and India and the government-authority preference in India---reach conventional significance in the unweighted analysis. Reporting weighted estimates in the main text therefore provides a more conservative test of the patterns documented above.

\clearpage

\subsection{Robustness: Attention Check Subsample}
\label{sec:F}

Survey respondents may not always read question prompts carefully, which could attenuate estimated effects (\textit{61}). As described in Section~1.4.2, our attention check was relatively demanding, requiring respondents to read the full question text before selecting the instructed options. Approximately 61\% passed, with rates varying across countries (see table~\ref{tab:B5}). To assess whether inattentive respondents bias the results, we re-estimate all marginal means restricting the sample to the 8,631 respondents who passed.

\begin{figure}[H]
\centering
\includegraphics[width=\textwidth]{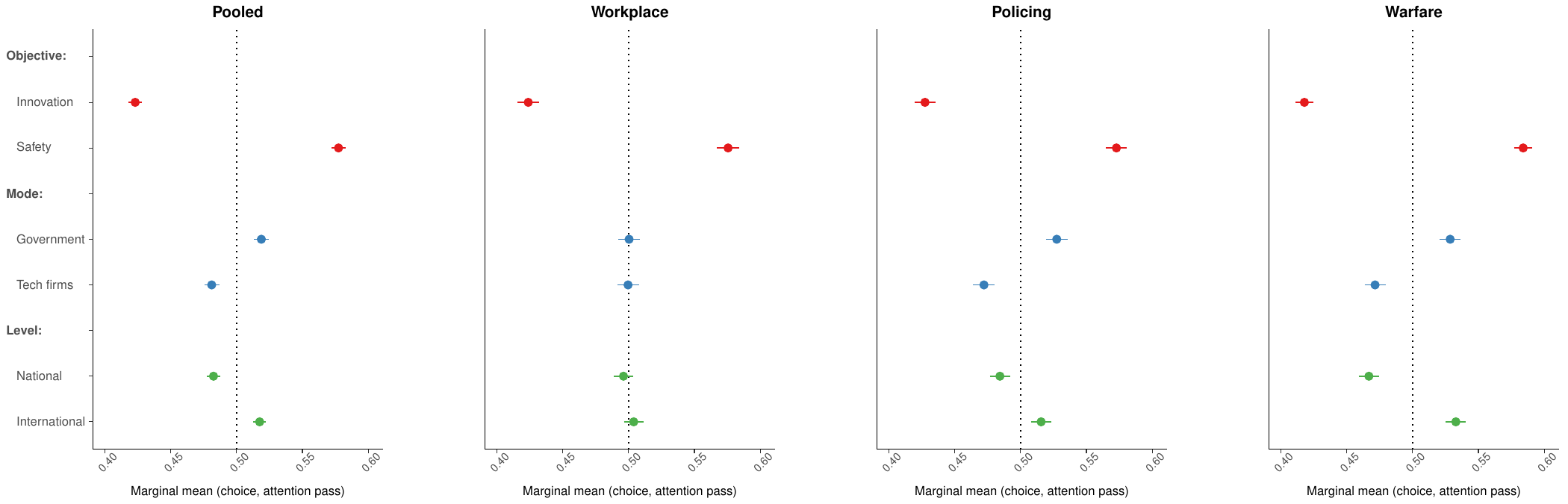}
\caption{\textbf{Marginal means for profile choice, restricted to respondents passing the attention check (N = 8,631).} Pooled and by domain of application.}
\label{fig:F1}
\end{figure}

\begin{figure}[H]
\centering
\includegraphics[width=\textwidth]{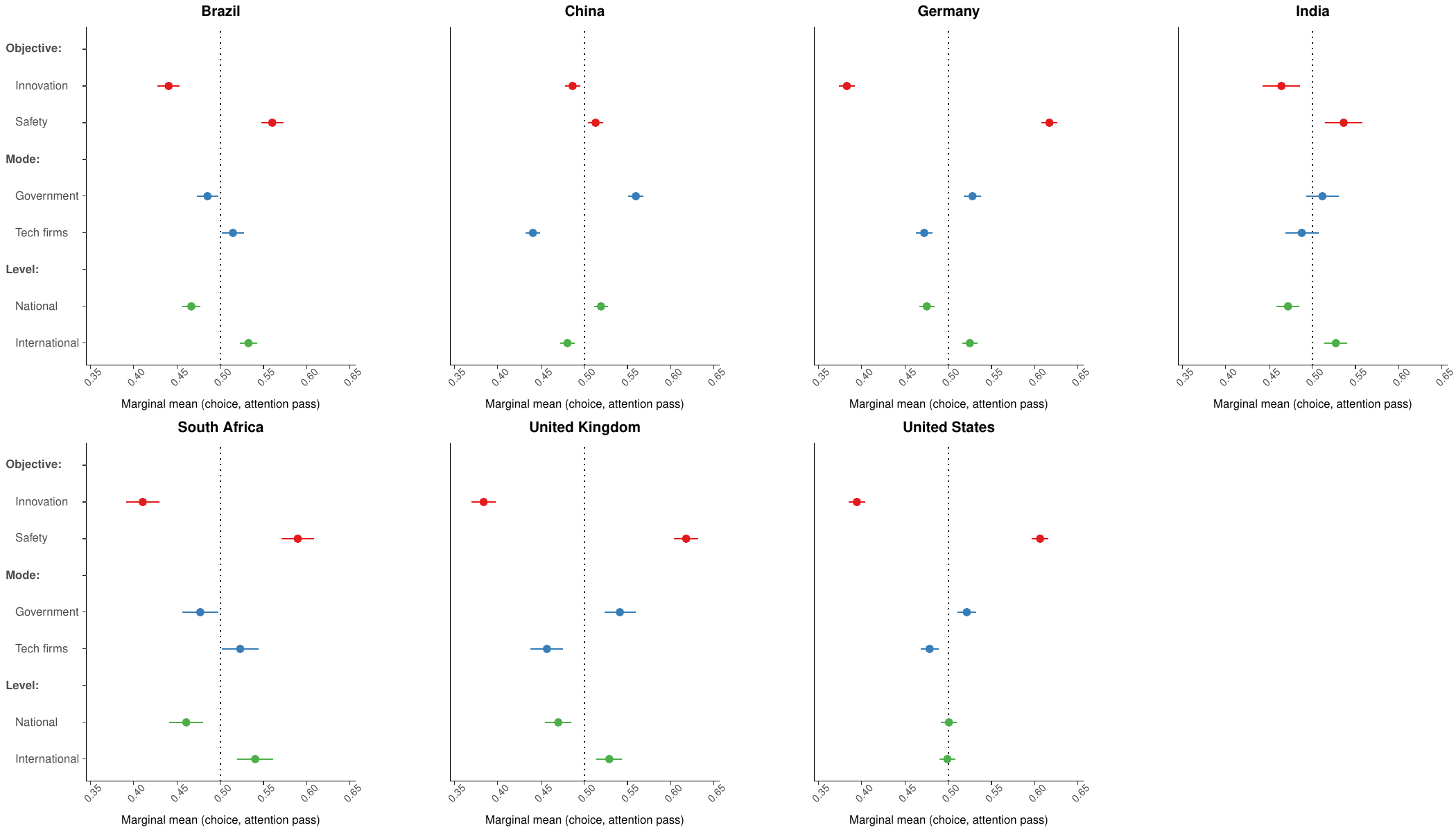}
\caption{\textbf{Marginal means for profile choice, restricted to respondents passing the attention check, by country.}}
\label{fig:F2}
\end{figure}

As shown in figs.~\ref{fig:F1} and~\ref{fig:F2}, the results are substantively unchanged. If anything, estimated preferences are slightly more pronounced among attentive respondents, suggesting that inattentiveness introduces noise rather than systematic bias.

\clearpage

\clearpage

\subsection{Robustness: Block Order Effects}
\label{sec:G}

The three thematic blocks (workplace, policing, warfare) were presented in randomized order across respondents. Earlier blocks might shape responses to later ones through priming, anchoring, or fatigue. To test for such carryover effects, we estimate attribute effects separately for each block depending on whether it was shown first, second, or third, and formally test for interactions between attribute effects and block position.

\begin{table}[H]
\centering
\caption{\textbf{Block order interaction tests.} Each row tests whether the attribute effect differs when the block is shown in a later versus earlier position. Estimates from weighted OLS with respondent-clustered standard errors.}
\label{tab:H2}
\small
\scriptsize
\begin{tabular}{llrrrr}
\toprule
\textbf{Interaction} & \textbf{Term} & \textbf{Estimate} & \textbf{Std. error} & \textbf{$t$-value} & \textbf{$p$-value} \\
\midrule
Objective $\times$ Position & Safety $\times$ Position 2 & $-$0.004 & 0.009 & $-$0.40 & 0.690 \\
 & Safety $\times$ Position 3 & 0.015 & 0.009 & 1.64 & 0.101 \\
\addlinespace
Mode $\times$ Position & Government $\times$ Position 2 & 0.040 & 0.009 & 4.26 & $<$0.001 \\
 & Government $\times$ Position 3 & 0.029 & 0.009 & 3.11 & 0.002 \\
\addlinespace
Level $\times$ Position & International $\times$ Position 2 & $-$0.006 & 0.010 & $-$0.62 & 0.534 \\
 & International $\times$ Position 3 & 0.028 & 0.009 & 3.07 & 0.002 \\
\bottomrule
\end{tabular}

\end{table}

Table~\ref{tab:H2} reports the results. For the Objective dimension, neither interaction with block position is significant. For the Mode dimension, the preference for government authority is significantly stronger when the block is shown second ($p < 0.001$) or third ($p = 0.002$). For the Level dimension, the preference for international governance is significantly stronger when the block is shown third ($p = 0.002$) but not second ($p = 0.534$). The pooled estimates reported in the main text average over all block orders by design.

\clearpage

\subsection{Robustness: Profile Order Effects}
\label{sec:H}

In paired conjoint tasks, respondents may disproportionately select the first or second proposal regardless of its attributes. Table~\ref{tab:I1} shows that Proposal 1 was chosen 55.8\% of the time, indicating a primacy bias toward the first-listed proposal. Left-profile primacy effects of this magnitude are common in paired conjoint designs (\textit{62}). Because attribute levels are balanced across proposal positions (Section~\ref{sec:B}), this bias does not affect the estimated marginal means or AMCEs.

\begin{table}[H]
\centering
\caption{\textbf{Profile order bias: mean choice rate by proposal position.} The expected rate under no positional bias is 0.50.}
\label{tab:I1}
\small
\scriptsize
\begin{tabular}{lrr}
\toprule
\textbf{Proposal} & \textbf{Mean Chosen} & \textbf{N} \\
\midrule
1 & 0.5582 & 128151 \\
2 & 0.4418 & 128151 \\
\bottomrule
\end{tabular}
\end{table}

\clearpage

\subsection{Attribute Interactions}
\label{sec:I}

The main analysis and the predicted-support profiles (Fig.~\ref{fig:predsupport}) assume that the effects of the three conjoint attributes combine additively. If this assumption is violated---for example, if respondents particularly favor safety when combined with government authority---the predicted rankings of regulatory profiles could differ from those implied by the additive model. We test for two-way and three-way interactions between attributes to assess the validity of this assumption.

\begin{table}[H]
\centering
\caption{\textbf{Two-way attribute interactions (interaction terms only).} From weighted OLS regression of profile choice on attribute indicators and all pairwise interactions, with respondent-clustered standard errors.}
\label{tab:J1}
\small
\scriptsize
\begin{tabular}{lrrrr}
\toprule
\textbf{Interaction term} & \textbf{Estimate} & \textbf{Std. error} & \textbf{$t$-value} & \textbf{$p$-value} \\
\midrule
Safety $\times$ Government & 0.013 & 0.008 & 1.77 & 0.077 \\
Safety $\times$ International & $-$0.004 & 0.007 & $-$0.61 & 0.541 \\
Government $\times$ International & 0.007 & 0.008 & 0.94 & 0.347 \\
\bottomrule
\end{tabular}

\end{table}

Table~\ref{tab:J1} reports the two-way interaction tests. None of the pairwise interactions reach statistical significance at the 0.05 level in this specification.

\begin{table}[H]
\centering
\caption{\textbf{Three-way attribute interaction (interaction terms only).} From weighted OLS regression of profile choice on all attribute indicators, pairwise interactions, and the three-way interaction, with respondent-clustered standard errors.}
\label{tab:J2}
\small
\scriptsize
\begin{tabular}{lrrrr}
\toprule
\textbf{Interaction term} & \textbf{Estimate} & \textbf{Std. error} & \textbf{$t$-value} & \textbf{$p$-value} \\
\midrule
Safety $\times$ Government & 0.023 & 0.010 & 2.21 & 0.027 \\
Safety $\times$ International & 0.005 & 0.010 & 0.54 & 0.592 \\
Government $\times$ International & 0.017 & 0.011 & 1.51 & 0.131 \\
Safety $\times$ Government $\times$ International & $-$0.019 & 0.015 & $-$1.29 & 0.196 \\
\bottomrule
\end{tabular}

\end{table}

In the three-way model (table~\ref{tab:J2}), the safety--government two-way interaction reaches significance ($p = 0.03$), suggesting a modest complementarity between safety objectives and government authority. However, the three-way interaction itself is not significant, and the effect size is small relative to the main effects. The additive specification provides a reasonable approximation.

\clearpage

\subsection{Extended Subgroup Analyses}
\label{sec:J}

The main paper reports heterogeneous treatment effects for five pre-registered moderators (H4--H8) using marginal means. Here we present the full set of subgroup analyses, including both marginal means and AMCEs for each moderator, as well as additional exploratory moderators (AI knowledge, trust in government, trust in the UN, trust in technology companies). Subgroup splits follow the procedure described in Section~1.6.

\subsubsection{Pre-registered moderators (H4--H8)}

\begin{figure}[H]
\centering
\includegraphics[width=\textwidth]{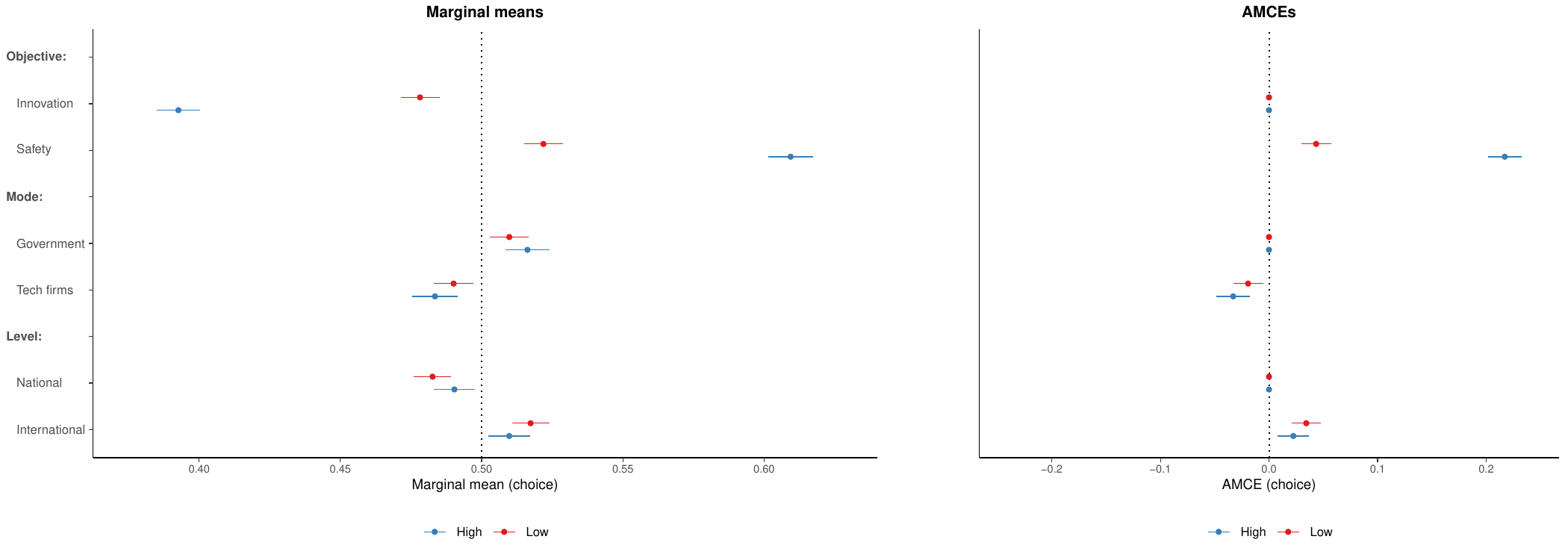}
\caption{\textbf{Marginal means (left) and AMCEs (right) for profile choice by perceived AI risk (H4).} ``Low'' denotes respondents scoring 1--3 on AI risk; ``High'' denotes respondents scoring 5--7. Respondents at the midpoint (4) are excluded.}
\label{fig:K_h4}
\end{figure}

\textbf{H4: Perceived AI risk.} We hypothesized that respondents who view AI as more of a risk than an opportunity would exhibit a stronger preference for safety-oriented regulation. Fig.~\ref{fig:K_h4} confirms this prediction: the safety--innovation gap in marginal means is substantially larger among high-risk respondents than among low-risk respondents. The difference operates primarily through the Objective dimension; preferences over Mode and Level are less affected by risk perceptions.

\begin{figure}[H]
\centering
\includegraphics[width=\textwidth]{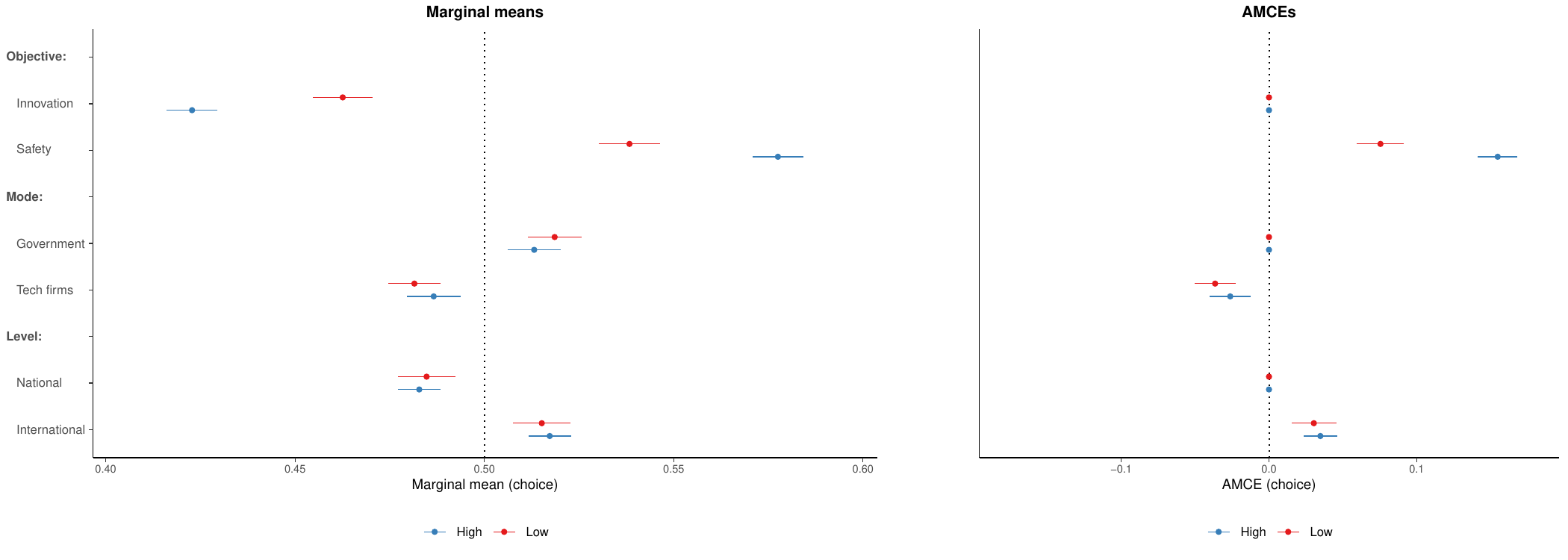}
\caption{\textbf{Marginal means (left) and AMCEs (right) for profile choice by perceived AI unpredictability (H5).} The unpredictability variable is reverse-coded from Q2 (8 $-$ Q2). ``Low'' = scores 1--3 (perceive AI as predictable); ``High'' = scores 5--7 (perceive AI as unpredictable).}
\label{fig:K_h5}
\end{figure}

\textbf{H5: Perceived AI unpredictability.} We hypothesized that respondents who perceive the consequences of AI as more unpredictable would prefer safety over innovation more strongly. Fig.~\ref{fig:K_h5} supports this hypothesis. Respondents who perceive AI as unpredictable show a stronger preference for safety compared to those who perceive AI as predictable.

\begin{figure}[H]
\centering
\includegraphics[width=\textwidth]{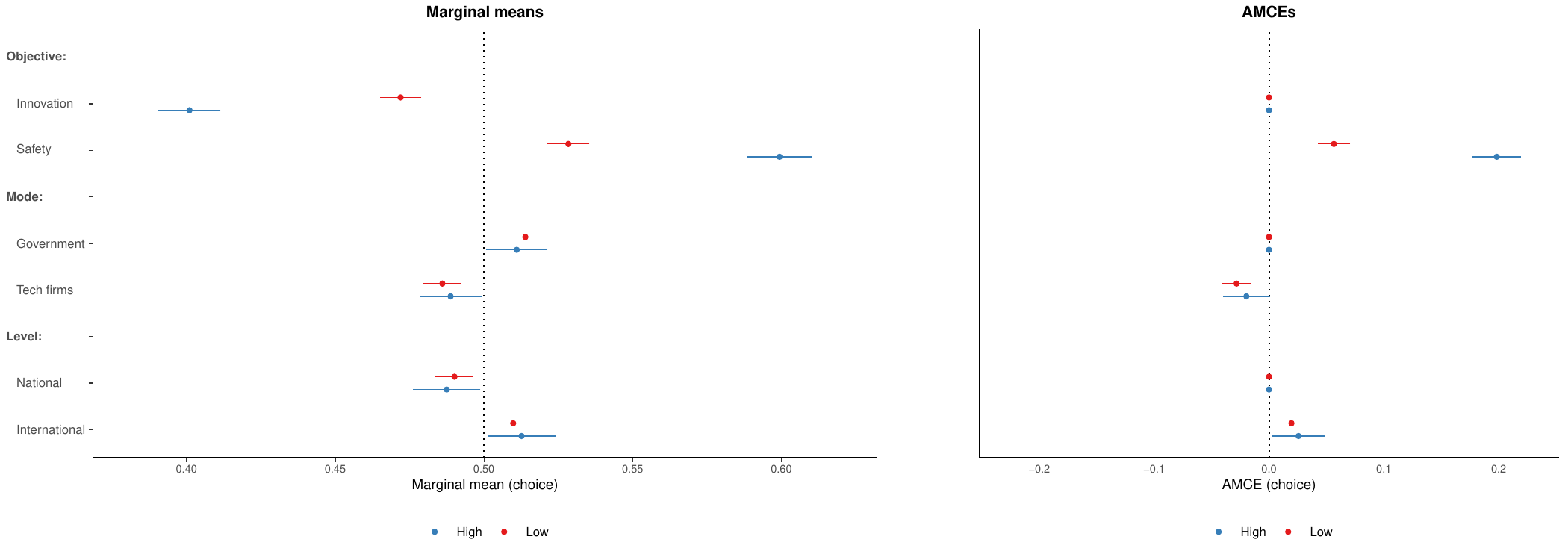}
\caption{\textbf{Marginal means (left) and AMCEs (right) by expected direction of AI impact, workplace domain (H6).} ``Low'' = expects positive impact (scores 1--3); ``High'' = expects negative impact (scores 5--7).}
\label{fig:K_h6a}
\end{figure}

\begin{figure}[H]
\centering
\includegraphics[width=\textwidth]{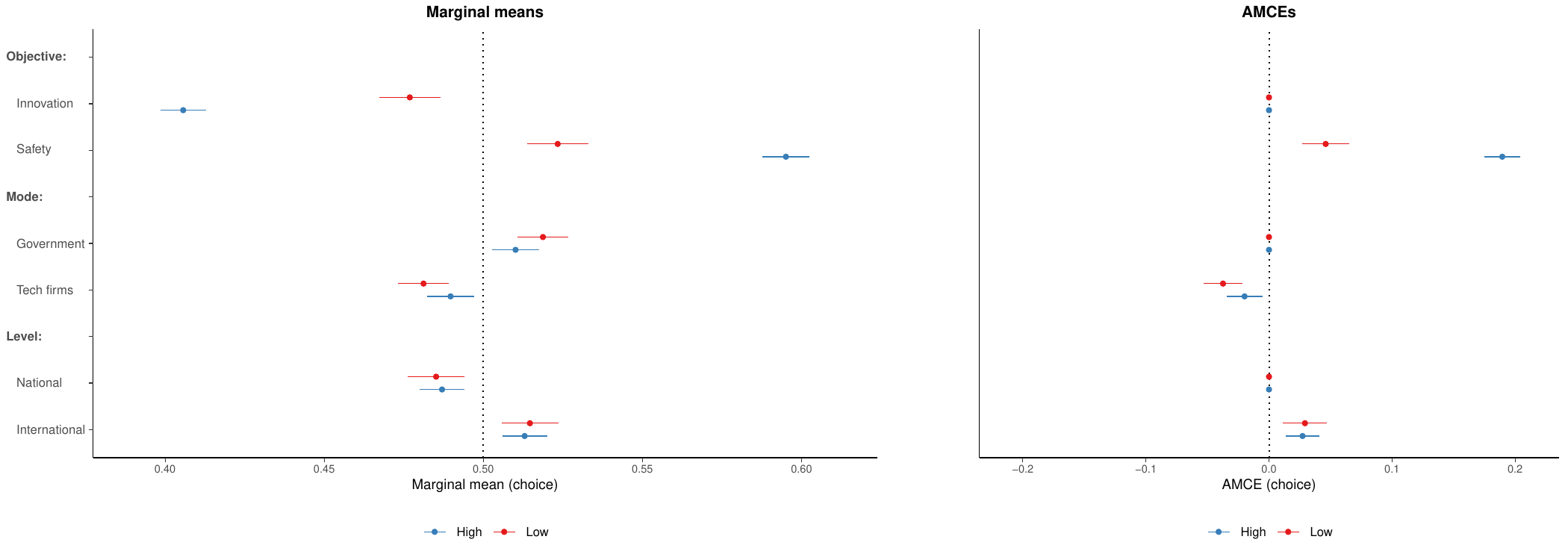}
\caption{\textbf{Marginal means (left) and AMCEs (right) by expected direction of AI impact, policing domain (H6).}}
\label{fig:K_h6b}
\end{figure}

\begin{figure}[H]
\centering
\includegraphics[width=\textwidth]{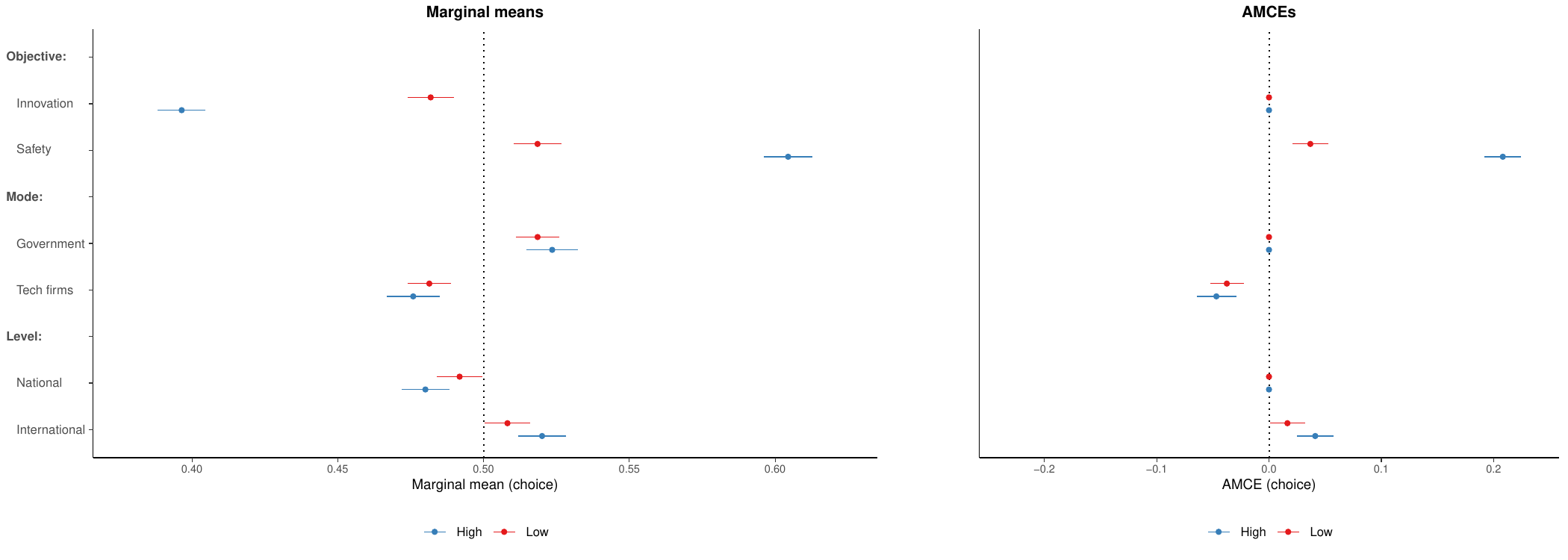}
\caption{\textbf{Marginal means (left) and AMCEs (right) by expected direction of AI impact, warfare domain (H6).}}
\label{fig:K_h6c}
\end{figure}

\textbf{H6: Personal affectedness.} We hypothesized that respondents who expect to be more negatively affected by AI would prefer safety more strongly. We measure the expected direction of AI's personal impact separately within each domain (Q12, Q20, Q28; 1 = always positive, 7 = always negative). Figs.~\ref{fig:K_h6a}--\ref{fig:K_h6c} present the results. Across all three domains, respondents who expect more negative personal consequences of AI show a stronger preference for safety-oriented regulation, supporting H6.

\begin{figure}[H]
\centering
\includegraphics[width=\textwidth]{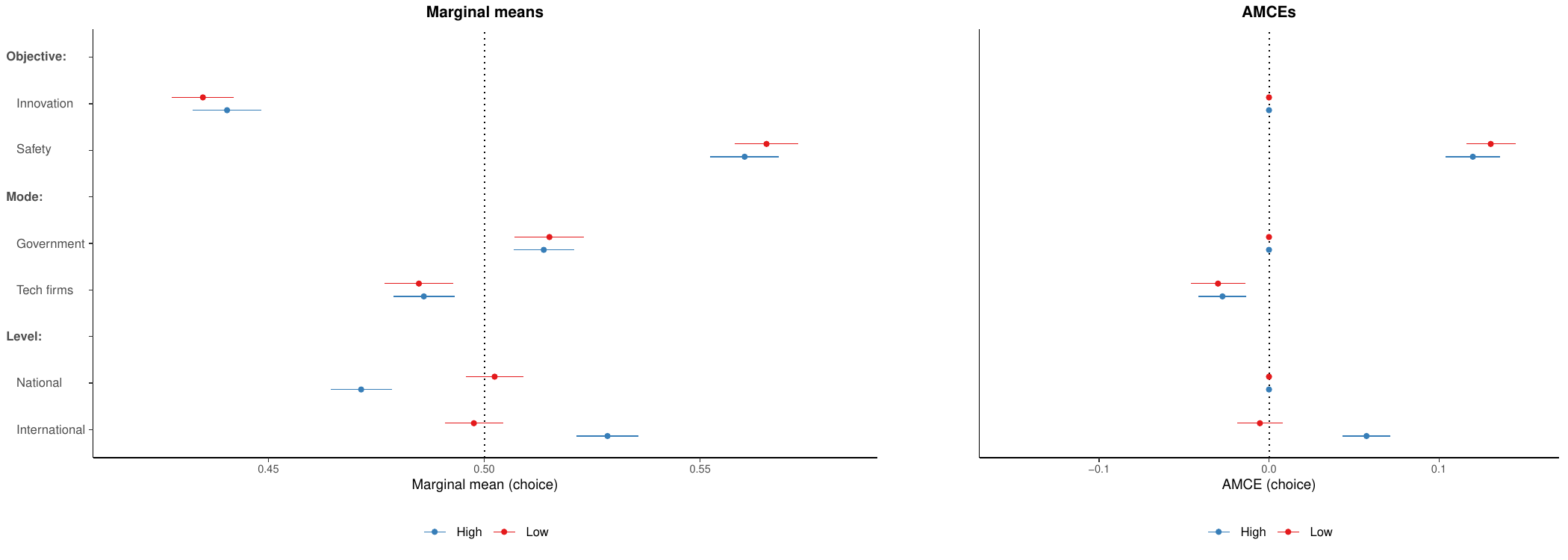}
\caption{\textbf{Marginal means (left) and AMCEs (right) for profile choice by internationalism (H7).} ``Low'' = scores 1--3; ``High'' = scores 5--7.}
\label{fig:K_h7}
\end{figure}

\textbf{H7: Internationalism.} We hypothesized that respondents with stronger internationalist attitudes would show a stronger preference for international over national regulation. Fig.~\ref{fig:K_h7} supports this: the international--national gap is positive and sizable among respondents with pro-internationalist attitudes, whereas it is essentially absent among those with more nationalist attitudes.

\begin{figure}[H]
\centering
\includegraphics[width=\textwidth]{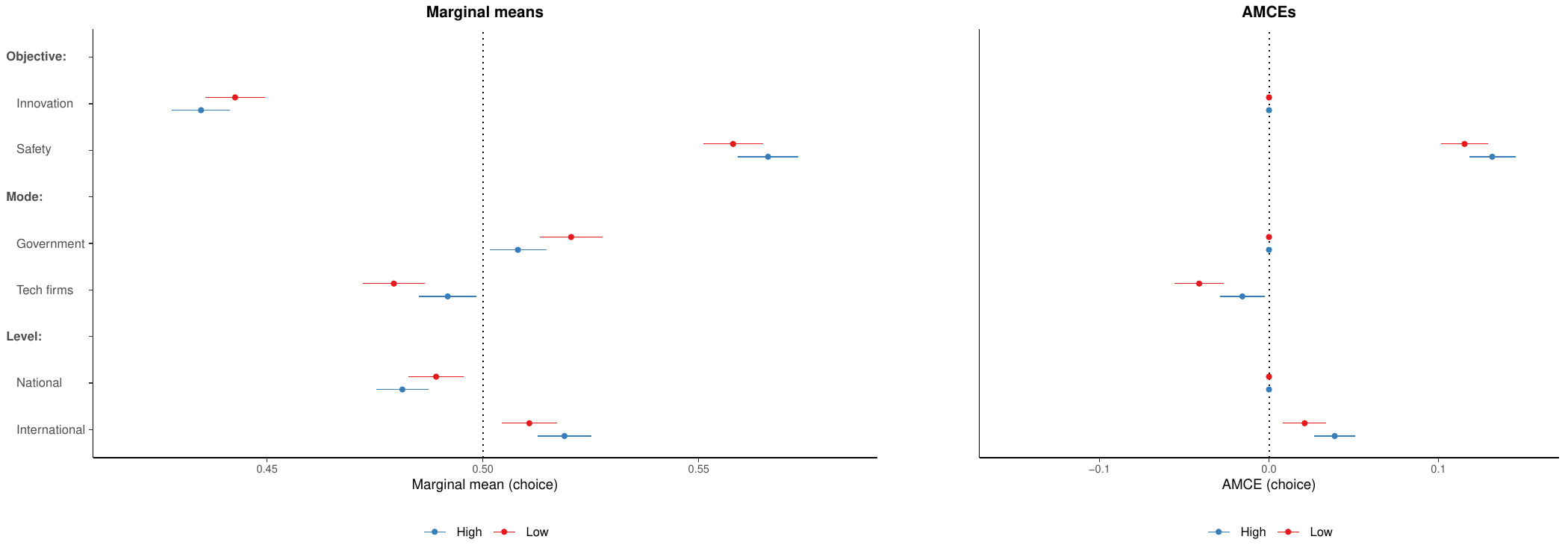}
\caption{\textbf{Marginal means (left) and AMCEs (right) for profile choice by ideology (H8).} ``Left'' = ideology index 1--3; ``Right'' = ideology index 5--7 (averaged across Q38--Q40).}
\label{fig:K_h8}
\end{figure}

\textbf{H8: Left--right ideology.} We hypothesized that left-leaning respondents would show a stronger preference for government regulation over private self-regulation. Fig.~\ref{fig:K_h8} provides only limited support for this hypothesis: the differences in the Mode dimension between left and right respondents are modest. More generally, ideological orientation does not appear to be a strong predictor of preferences along any of the three regulatory dimensions, suggesting that attitudes toward AI governance are not primarily structured by conventional left--right ideology.

\subsubsection{Exploratory moderators}

In addition to the pre-registered moderators, we examine several exploratory moderators that may structure preferences over AI governance but were not the subject of directional hypotheses. These analyses are not pre-registered, involve multiple comparisons, and should be interpreted as exploratory.

\begin{figure}[H]
\centering
\includegraphics[width=\textwidth]{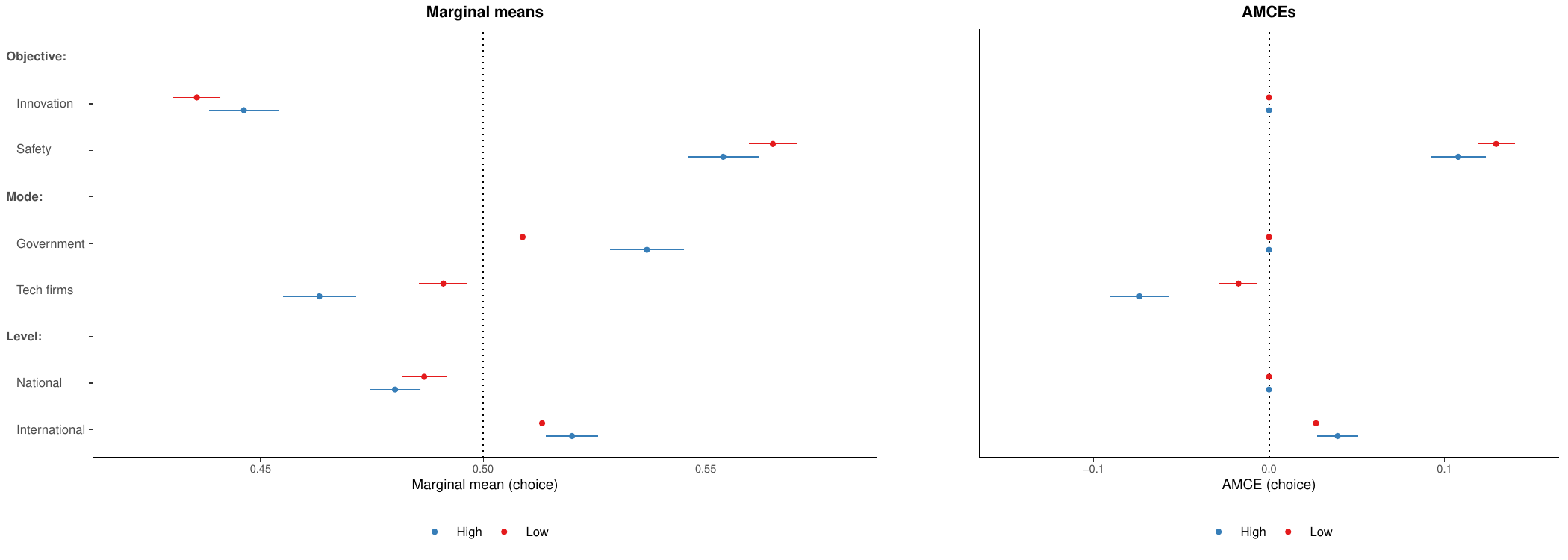}
\caption{\textbf{Marginal means (left) and AMCEs (right) for profile choice by AI knowledge.} ``High'' = all three knowledge questions answered correctly (N $\approx$ 3,274); ``Low'' = at least one incorrect (N $\approx$ 10,965).}
\label{fig:K_knowledge}
\end{figure}

\textbf{AI knowledge.} Respondents who correctly answered all three factual questions about AI (identifying face recognition as an AI application, Anthropic as an AI company, and deepfakes as a feared consequence of AI) are classified as having high AI knowledge. As shown in fig.~\ref{fig:K_knowledge}, high-knowledge respondents show a stronger preference for government over private authority compared to low-knowledge respondents. The preference for safety over innovation is similar across the two groups.

\begin{figure}[H]
\centering
\includegraphics[width=\textwidth]{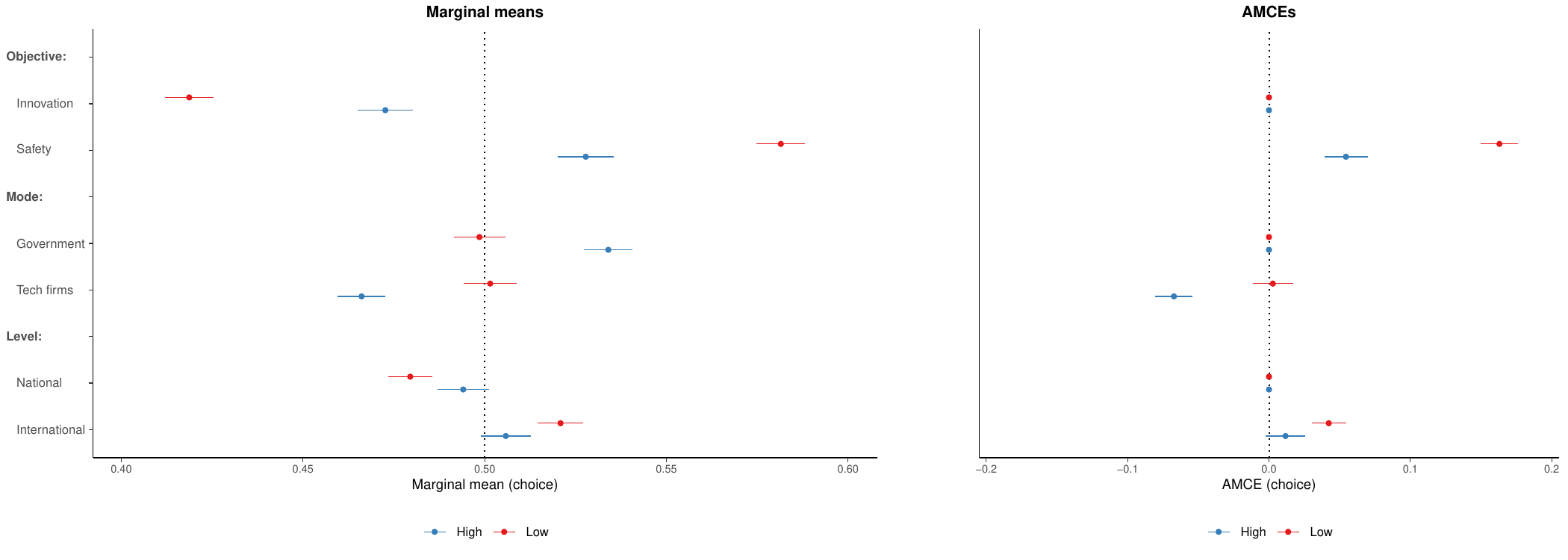}
\caption{\textbf{Marginal means (left) and AMCEs (right) for profile choice by trust in national government.} Subgroups based on median split of Q41 (confidence in government).}
\label{fig:K_trustgov}
\end{figure}

\textbf{Trust in government.} Respondents who express greater confidence in their national government might be expected to favor government-led regulation more strongly. Fig.~\ref{fig:K_trustgov} offers support for this expectation: high-trust respondents show a stronger preference for government over private authority compared to low-trust respondents.

\begin{figure}[H]
\centering
\includegraphics[width=\textwidth]{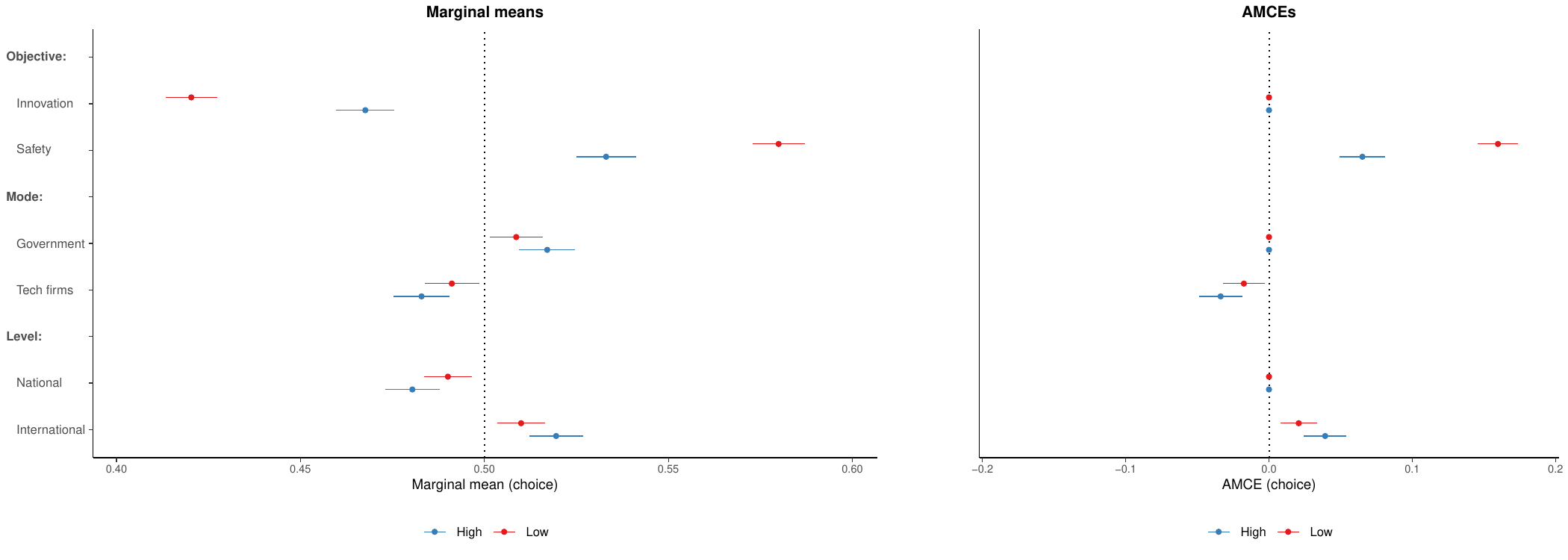}
\caption{\textbf{Marginal means (left) and AMCEs (right) for profile choice by trust in the United Nations.} Subgroups based on median split of Q42 (confidence in the UN).}
\label{fig:K_trustun}
\end{figure}

\textbf{Trust in the United Nations.} Respondents with greater confidence in the UN might be expected to favor international regulation. Fig.~\ref{fig:K_trustun} shows that high-UN-trust respondents do indeed exhibit a slightly stronger preference for international over national governance, though the difference is modest.

\begin{figure}[H]
\centering
\includegraphics[width=\textwidth]{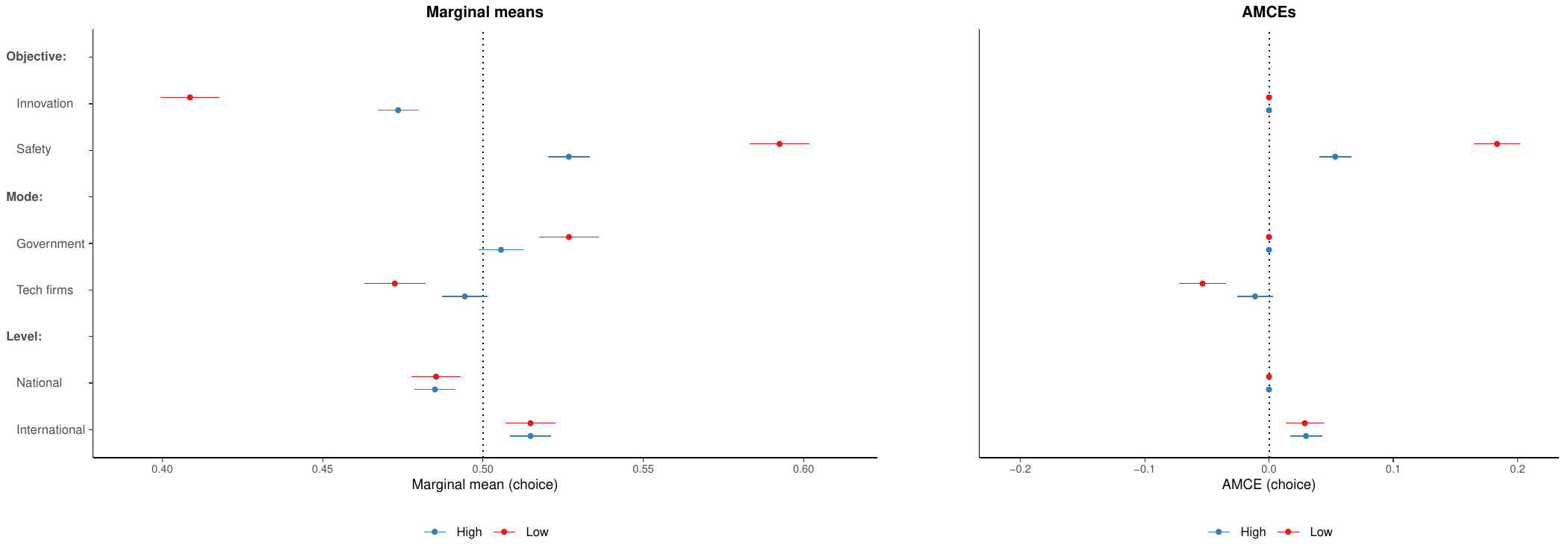}
\caption{\textbf{Marginal means (left) and AMCEs (right) for profile choice by trust in technology companies.} Subgroups based on median split of Q43 (confidence in technology companies).}
\label{fig:K_trusttech}
\end{figure}

\textbf{Trust in technology companies.} If respondents trust technology firms, they might be more willing to support private self-regulation over government authority. Fig.~\ref{fig:K_trusttech} provides some evidence for this pattern: high-trust respondents show a weaker preference for government authority compared to low-trust respondents. However, even among respondents with high trust in technology companies, government authority is still slightly preferred.

\clearpage

\subsection{Country $\times$ Domain Breakdowns}
\label{sec:K}

The main text presents country-level results pooled across all three application domains and domain-level results pooled across all countries. Here we examine the full cross-classification: preferences within each country separately for each domain. This allows us to assess whether the country-level patterns identified in the main text are consistent across application areas, or whether specific country--domain combinations produce distinctive patterns.

\begin{figure}[H]
\centering
\includegraphics[width=\textwidth]{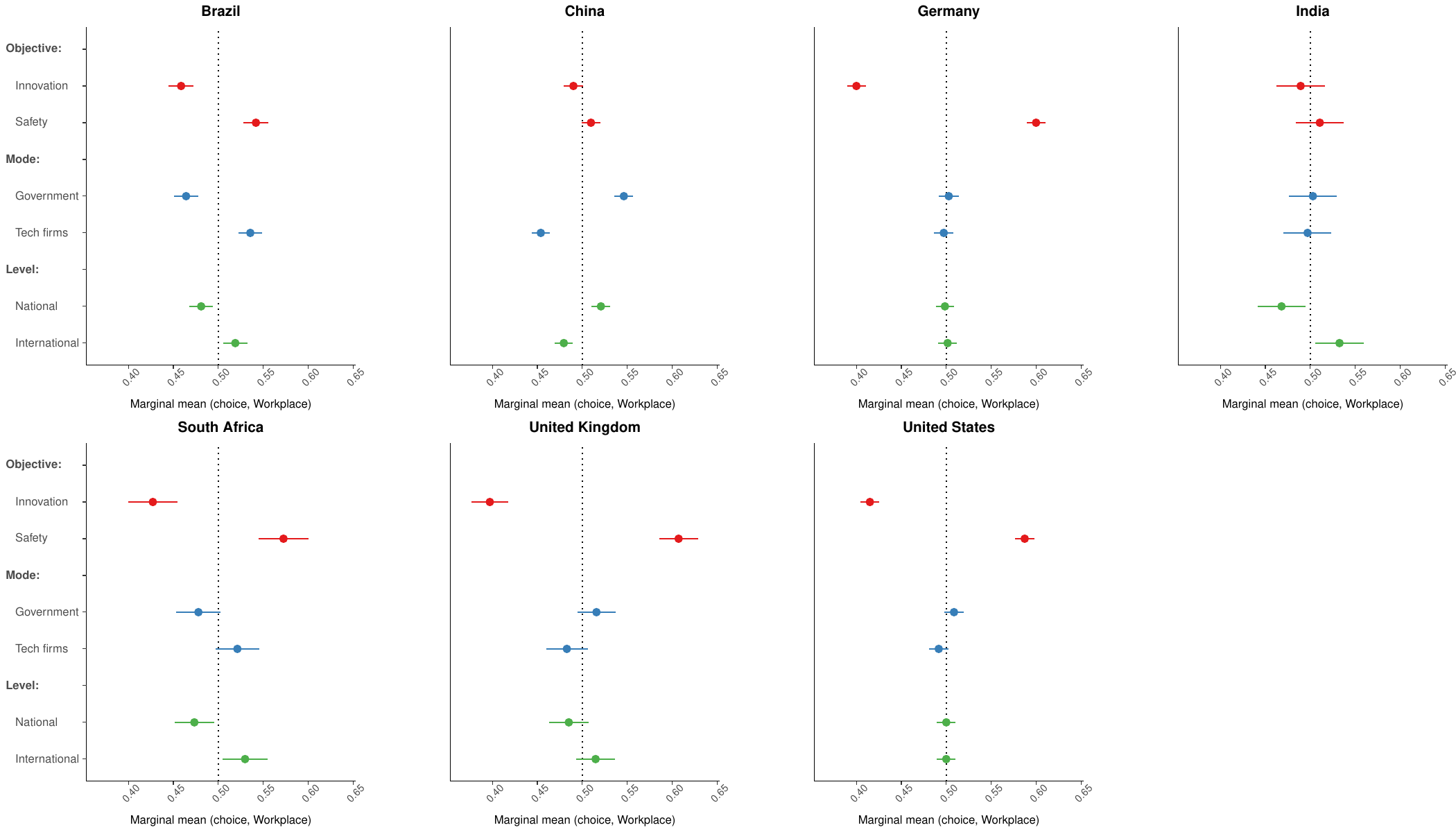}
\caption{\textbf{Marginal means for profile choice in the workplace domain, by country.} Each point is the weighted mean probability that a profile is chosen when it includes the indicated attribute level. Survey weights and respondent-clustered standard errors; error bars indicate 95\% confidence intervals.}
\label{fig:L_work}
\end{figure}

Figs.~\ref{fig:L_work}--\ref{fig:L_war} present marginal means by country for each of the three domains. The patterns are similar to the main findings. The preference for safety over innovation is evident in all countries across all domains, though it is weakest in China and India regardless of the application area. The preference for government authority is strongest in China across all three domains. The preference for international governance is most pronounced in the warfare domain across most countries, in line with the pooled results.

Certain country--domain combinations stand out. In the workplace domain, Brazil and South Africa show a slight preference for private over government authority---a pattern that does not appear in the policing or warfare domains for these countries. This may reflect the perception that workplace regulation falls more naturally within the purview of the private sector. In the warfare domain, China shows a strong preference for national governance, whereas the United States shows no significant difference between national and international approaches.

\begin{figure}[H]
\centering
\includegraphics[width=\textwidth]{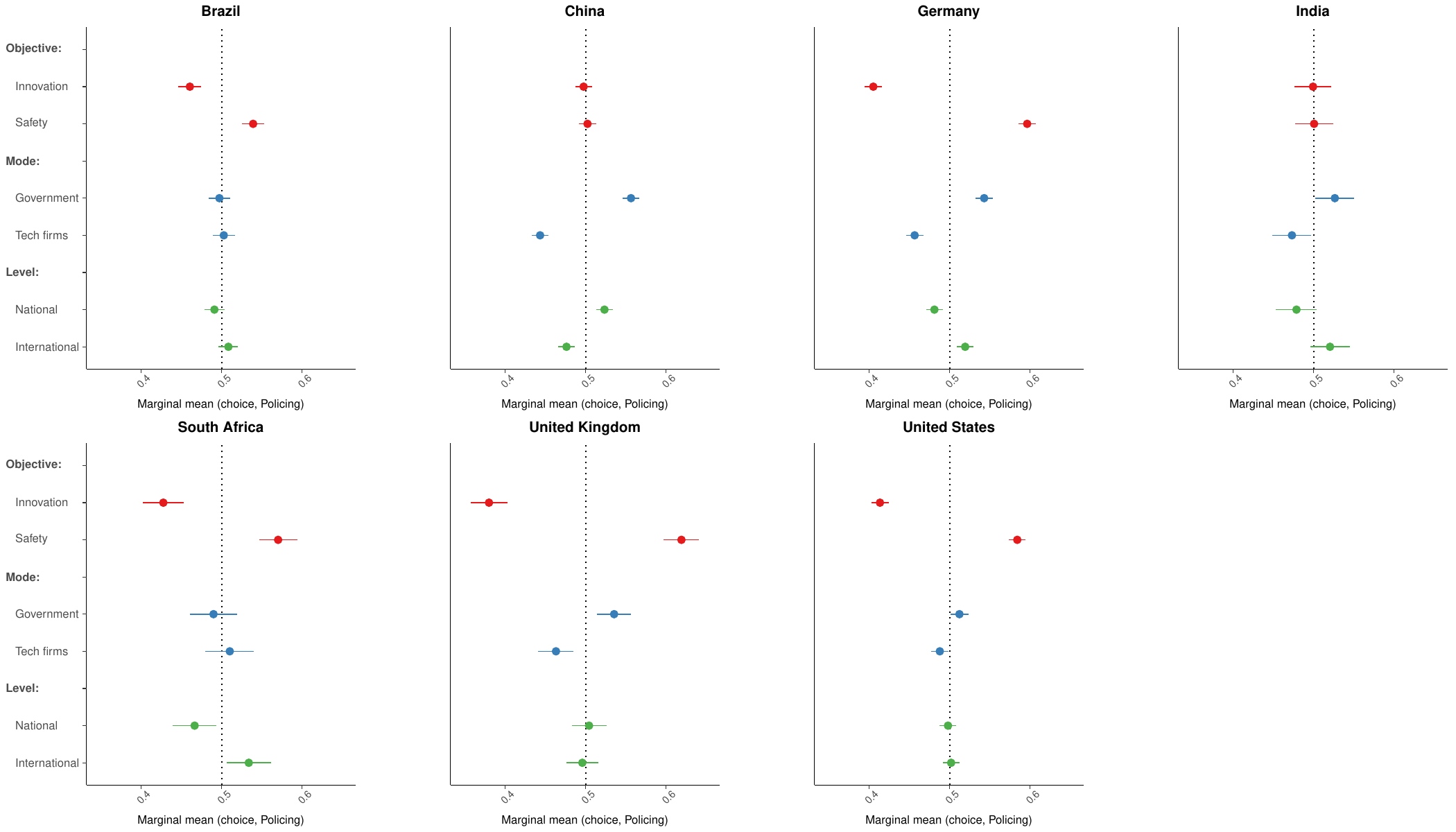}
\caption{\textbf{Marginal means for profile choice in the policing domain, by country.} Estimation as in fig.~\ref{fig:L_work}.}
\end{figure}

\begin{figure}[H]
\centering
\includegraphics[width=\textwidth]{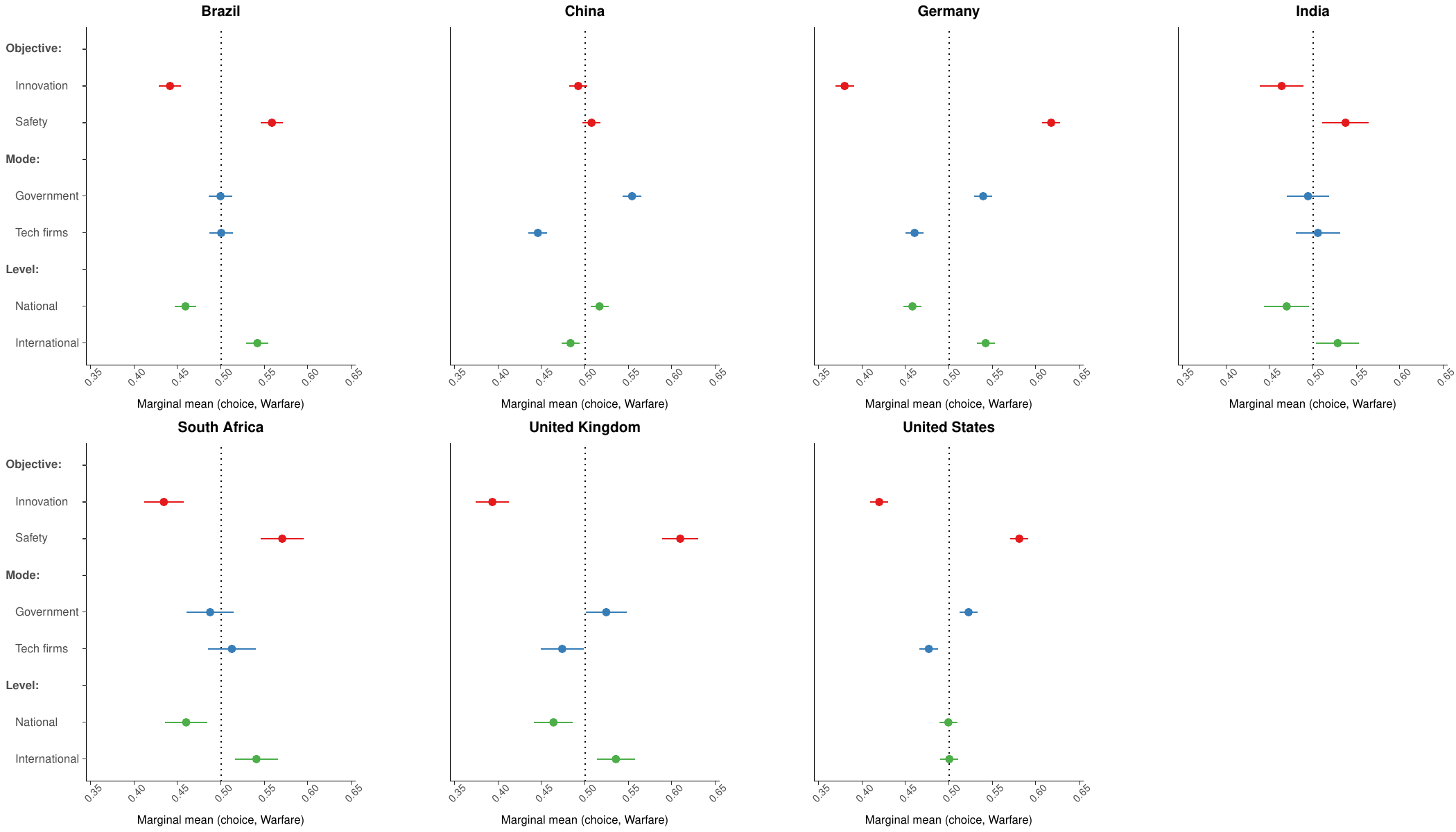}
\caption{\textbf{Marginal means for profile choice in the warfare domain, by country.} Estimation as in fig.~\ref{fig:L_work}.}
\label{fig:L_war}
\end{figure}

\clearpage

\subsection{Predicted Support}
\label{sec:L}

\begin{figure}[H]
\centering
\includegraphics[width=\textwidth]{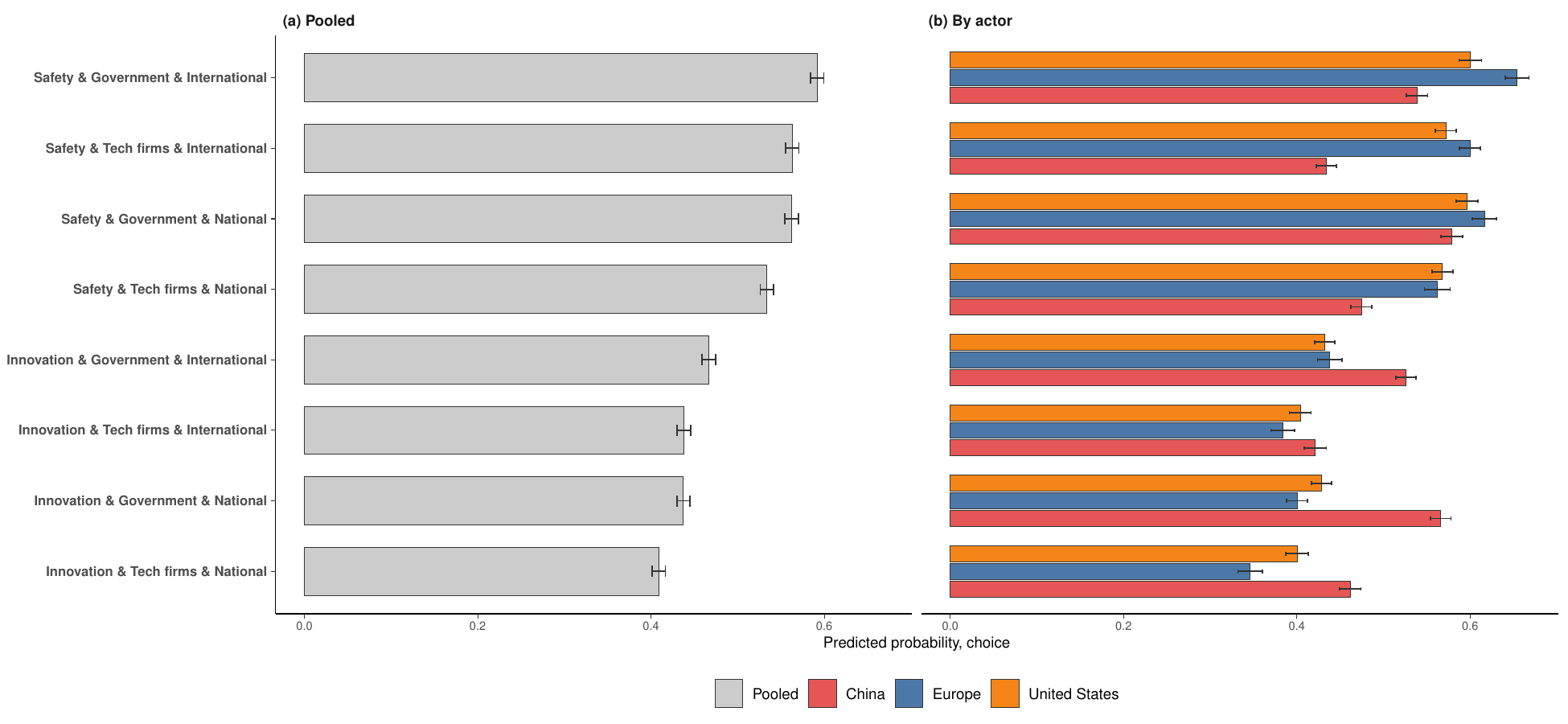}
\caption{\textbf{Predicted levels of support.} Bars show predicted probabilities of choosing each of the eight possible regulatory profiles, derived from a weighted linear probability model of profile choice on the three conjoint attributes (objective, mode, level). Panel (a) shows pooled estimates across all seven countries; panel (b) shows estimates for China, Europe (Germany and the UK), and the US. Error bars indicate 95\% confidence intervals based on respondent-clustered standard errors.}
\label{fig:predsupport}
\end{figure}

We examine the substantive implications of these preferences for real-world governance models. If states were to combine the three design features in different ways, which resulting models would attract greater or lesser support? Fig.~\ref{fig:predsupport} presents the predicted support for the eight possible combinations of the three design dimensions.

The pooled estimates (panel a) show that safety-oriented regulatory designs are consistently preferred to innovation-oriented designs, regardless of how governance arrangements combine mode (government vs.\ tech firms) and level (international vs.\ national). The most preferred model combines safety-oriented objectives, government authority, and international collaboration. In contemporary AI governance, this model aligns closely with the EU's AI Act and the OECD's AI Principles (\textit{4,63}). By contrast, the least preferred model combines innovation-oriented objectives, private authority, and national solutions, approximating the current US approach, which relies heavily on self-regulation by the tech sector (\textit{5}). Other initiatives, such as the G7 Hiroshima AI Process, align with citizen preferences on safety and international scope but rely on voluntary self-regulation rather than the government authority that citizens favor (\textit{64}).

Panel (b) reports support for these models among three major AI actors: China, Europe (Germany and the UK), and the US. There is a relatively high degree of convergence across respondents in all three on the model combining safety, government authority, and international collaboration. At the same time, important differences emerge. Chinese respondents most strongly favor models centered on government authority at the national level, whether safety- or innovation-oriented. European and US respondents, by contrast, consistently prefer safety-oriented arrangements across combinations of authority and level, in line with the pooled results.

We further examine whether the ranking of regulatory profiles varies across the three application domains (workplace, policing, warfare).

\begin{figure}[H]
\centering
\includegraphics[width=0.85\textwidth]{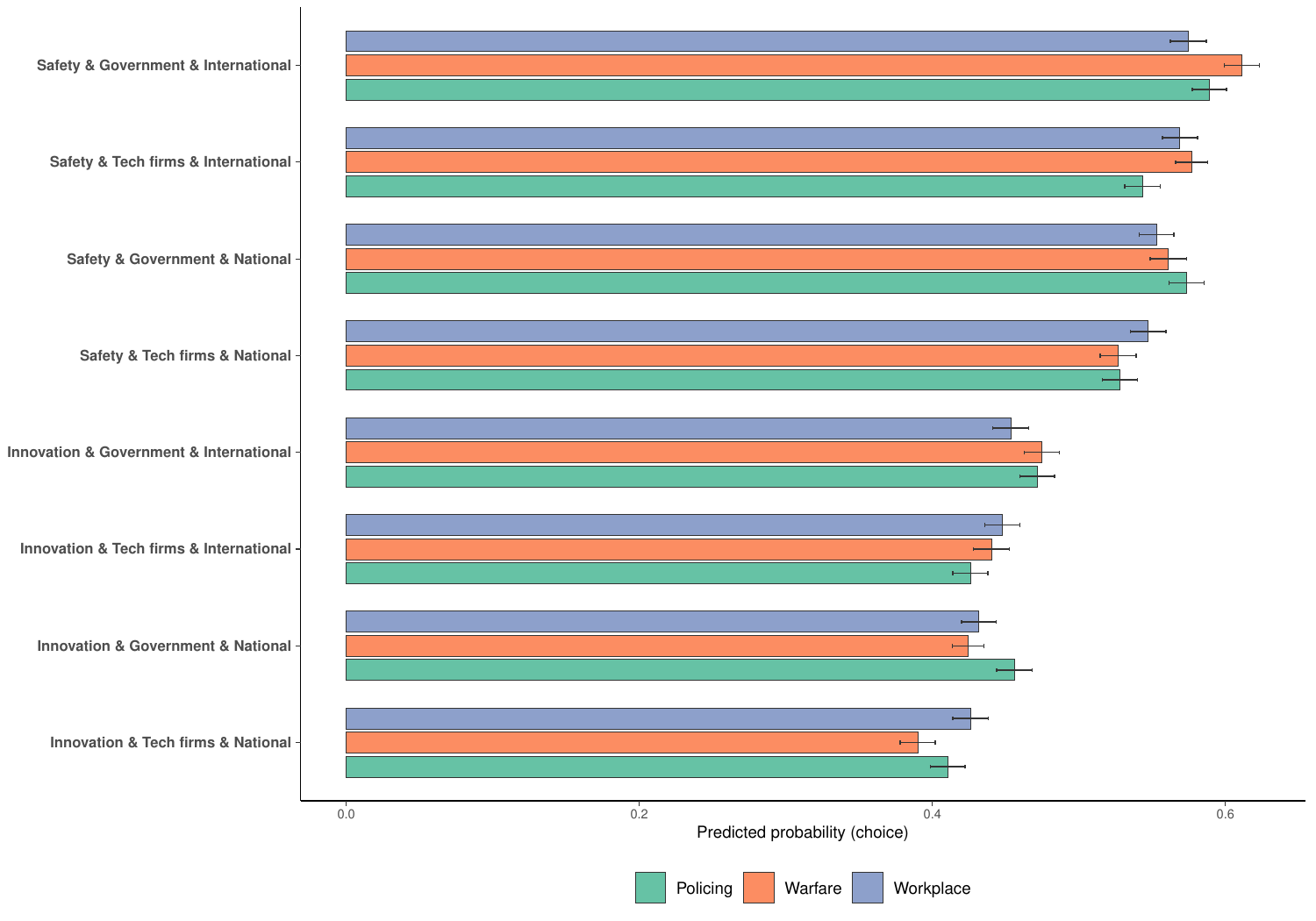}
\caption{\textbf{Predicted probability of choosing each regulatory profile, by domain of application.} Predictions derived from separate weighted linear probability models within each domain. Error bars indicate 95\% confidence intervals based on respondent-clustered standard errors. Profile ordering follows the pooled ranking in Fig.~\ref{fig:predsupport}.}
\label{fig:M1}
\end{figure}

Fig.~\ref{fig:M1} shows that the ranking of profiles is broadly stable across domains. Safety-oriented profiles consistently outperform innovation-oriented ones, regardless of the domain. The model combining safety, government authority, and international collaboration ranks highest in all three domains, while the model combining innovation, private authority, and national solutions ranks lowest. The main variation across domains concerns the magnitude of the gaps: preferences are more differentiated in the warfare domain, where the spread between the most and least preferred profiles is largest.

\clearpage

\subsection{Elite--Mass Analysis}
\label{sec:M}

We conduct an exploratory analysis of whether socioeconomic status moderates regulatory preferences. We define ``elite'' respondents as those who simultaneously have high education, high income, and professional/managerial occupations (N = 1,632), and ``non-elite'' respondents as those with none of these characteristics (N = 6,446). Respondents with intermediate profiles are excluded.

\begin{figure}[H]
\centering
\includegraphics[width=\textwidth]{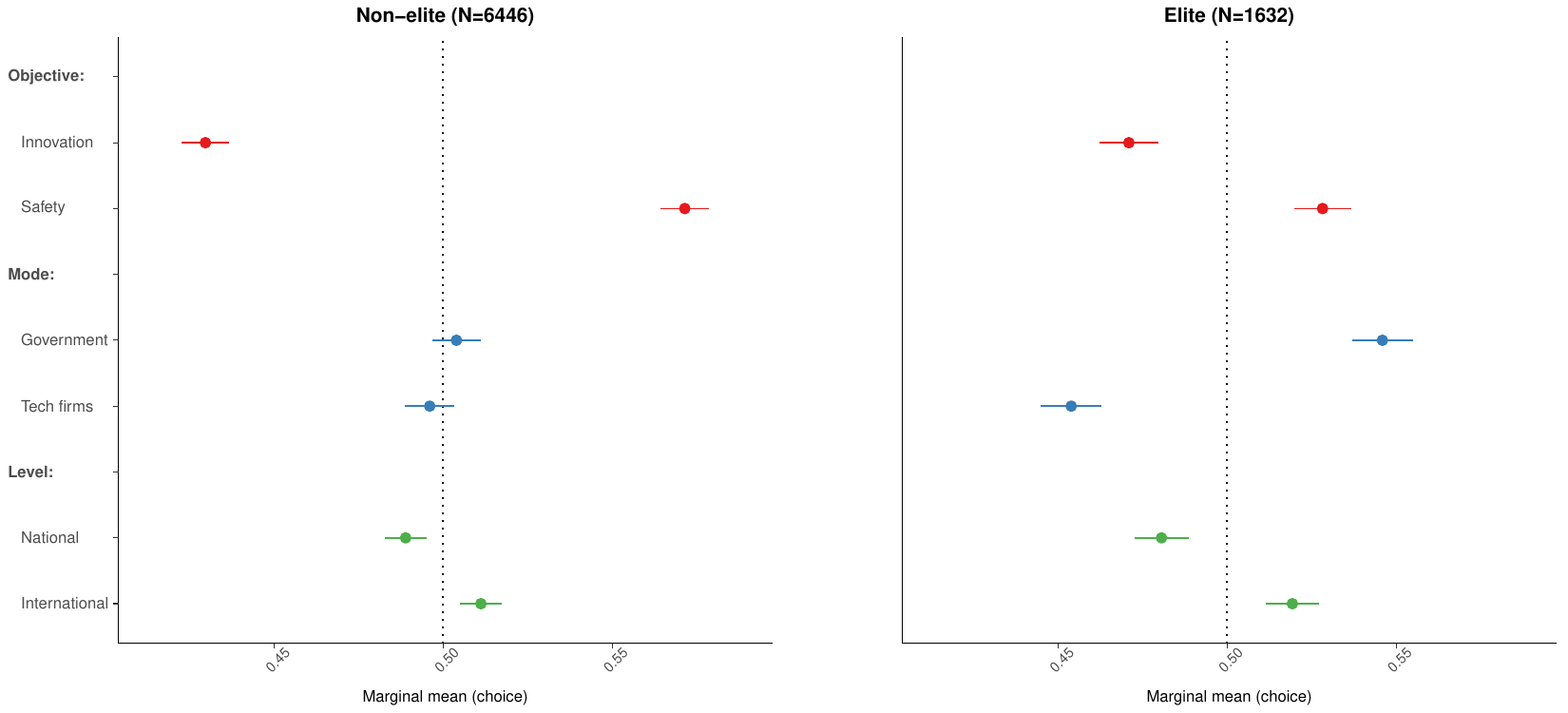}
\caption{\textbf{Marginal means for profile choice, by elite status.}}
\label{fig:N1}
\end{figure}

\begin{table}[H]
\centering
\caption{\textbf{Elite--mass interaction tests.} Positive coefficients indicate that elite respondents show a stronger effect of the attribute on choice probability.}
\label{tab:N1}
\scriptsize
\begin{tabular}{lrrrr}
\toprule
\textbf{Interaction term} & \textbf{Estimate} & \textbf{Std. error} & \textbf{$t$-value} & \textbf{$p$-value} \\
\midrule
Safety $\times$ Elite & $-$0.085 & 0.011 & $-$7.55 & $<$0.001 \\
Government $\times$ Elite & 0.086 & 0.012 & 7.23 & $<$0.001 \\
International $\times$ Elite & 0.015 & 0.010 & 1.50 & 0.135 \\
\bottomrule
\end{tabular}

\end{table}

We find that elite respondents are significantly less likely to prioritize safety over innovation (interaction $p < 0.001$), significantly more likely to prefer government over private authority ($p < 0.001$), and show no significant difference on the national--international dimension ($p = 0.13$; table~\ref{tab:N1}). These results suggest that while the broad preference structure is shared across socioeconomic groups, the intensity of preferences varies: elite respondents show a weaker safety preference but a stronger preference for public authority.

\clearpage

\subsection{Full Regression Output}
\label{sec:N}

This section reports the full regression output underlying the figures in the main text.

\subsubsection{Determinants of general support for AI regulation (Figure 1b)}

\begin{figure}[H]
\centering
\includegraphics[width=0.7\textwidth]{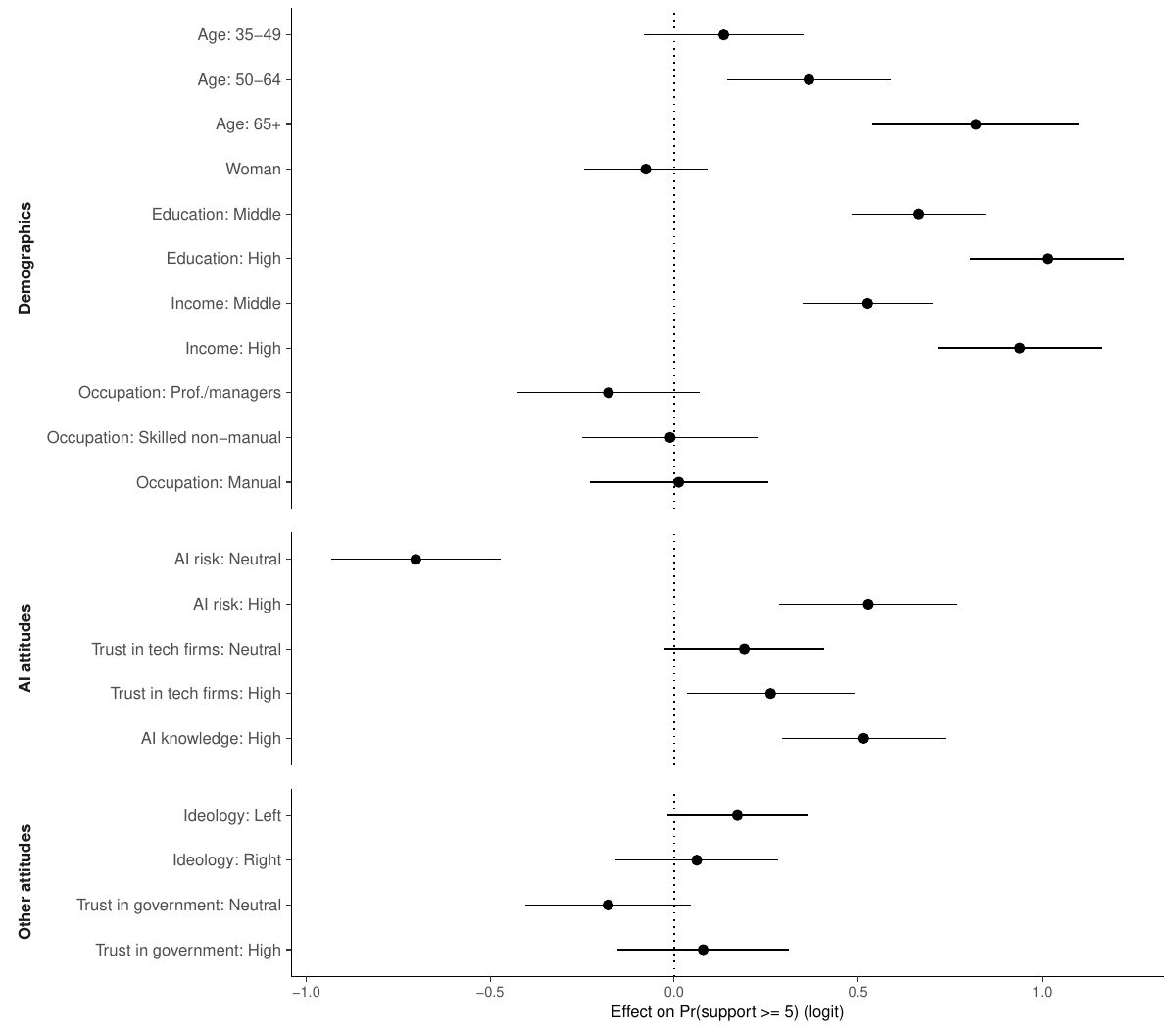}
\caption{\textbf{Logit model: Predictors of support for AI regulation ($\text{Pr}(\text{support} \geq 5)$).} Coefficient plot with 95\% confidence intervals. Reference categories same as Figure 1b in main text.}
\label{fig:O1}
\end{figure}

Figure 1b in the main text presents a coefficient plot summarizing the determinants of general support for AI regulation. Fig.~\ref{fig:O1} presents an alternative logit specification with a dichotomized outcome (support $\geq$ 5). Both models use survey weights, country fixed effects, and robust (HC1) standard errors. The two specifications produce consistent results: all variables that are significant in the OLS model are also significant in the logit model, with the same sign. Table~\ref{tab:O1} reports the full output for both specifications.

\begin{table}[H]
\centering
\caption{\textbf{Determinants of support for AI regulation (Figure 1b).} Reference categories: Age 18--34, Man, Education (Low), Income (Low), Not in labor force, AI risk (Low), Trust in tech firms (Low), AI knowledge (Low), Ideology (Center), Trust in government (Low), United Kingdom. N = 14,239. Standard errors in parentheses. $^{*}p<0.05$, $^{**}p<0.01$, $^{***}p<0.001$.}
\label{tab:O1}

\begingroup
\centering
\scriptsize
\begin{tabular}{lcc}
   \toprule
                                & Support for regulation & Pr(support $\geq$ 5)\\
                                & OLS                    & Logit \\
                                & (1)                    & (2)\\
   \midrule
   Constant                     & 4.409$^{***}$          & 0.0725\\
                                & (0.1330)               & (0.2207)\\
   \addlinespace
   Brazil                       & $-$0.180$^{**}$        & $-$0.345$^{**}$\\
                                & (0.086)                & (0.165)\\
   China                        & $-$0.296$^{***}$       & 0.311$^{*}$\\
                                & (0.087)                & (0.180)\\
   Germany                      & $-$0.369$^{***}$       & $-$0.426$^{***}$\\
                                & (0.073)                & (0.145)\\
   India                        & $-$0.567$^{***}$       & $-$0.615$^{***}$\\
                                & (0.132)                & (0.210)\\
   South Africa                 & $-$0.242$^{*}$         & $-$0.318\\
                                & (0.127)                & (0.207)\\
   United States                & $-$0.293$^{***}$       & $-$0.482$^{***}$\\
                                & (0.073)                & (0.144)\\
   \addlinespace
   Age: 35--49                  & 0.181$^{**}$           & 0.135\\
                                & (0.076)                & (0.110)\\
   Age: 50--64                  & 0.445$^{***}$          & 0.366$^{***}$\\
                                & (0.069)                & (0.114)\\
   Age: 65+                     & 0.601$^{***}$          & 0.820$^{***}$\\
                                & (0.079)                & (0.143)\\
   Woman                        & 0.012                  & $-$0.077\\
                                & (0.053)                & (0.086)\\
   Education: Middle            & 0.500$^{***}$          & 0.665$^{***}$\\
                                & (0.070)                & (0.093)\\
   Education: High              & 0.661$^{***}$          & 1.014$^{***}$\\
                                & (0.073)                & (0.107)\\
   Income: Middle               & 0.392$^{***}$          & 0.526$^{***}$\\
                                & (0.056)                & (0.090)\\
   Income: High                 & 0.623$^{***}$          & 0.940$^{***}$\\
                                & (0.068)                & (0.113)\\
   Occupation: Prof./managers   & $-$0.084               & $-$0.179\\
                                & (0.082)                & (0.127)\\
   Occupation: Skilled non-manual & $-$0.006             & $-$0.011\\
                                & (0.079)                & (0.121)\\
   Occupation: Manual           & $-$0.085               & 0.013\\
                                & (0.086)                & (0.124)\\
   \addlinespace
   AI risk: Neutral             & $-$0.212$^{***}$       & $-$0.702$^{***}$\\
                                & (0.071)                & (0.117)\\
   AI risk: High                & 0.634$^{***}$          & 0.528$^{***}$\\
                                & (0.074)                & (0.123)\\
   Trust in tech firms: Neutral & 0.062                  & 0.191$^{*}$\\
                                & (0.070)                & (0.111)\\
   Trust in tech firms: High    & 0.133$^{*}$            & 0.262$^{**}$\\
                                & (0.080)                & (0.116)\\
   AI knowledge: High           & 0.257$^{***}$          & 0.515$^{***}$\\
                                & (0.047)                & (0.114)\\
   \addlinespace
   Ideology: Left               & 0.088                  & 0.172$^{*}$\\
                                & (0.058)                & (0.097)\\
   Ideology: Right              & $-$0.044               & 0.062\\
                                & (0.068)                & (0.112)\\
   Trust in government: Neutral & $-$0.068               & $-$0.180\\
                                & (0.069)                & (0.114)\\
   Trust in government: High    & 0.123                  & 0.079\\
                                & (0.079)                & (0.119)\\
    \\
   Observations                 & 14,239                 & 14,239\\
   R$^2$                        & 0.149                  & \\
   \bottomrule
\end{tabular}
\par\endgroup

\end{table}

\subsubsection{Conjoint AMCE regressions}

Tables~\ref{tab:O_amce_pooled}--\ref{tab:O_amce_country} report the regression output underlying the conjoint analyses in the main text. All models are weighted linear probability models with respondent-clustered standard errors.

\begin{table}[H]
\centering
\caption{\textbf{AMCE regressions, pooled sample.} Choice: binary forced-choice outcome. Rating: 7-point support scale.}
\label{tab:O_amce_pooled}

\begingroup
\centering
\scriptsize
\begin{tabular}{lcc}
   \toprule
                           & Choice          & Rating \\
                           & (1)             & (2)\\
   \midrule
   Constant                & 0.4374$^{***}$  & 4.531$^{***}$\\   
                           & (0.0038)        & (0.0211)\\   
   Objective $=$ Safety    & 0.1251$^{***}$  & 0.2176$^{***}$\\   
                           & (0.0046)        & (0.0149)\\   
   Mode $=$ Tech firms     & -0.0286$^{***}$ & -0.0741$^{***}$\\   
                           & (0.0047)        & (0.0148)\\   
   Level $=$ International & 0.0292$^{***}$  & 0.0614$^{***}$\\   
                           & (0.0043)        & (0.0131)\\   
    \\
   Observations             & 256,302         & 256,302\\
   Respondents              & 14,239          & 14,239\\
   \bottomrule
\end{tabular}
\par\endgroup

\end{table}

\begin{table}[H]
\centering
\caption{\textbf{AMCE regressions by domain of application (choice outcome).}}
\label{tab:O_amce_theme}

\begingroup
\centering
\scriptsize
\begin{tabular}{lccc}
   \toprule
                           & Workplace      & Policing        & Warfare \\
                           & (1)            & (2)             & (3)\\
   \midrule
   Constant                & 0.4318$^{***}$ & 0.4560$^{***}$  & 0.4244$^{***}$\\   
                           & (0.0060)       & (0.0062)        & (0.0055)\\   
   Objective $=$ Safety    & 0.1212$^{***}$ & 0.1175$^{***}$  & 0.1366$^{***}$\\   
                           & (0.0073)       & (0.0070)        & (0.0068)\\   
   Mode $=$ Tech firms     & -0.0057        & -0.0457$^{***}$ & -0.0344$^{***}$\\   
                           & (0.0070)       & (0.0072)        & (0.0073)\\   
   Level $=$ International & 0.0217$^{***}$ & 0.0155$^{**}$   & 0.0502$^{***}$\\   
                           & (0.0069)       & (0.0070)        & (0.0068)\\   
    \\
   Observations             & 85,434         & 85,434          & 85,434\\
   Respondents              & 14,239         & 14,239          & 14,239\\
   \bottomrule
\end{tabular}
\par\endgroup

\end{table}

\begin{table}[H]
\centering
\caption{\textbf{AMCE regressions by country (choice outcome).}}
\label{tab:O_amce_country}

\begingroup
\centering
\scriptsize
\begin{tabular}{lccccccc}
   \toprule
                           & Brazil         & China           & Germany         & India          & South Africa   & United Kingdom  & United States \\
                           & (1)            & (2)             & (3)             & (4)            & (5)            & (6)             & (7)\\
   \midrule
   Constant                & 0.4175$^{***}$ & 0.5661$^{***}$  & 0.4030$^{***}$  & 0.4655$^{***}$ & 0.3804$^{***}$ & 0.3981$^{***}$  & 0.4284$^{***}$\\   
                           & (0.0076)       & (0.0061)        & (0.0060)        & (0.0129)       & (0.0151)       & (0.0106)        & (0.0061)\\   
   Objective $=$ Safety    & 0.0930$^{***}$ & 0.0128$^{*}$    & 0.2098$^{***}$  & 0.0304$^{*}$   & 0.1421$^{***}$ & 0.2223$^{***}$  & 0.1680$^{***}$\\   
                           & (0.0093)       & (0.0073)        & (0.0071)        & (0.0163)       & (0.0175)       & (0.0124)        & (0.0074)\\   
   Mode $=$ Tech firms     & 0.0262$^{***}$ & -0.1045$^{***}$ & -0.0570$^{***}$ & -0.0160        & 0.0296$^{*}$   & -0.0519$^{***}$ & -0.0283$^{***}$\\   
                           & (0.0091)       & (0.0074)        & (0.0077)        & (0.0157)       & (0.0180)       & (0.0154)        & (0.0076)\\   
   Level $=$ International & 0.0460$^{***}$ & -0.0403$^{***}$ & 0.0417$^{***}$  & 0.0542$^{***}$ & 0.0702$^{***}$ & 0.0337$^{**}$   & 0.0038\\   
                           & (0.0082)       & (0.0068)        & (0.0067)        & (0.0143)       & (0.0158)       & (0.0146)        & (0.0067)\\   
    \\
   Observations             & 37,188         & 36,648          & 36,180          & 37,080         & 36,468         & 36,522          & 36,216\\
   Respondents              & 2,066          & 2,036           & 2,010           & 2,060          & 2,026          & 2,029           & 2,012\\
   \bottomrule
\end{tabular}
\par\endgroup

\end{table}

\clearpage

\subsection{Moderator Variable Distributions}
\label{sec:O}

This section documents the distributions of key pre-treatment and post-treatment moderator variables used in the subgroup analyses (Section~\ref{sec:J}) and in the regression models (Section~\ref{sec:N}). Understanding these distributions is important for two reasons. First, highly skewed distributions can reduce statistical power for detecting subgroup differences. Second, cross-country variation in these distributions helps interpret country-level differences in conjoint preferences.

Table~\ref{tab:P2} reports pairwise Pearson correlations between the key moderator variables, pooled across all countries. Most correlations are modest ($|r| < 0.3$), indicating that the moderators capture distinct dimensions of respondent attitudes and are not redundant. The strongest correlation is between trust in government and trust in technology companies ($r = 0.44$), suggesting that institutional trust has a common component across targets. AI risk and AI unpredictability are positively but modestly correlated ($r = 0.21$), confirming that these two perceptions are related but distinct: respondents can view AI as risky without necessarily viewing its consequences as unpredictable, and vice versa.

\begin{figure}[H]
\centering
\includegraphics[width=\textwidth]{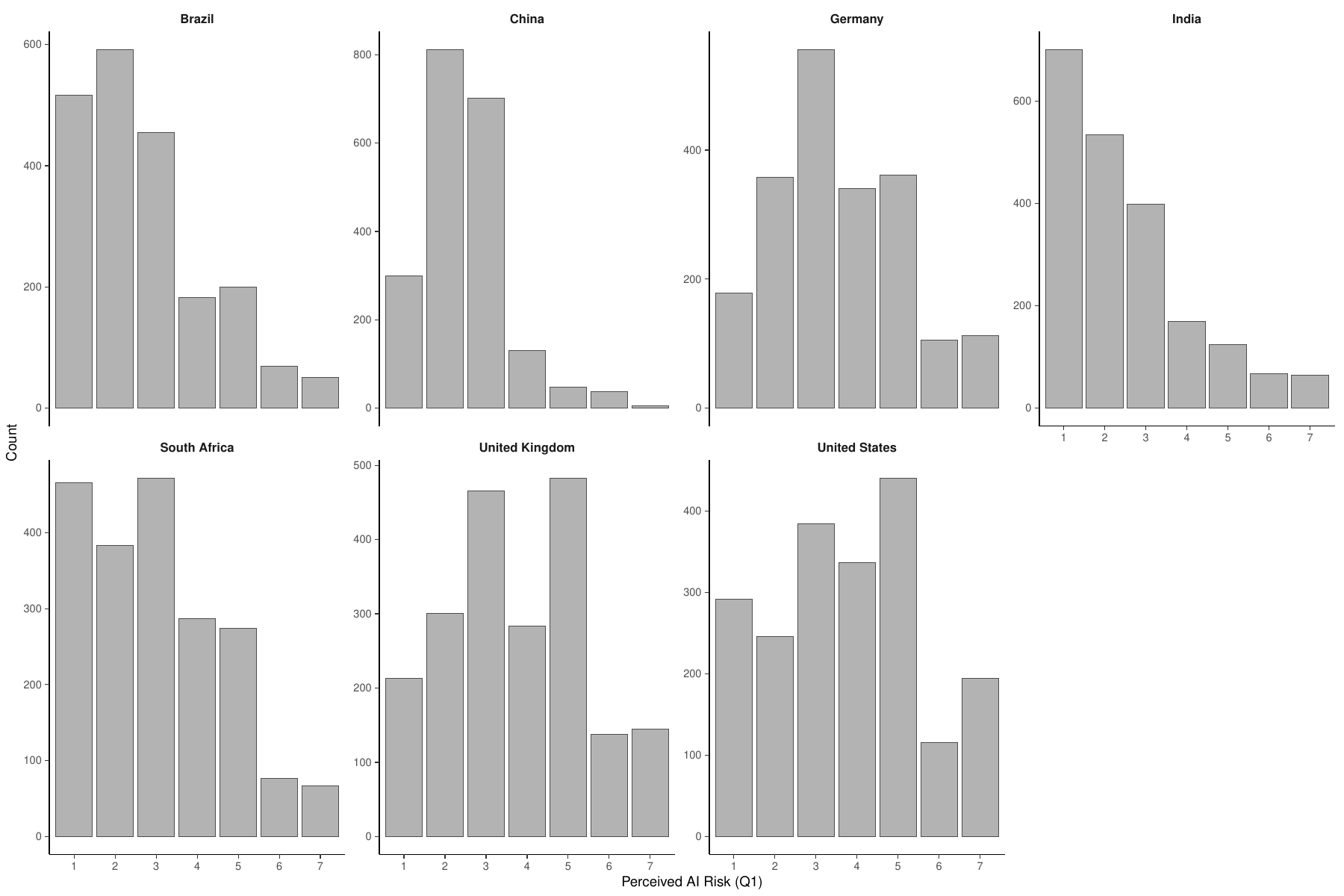}
\caption{\textbf{Distribution of perceived AI risk (Q1), by country.}}
\label{fig:P_risk}
\end{figure}

Figs.~\ref{fig:P_risk}--\ref{fig:P_trust_tech} present the country-level distributions of each moderator.  Perceived AI risk is skewed toward the opportunity end of the scale in most countries, with China standing out as the most optimistic. Internationalism varies considerably across countries, with Brazil and India showing relatively high levels and the United States showing a more even distribution. Trust in technology companies is generally higher than trust in government, a pattern that is especially pronounced in India and South Africa.

\begin{table}[H]
\centering
\caption{\textbf{Pairwise correlations between key moderator variables (Pearson $r$, pooled across countries).}}
\label{tab:P2}
\scriptsize
\begin{tabular}{lrrrrrrr}
\toprule
\textbf{} & \textbf{AI risk} & \textbf{AI unpred.} & \textbf{Support reg.} & \textbf{Internationalism} & \textbf{Trust gov.} & \textbf{Trust tech.} & \textbf{Ideology} \\
\midrule
AI risk & 1.00 & 0.21 & 0.11 & -0.09 & -0.26 & -0.37 & 0.10 \\
AI unpred. & 0.21 & 1.00 & -0.07 & -0.12 & -0.25 & -0.27 & 0.01 \\
Support reg. & 0.11 & -0.07 & 1.00 & 0.23 & 0.06 & 0.04 & 0.02 \\
Internationalism & -0.09 & -0.12 & 0.23 & 1.00 & 0.01 & 0.15 & -0.07 \\
Trust gov. & -0.26 & -0.25 & 0.06 & 0.01 & 1.00 & 0.44 & -0.07 \\
Trust tech. & -0.37 & -0.27 & 0.04 & 0.15 & 0.44 & 1.00 & 0.06 \\
Ideology & 0.10 & 0.01 & 0.02 & -0.07 & -0.07 & 0.06 & 1.00 \\
\bottomrule
\end{tabular}
\end{table}

\begin{figure}[H]
\centering
\includegraphics[width=\textwidth]{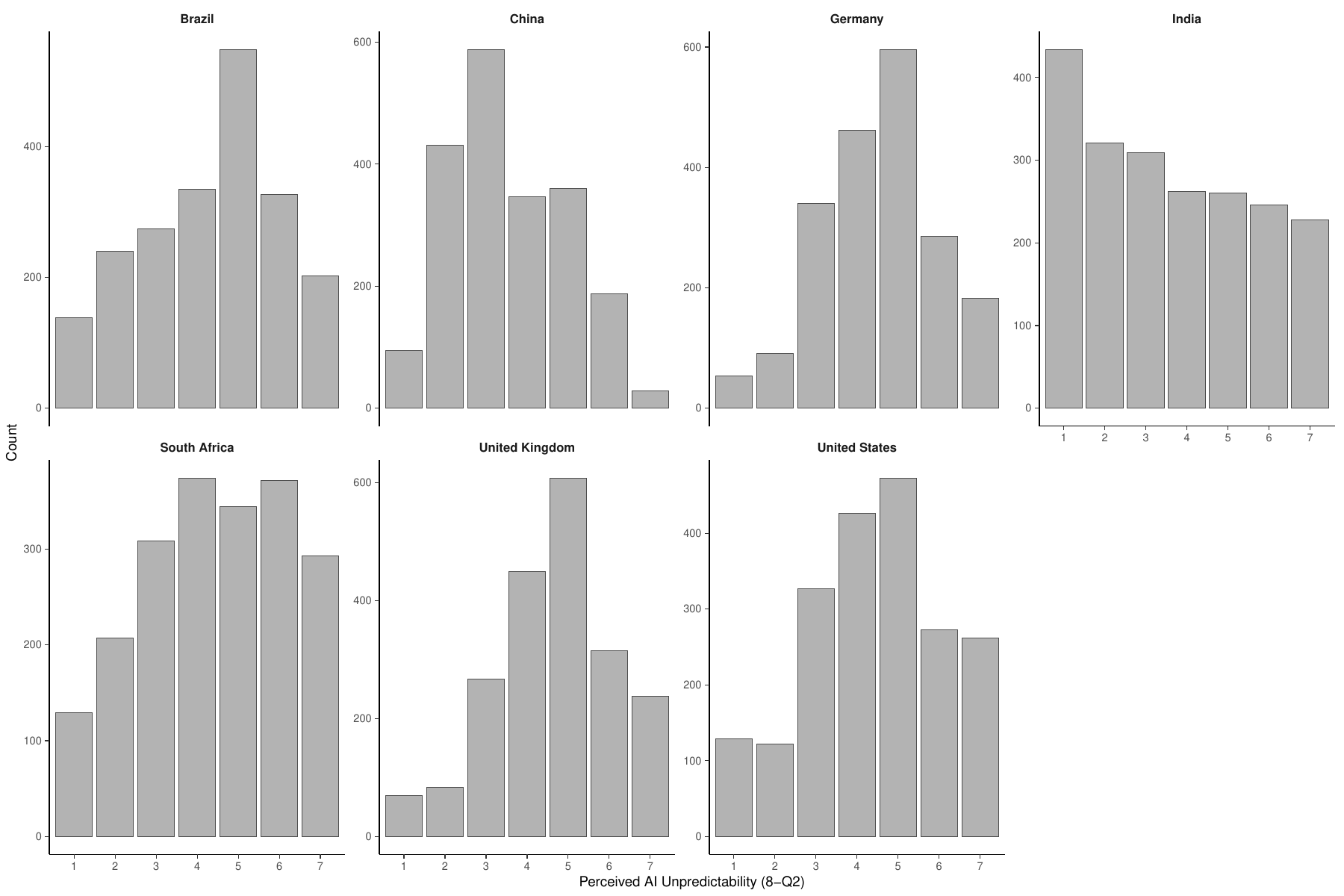}
\caption{\textbf{Distribution of perceived AI unpredictability (8$-$Q2), by country.}}
\end{figure}

\begin{figure}[H]
\centering
\includegraphics[width=\textwidth]{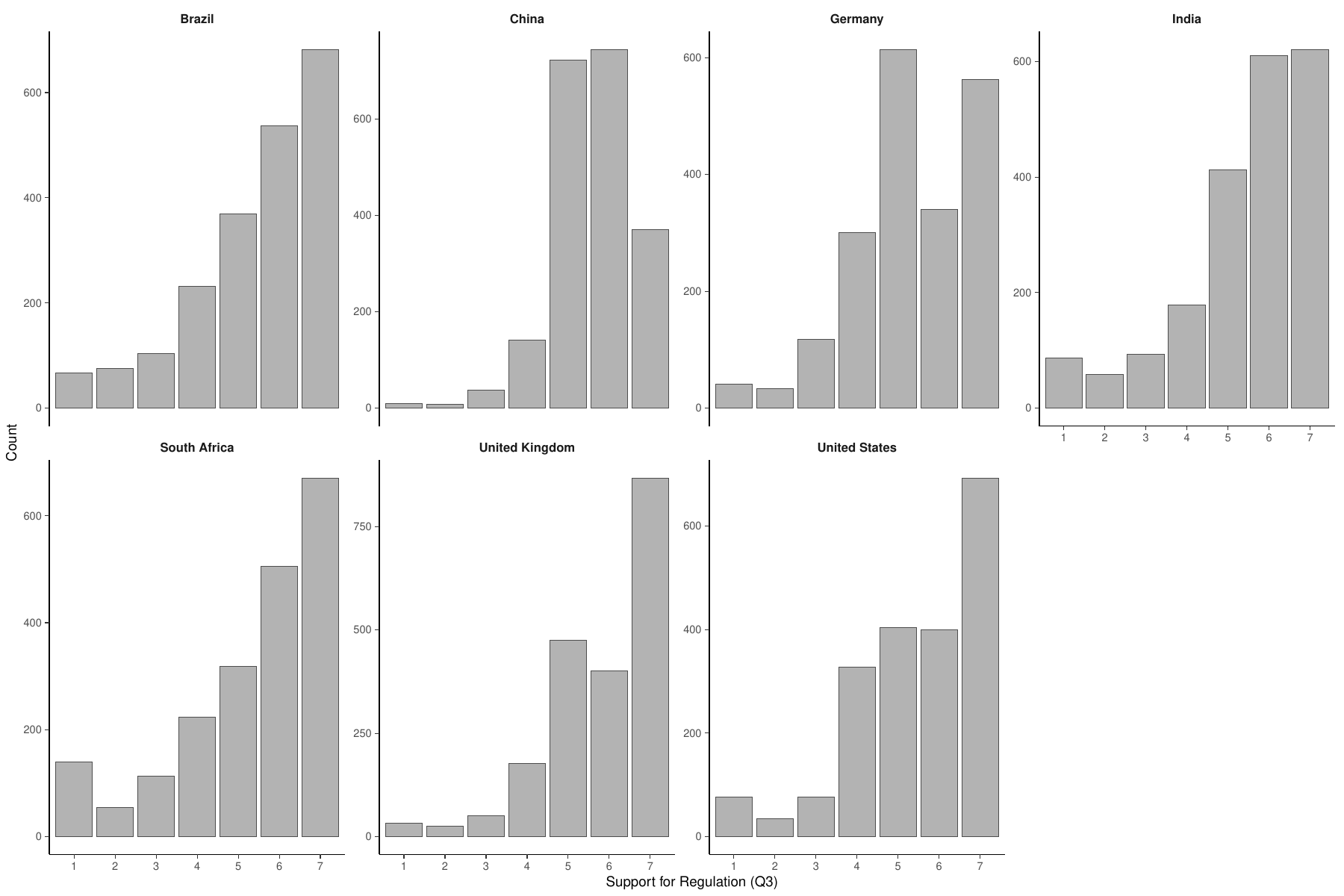}
\caption{\textbf{Distribution of support for AI regulation (Q3), by country.}}
\end{figure}

\begin{figure}[H]
\centering
\includegraphics[width=\textwidth]{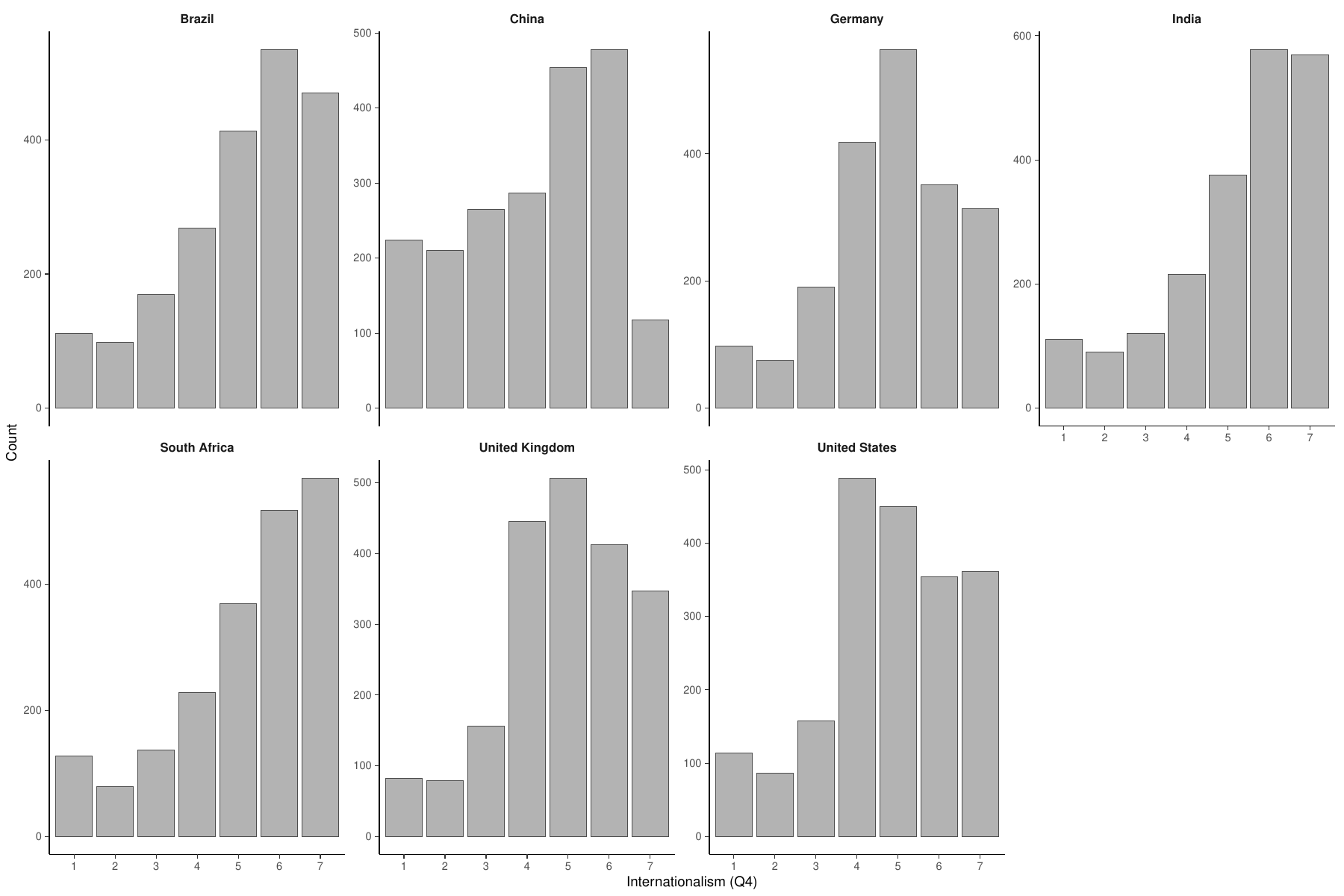}
\caption{\textbf{Distribution of internationalism (Q4), by country.}}
\end{figure}

\begin{figure}[H]
\centering
\includegraphics[width=\textwidth]{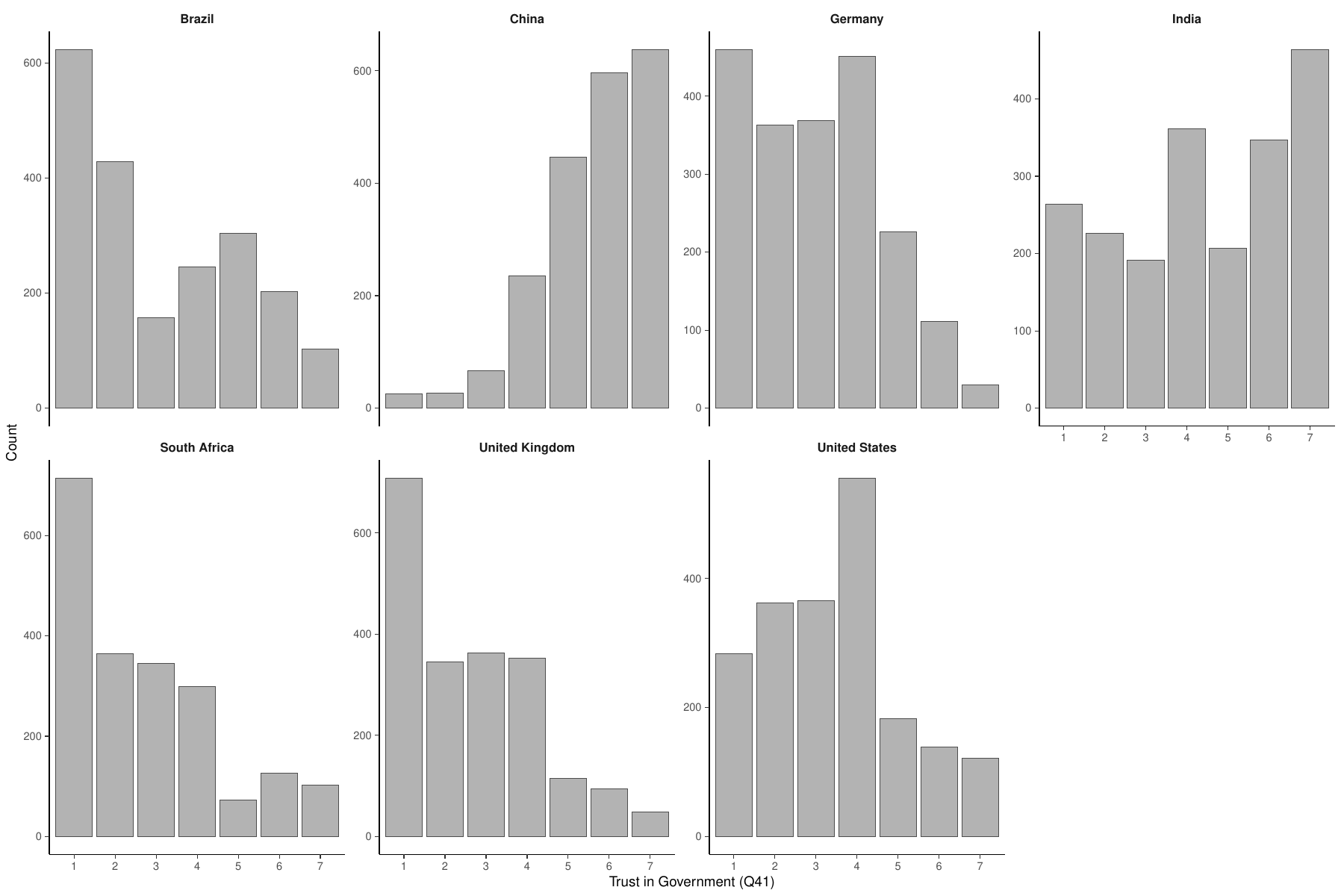}
\caption{\textbf{Distribution of trust in government (Q41), by country.}}
\end{figure}

\begin{figure}[H]
\centering
\includegraphics[width=\textwidth]{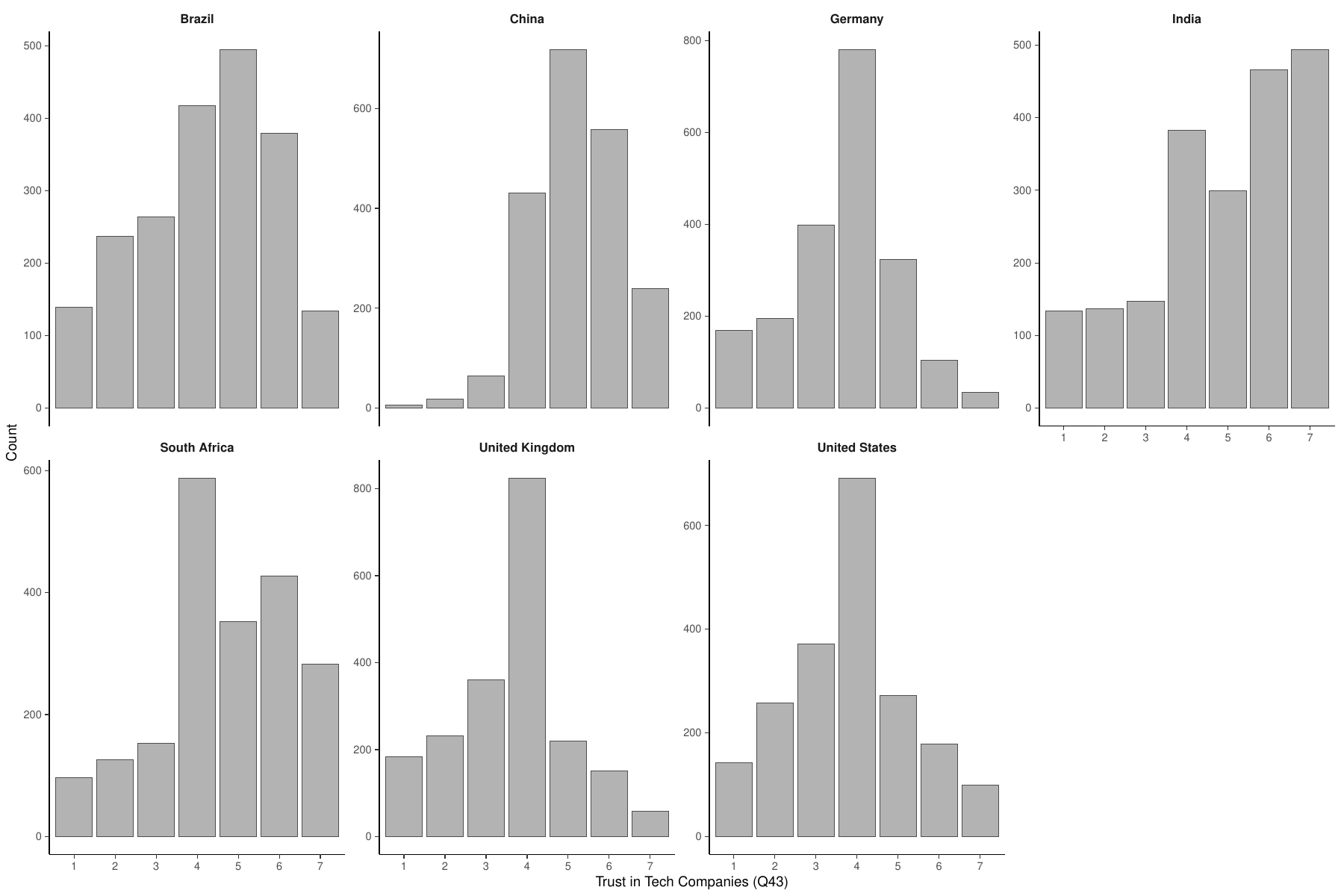}
\caption{\textbf{Distribution of trust in technology companies (Q43), by country.}}
\label{fig:P_trust_tech}
\end{figure}

\clearpage

\clearpage

\subsection{Full Survey Instrument}
\label{sec:survey_full}

The complete English-language questionnaire is reproduced below, following the pre-registered question numbering (Q1--Q48). Text in square brackets indicates country-specific adaptations (e.g., [COUNTRY] was replaced with the respondent's country name) or thematic area adaptations. Response options were presented on 7-point scales unless otherwise noted. The order of thematic blocks (workplace, policing, warfare) and the order of conjoint attributes were randomized across respondents but held constant within each respondent.

\small


\subsection*{Introduction and consent}

\noindent\textit{The following text was displayed to respondents at the start of the survey.}

\medskip
\noindent\textbf{Research project on the regulation of artificial intelligence (AI)}

\medskip
\noindent This survey is part of a research project on the regulation of artificial intelligence (AI) conducted by researchers at Stockholm University and the University of Gothenburg, Sweden. It is conducted in collaboration with Dynata and its partners. It has received ethics clearance by the Swedish Ethical Review Authority.

We are interested in what you think about AI and how this technology could be regulated. AI allows machines to do things that usually need human intelligence, like understanding language or making decisions. For instance, AI can guide self-driving cars, analyze medical data, and automate manufacturing.

The results of this survey will only be used for academic research. Your input will be treated as strictly confidential. There are no right or wrong answers---we only ask that you respond honestly and to the best of your knowledge.

\medskip
\noindent\textit{Consent:} Do you consent to the collection of this information? [Yes, I consent / No, I do not consent]


\subsection*{Pre-treatment items (Q1--Q5)}

\noindent\textbf{Q1.} Generally speaking, do you think that AI presents an opportunity or a risk to human society?
\begin{enumerate}
\item Always an opportunity
\item Usually an opportunity
\item More often an opportunity
\item Neither an opportunity nor a risk
\item More often a risk
\item Usually a risk
\item Always a risk
\end{enumerate}

\noindent\textbf{Q2.} Generally speaking, how predictable or unpredictable do you think that the consequences of AI are for human society?
\begin{enumerate}
\item Very unpredictable
\item Unpredictable
\item Rather unpredictable
\item Neither predictable nor unpredictable
\item Rather predictable
\item Predictable
\item Very predictable
\end{enumerate}

\noindent\textbf{Q3.} Generally speaking, to what extent do you agree or disagree that AI requires regulation?
\begin{enumerate}
\item Strongly disagree
\item Disagree
\item Somewhat disagree
\item Neither agree nor disagree
\item Somewhat agree
\item Agree
\item Strongly agree
\end{enumerate}

\noindent\textbf{Q4.} Generally speaking, to what extent do you agree or disagree that [COUNTRY] should work together with other countries to solve global problems, even if this reduces [COUNTRY'S] national sovereignty?

\noindent\textit{[Same 7-point agree/disagree scale as Q3]}

\medskip
\noindent\textbf{Q5. Attention check.} Some people get most of their news from traditional sources like TV programs or newspapers, whereas others go online or receive their news from social media. However, this question is actually not designed to tell us where you get your news. It is a standard attention check to determine if participants read questions until the end. Please only select radio and TikTok, and ignore the following question. What are your most important news sources? \textit{[Multiple selection: Television, Radio, Newspapers, News websites or apps, YouTube, Facebook, Instagram, WhatsApp, Telegram, TikTok, Other]}


\subsection*{Demographics (Q6--Q10)}

\noindent\textbf{Q6.} How old are you? \_\_\_ years

\medskip
\noindent\textbf{Q7.} What is your gender?

\noindent\textit{[Woman / Man / Other / Prefer not to say]}

\medskip
\noindent\textbf{Q8.} What is the highest educational level that you have attained (or expect to attain, if you have not yet completed your education)?

\noindent\textit{[Answer choices adapted to each country's education system and mapped onto ISCED levels 0--8, then recoded into three categories: Low (ISCED 0--2), Middle (ISCED 3--5), High (ISCED 6--8). Country-specific options available upon request.]}

\medskip
\noindent\textbf{Q9.} What is your approximate yearly income before taxes?

\noindent\textit{[Ten brackets denominated in local currency, adapted to each country's income distribution. Country-specific brackets available upon request.]}

\medskip
\noindent\textbf{Q10.} What is your primary occupation?

\noindent\textit{[Respondents selected from: Manager, Professional, Technician or associate professional, Clerical support worker, Service and sales worker, Skilled agricultural/forestry/fishery worker, Craft and related trades worker, Plant and machine operator/assembler, Elementary occupation, Armed forces occupation, Student, No occupation. Examples were provided for each category.]}


\subsection*{Conjoint experiment (Q11--Q34)}

\noindent\textit{The experiment was administered in three thematic blocks (workplace, policing, warfare), with block order randomized across respondents. Each block contained two pre-treatment questions on personal affectedness, followed by three paired conjoint tasks.}

\medskip
\noindent\textbf{General introduction shown to all respondents:}

\medskip
\noindent Countries around the world are currently discussing whether and how to regulate the development and use of AI. We are interested in what you think about regulating AI.

In the following, we present several ideas for how AI could be regulated. These proposals vary along three features:
\begin{itemize}
\item \textbf{Purpose of regulating AI:} Some proposals aim to strengthen people's safety while other proposals aim to strengthen technological innovation.
\item \textbf{Who is responsible for regulating AI:} Some proposals involve public regulation by government authorities while other proposals involve private self-regulation by technology firms.
\item \textbf{At what level AI would be regulated:} Some proposals involve rules set nationally by each country while other proposals involve rules agreed internationally by many countries.
\end{itemize}

\noindent We will ask you to compare proposals in three thematic areas: the workplace, policing, and warfare. In each area, we will first ask two questions about whether and how you are personally affected. You will then see three comparisons between different regulatory proposals.

\medskip
\noindent\textbf{Thematic block introductions:}

\medskip
\noindent\textit{AI in the workplace.} AI is increasingly used to automate tasks in the workplace. Examples include AI systems replacing human workers in call centers, warehouses, and areas like accounting, translation, and computer programming. While some see these technologies as a path to greater productivity and economic growth, others are concerned that AI could replace many jobs and disrupt labor markets.

\medskip
\noindent\textit{AI in policing.} AI is increasingly used in policing to detect and predict crime. Examples include facial recognition in public spaces, tools that predict where crime might occur, and automated systems for identifying suspicious behavior. While some see these technologies as a way to improve efficiency and security, others raise concerns about bias, over-policing, and violations of privacy.

\medskip
\noindent\textit{AI in warfare.} AI is increasingly used in warfare to identify and attack targets. Examples include attack drones, automated defenses, and missile systems that can operate with limited or no human oversight. While some see these technologies as a way to gain advantages on the battlefield, others raise concerns about human responsibility, arms races, and compliance with international laws of war.

\medskip
\noindent\textbf{Pre-conjoint questions (asked within each thematic block):}

\medskip
\noindent\textbf{Q11/Q19/Q27.} Generally speaking, do you think you will be personally affected by AI in [THEMATIC AREA]?
\begin{enumerate}
\item Not affected at all
\item Very little affected
\item Little affected
\item Rather little affected
\item Rather much affected
\item Much affected
\item Very much affected
\end{enumerate}

\noindent\textit{If respondents indicated options 2--7:}

\medskip
\noindent\textbf{Q12/Q20/Q28.} If you are personally affected by AI in [THEMATIC AREA], do you think the effect will be primarily positive or negative?
\begin{enumerate}
\item Always positive
\item Usually positive
\item More often positive
\item Just as much positive as negative
\item More often negative
\item Usually negative
\item Always negative
\end{enumerate}

\medskip
\noindent\textbf{Conjoint task format:}

\noindent\textit{Each task presented two hypothetical regulatory proposals side by side. Attribute levels were independently randomized (subject to the constraint described in Section~1.4.3).}

\medskip
\begin{tabular}{p{5cm}p{4cm}p{4cm}}
\toprule
\textbf{Features} & \textbf{Proposal 1} & \textbf{Proposal 2} \\
\midrule
The purpose of the proposed regulation would be to strengthen\ldots & \ldots people's safety against AI harms & \ldots technological innovation in AI \\
\addlinespace
The actors developing the proposed regulation would be\ldots & \ldots government authorities & \ldots technology firms \\
\addlinespace
The proposed regulation would regulate AI at the\ldots & \ldots national level & \ldots international level \\
\bottomrule
\end{tabular}

\medskip
\noindent\textbf{Q13/Q15/Q17/Q21/Q23/Q25/Q29/Q31/Q33 (Choice).} Which of these two proposals for regulating AI in [THEMATIC AREA] do you prefer?

\noindent\textit{[Proposal 1 / Proposal 2]}

\medskip
\noindent\textbf{Q14/Q16/Q18/Q22/Q24/Q26/Q30/Q32/Q34 (Rating).} To what extent do you support or oppose each of these two proposals for regulating AI in [THEMATIC AREA]?

\noindent\textit{Rated separately for each proposal:}
\begin{enumerate}
\item Strongly oppose
\item Oppose
\item Somewhat oppose
\item Neither support nor oppose
\item Somewhat support
\item Support
\item Strongly support
\end{enumerate}


\subsection*{Post-treatment items: AI knowledge (Q35--Q37)}

\noindent\textit{Here are some more questions about AI. Many people don't know the answers to these questions, but if you do, please indicate the answer that you believe is most correct.}

\medskip
\noindent\textbf{Q35.} Which of the following is an AI application?

\noindent\textit{[Fuel-injected internal combustion engine / USB flash-drive storage / \textbf{Face recognition used to unlock smartphone} / Electric pencil sharpener / I don't know]}

\medskip
\noindent\textbf{Q36.} Which of the following is a company known for developing AI?

\noindent\textit{[\textbf{Anthropic} / Maersk / RioTinto / Allianz / I don't know]}

\medskip
\noindent\textbf{Q37.} Which of the following is a feared consequence of AI?

\noindent\textit{[Sudden depletion of global helium reserves / \textbf{Mass production of convincing synthetic videos (``deepfakes'')} / Falling handwriting skills / Reversal of Earth's magnetic poles / I don't know]}

\subsection*{Post-treatment items: ideology (Q38--Q40)}

\noindent\textbf{Q38.} Generally speaking, do you think incomes in [COUNTRY] should be made more equal among people or depend more on individual effort?

\noindent\textit{[7-point scale from ``Be made much more equal'' to ``Depend much more on individual effort'']}

\medskip
\noindent\textbf{Q39.} Generally speaking, do you think ownership of business and industry should be public or private?

\noindent\textit{[7-point scale from ``Always public'' to ``Always private'']}

\medskip
\noindent\textbf{Q40.} Generally speaking, do you think that governments should make sure that everyone is provided for or that people should take responsibility to provide for themselves?

\noindent\textit{[7-point scale from ``Always governments'' to ``Always people themselves'']}

\subsection*{Post-treatment items: trust and attitudes (Q41--Q48)}

\noindent\textbf{Q41.} Generally speaking, how much or little confidence do you have in [COUNTRY'S] government?
\begin{enumerate}
\item Very little confidence
\item Little confidence
\item Rather little confidence
\item Some confidence
\item Rather much confidence
\item Much confidence
\item Very much confidence
\end{enumerate}

\noindent\textbf{Q42.} Generally speaking, how much or little confidence do you have in the United Nations (UN)? \textit{[Same scale as Q41]}

\medskip
\noindent\textbf{Q43.} Generally speaking, how much or little confidence do you have in technology companies? \textit{[Same scale as Q41]}

\medskip
\noindent\textbf{Q44.} Do you consider yourself more a citizen of your country or a citizen of the world?

\noindent\textit{[7-point scale from ``Much more a citizen of my country'' to ``Much more a citizen of the world'']}

\medskip
\noindent\textbf{Q45.} Generally speaking, how much or little trust do you have in other people?

\noindent\textit{[7-point scale from ``Very little trust'' to ``Very much trust'']}

\medskip
\noindent\textbf{Q46.} How satisfied or dissatisfied are you with the economic situation of [COUNTRY]?

\noindent\textit{[7-point scale from ``Very dissatisfied'' to ``Very satisfied'']}

\medskip
\noindent\textbf{Q47.} How satisfied or dissatisfied are you with the financial situation of your household? \textit{[Same scale as Q46]}

\medskip
\noindent\textbf{Q48.} How satisfied or dissatisfied are you with how the political system is functioning in your country these days? \textit{[Same scale as Q46]}

\subsection*{Debrief}

\noindent These are all the questions. Thank you very much for contributing to our research! This survey was part of an academic study conducted by researchers at the University of Gothenburg and Stockholm University, investigating attitudes toward AI policy and regulation across different countries.

\end{document}